\documentclass[fleqn,usenatbib]{mnras}
\usepackage{newtxtext,newtxmath, hyperref}
\usepackage{graphicx,amsmath}	
\usepackage[T1]{fontenc}
\usepackage{appendix}
\usepackage{amsmath}

\DeclareRobustCommand{\VAN}[3]{#2}
\let\VANthebibliography\thebibliography
\def\thebibliography{\DeclareRobustCommand{\VAN}[3]{##3}\VANthebibliography}

\def\xmm{\textit{XMM-Newton}\,}
\def\xmms{\textit{XMM-Newton}}

\title[XMM SuperRes]{Deep Learning-Based Super-Resolution and De-Noising for XMM-Newton Images}

\author[S.F. Sweere et al.]{
Sam F. Sweere,$^{1,2}$ \thanks{E-mail: samsweere@gmail.com}
Ivan Valtchanov,$^{3}$
Maggie Lieu,$^{4}$
Antonia Vojtekova,$^{2}$
Eva Verdugo,$^{2}$
Maria Santos-Lleo,$^{2}$
\newauthor
Florian Pacaud,$^{5}$
Alexia Briassouli$^{1}$
and Daniel Cámpora Pérez$^{1}$
\\
$^{1}$Faculty of Science and Engineering, Maastricht University, Maastricht, Netherlands \\
$^{2}$European Space Agency, ESAC, Camino Bajo del Castillo, 28692, Villanueva de la Ca\~nada, Madrid, Spain \\
$^{3}$Telespazio UK for European Space Agency, ESAC, Camino Bajo del Castillo, 28692, Villanueva de la Ca\~nada, Madrid, Spain \\
$^{4}$ School of Physics $\&$ Astronomy, University of Nottingham, University Park,
Nottingham, NG7 2RD, UK \\
$^{5}$ University of Bonn, 
Argelander Institut für Astronomie (AIFA), Auf dem Huegel 71, D-53121, Bonn, Germany
}

\date{Accepted XXX. Received YYY; in original form ZZZ}

\pubyear{2022}

\begin{document}
\label{firstpage}
\pagerange{\pageref{firstpage}--\pageref{lastpage}}
\maketitle

\begin{abstract}
The field of artificial intelligence based image enhancement has been rapidly evolving over the last few years and is able to produce impressive results on non-astronomical images. In this work we present the first application of Machine Learning based super-resolution (SR) and de-noising (DN) to enhance X-ray images from the European Space Agency's \xmm telescope. Using \xmm images in band [0.5,2] keV  from the European Photon Imaging Camera pn detector (EPIC-pn), we develop \textit{XMM-SuperRes} and \textit{XMM-DeNoise} --- deep learning-based models that can generate enhanced SR and DN images from real observations. The models are trained on realistic \xmm simulations such that \textit{XMM-SuperRes} will output images with two times smaller point-spread function and with improved noise characteristics. The \textit{XMM-DeNoise} model is trained to produce images with 2.5$\times$ the input exposure time from 20 to 50 ks. When tested on real images, DN improves the image quality by $8.2\%$, as quantified by the global peak-signal-to-noise ratio. These enhanced images allow identification of features that are otherwise hard or impossible to perceive in the original or in filtered/smoothed images with traditional methods. We demonstrate the feasibility of using our deep learning models to enhance \textit{XMM-Newton} X-ray images to increase their scientific value in a way that could benefit the legacy of the \xmm archive. 
\end{abstract}




\section{Introduction}

Over the last two decades, the European Space Agency \textit{XMM-Newton} \citep{jansen2001} X-ray space observatory has been continuously advancing our understanding of the cosmos through detailed observations of black holes, the formation of galaxies and many other phenomena in our X-ray sky \citep{santos2009, wilkins2021}. The 3 X-ray telescopes on-board are equipped with a set of imaging CCD detectors: European Photon Imaging Cameras (EPIC), with two MOS-CCD arrays \cite{Turner2001} and one pn-CCD \cite{Struder2001}. EPIC-pn has an effective area on average $\sim$2-3 times that of MOS, depending on the energy band. Concerning the characteristics of the cameras for imaging, they have comparable point-spread function (PSF) with Full Width at Half Maximum (FWHM) of $\sim$5-6\arcsec\ on axis (and half-energy width HEW of $\sim$14-15\arcsec), comparable field of view of $\sim$14-15\arcmin\ radius, and MOS detectors have pixel physical size of 1.1\arcsec, compared to 4.1\arcsec\ for pn (see e.g. \href{https://xmm-tools.cosmos.esa.int/external/xmm_user_support/documentation/uhb/}{the \xmm User Handbook}). The NASA's \textit{Chandra} X-ray telescope \citep{weisskopf2000} has a spatial resolution far superior to \xmm, with PSF HEW of the ACIS detector of $\sim0.5$\arcsec, limited by the physical pixel size, with a drawback of having much smaller effective area. It is desirable to have both good sensitivity and spatial resolutions: longer exposures allow collecting more photons and hence pick up fainter sources. However, the noise will also increase, thus there is a need for better sensitivity in order to detect extended emission. 

Most observations need to achieve a certain signal-to-noise (SNR) for the targets under study, in order to be able to draw scientific conclusions. This threshold dictates the exposure time that the observers request to the time allocation committees, and for observatories with high over-subscription rate asking too much time has its drawbacks, i.e. less chances for approval. Therefore, methods to enhance the SNR for a given observing time can be used to increase the science quality of the data.  Image enhancement through noise level reduction is a popular way to improve the SNR \citep[e.g.][]{vojtekova2020learning}. 

In X-ray observations, photon counts are subject to a Poissonian noise that is dependent on the count rate itself. Therefore, the SNR is smaller in low count rate areas, limiting the detection of faint sources. Binning of X-ray photons is one way to increase the total SNR, albeit at the cost of reducing spatial resolution. \cite{Sanders2001} use an adaptive binning method on Chandra observations of the Perseus cluster to reveal structure in the central region. \cite{bourdin2001} introduced a multi-scale wavelet transform approach to de-noise MOS1 and MOS2 images and were able to successfully recover the total flux and signal shape of toy-model sources, demonstrating that de-noising methods like these can be used to provide more accurate brightness mapping.

In addition to the noise, the PSF can lead to blending and sources confusion, when sources closer than a certain fraction of the PSF's FWHM can no longer be separated. Resolving and deblending such sources can be achieved with super-resolution. 

Super-resolution (SR) describes a class of methods that can upscale video or images from lower resolutions to higher ones. Such methods have been successfully demonstrated on astronomical imaging, e.g. \citep{starck2002deconvolution, puschmann2005, li2018}. Many methods for SR exist \citep[see  e.g.][]{zhou2012,siu2012}. Alternatively, edge-detection methods (see e.g. \citealt{sanders2016}) are being used for enhancing features and identifications of structures, although in general they do not increase the spatial resolution.

Traditionally, interpolation methods such as bilinear and nearest neighbour interpolation are used for upscaling. However, these methods often introduce side effects such as noise amplification and blurring. Furthermore, super-resolution on X-ray images imposes additional challenges since X-ray images are typically sparse, and the data are poisson distributed. Nevertheless, \cite{Feng2003} demonstrate using a direct demodulation (DD) method, the spatial resolution of \xmm EPIC images can be improved by a factor of 5 whilst adhering to the requirements for spectral studies. 

Super-resolution and de-noising is fundamentally an ill-posed problem since given a noisy/low-resolution input image, there are an infinite number of possible enhanced (high resolution) images that it could correspond to. The noisy, low resolution, input image inherently does not contain all the information of an enhanced image. However, in recent years significant progress has been made in the field of de-noising and super-resolution using machine learning methods \citep{yang2019,zhang2019zoom,wang2020deep,wang2018esrgan,chen2018learning,lugmayr2020srflow,dong2014learning,jain2008natural}. In these learning-based approaches, a network is trained with data to learn the mapping between an image and an enhanced image, where the enhanced image in our case is a higher SNR image and/or a higher resolution image. 

These models primarily make use of Fully Convolutional Networks (FCN), trained using a relevant quantitative metric used as a loss function. Similar to traditional convolutional neural networks \citep[CNNs,][]{LeCun1989, LeCun1998}, FCNs comprise of convolutional, pooling and layers, however they do not have dense layers \citep[see e.g.][for an in depth introduction to these components]{su2020deep} and their output size are typically the same or larger than the input. For this reason, FCNs are often used for computer vision tasks such as semantic segmentation, de-noising and super-resolution \citep{jain2008natural,dong2014learning,chen2018learning}.
Images generated this way tend to lack clarity as they often minimise a \textit{simple} loss function such as the mean absolute error (L1), that favors predicting the average over all plausible enhanced images. This leads to fewer finer details in the generated images. To address this, more recent approaches make use of more complex loss functions. The perceptual loss function \citep{ johnson2016perceptual,zhang2019zoom} incorporates style transfer through pre-training on a target dataset with a particular style or content. Generative Adversarial Networks (GANs) use two competing models - a generator to produce enhanced images from a given input image and a discriminator to differentiate between the real and generated images. Such networks make use of an adversarial loss \citep{wang2018esrgan} to obtain photorealistic images. In astronomy, these methods have been used to improve observations. \cite{Schawinski_2017} showed that using a GAN, they were able to recover features from artificially degraded optical observations. \cite{vojtekova2020learning} used a FCN and perceptual loss to de-noise Hubble Space Telescope images, improving the signal-to-noise ratio by a factor of 1.3-1.5, and \cite{lauritsen2021super} use an auto-encoder to obtain super-resolution of \textit{Herschel} observations in the sub-millimetre wavelength range.

This paper aims to apply these ideas and develop deep learning-based methods for super-resolution and de-noising of images from \xmm to increase their scientific value. The \xmm Science Archive contains observations spanning over 20 years and therefore there is ample data to satiate the training of a machine learning model. Improving the quality of this existing data is of great interest to the astronomical community and the lasting legacy of \xmms. In \autoref{sec:Data} we introduce the real and simulated data that are used to train and validate our method, we describe the different components in the simulated data sets and the pre-processing techniques. In \autoref{sec:Model} we define the models and the model architecture and we detail the loss functions and the evaluation metrics. The optimisation process of the models is presented in \autoref{sec:model_tuning} and the results on simulated and real observations are given in \autoref{sec:Results}. We discuss the results and put forward some caveats and limitations in \autoref{sec:Discussion} and we summarise with our conclusions in \autoref{sec:Conclusions}. More technical details on particular aspects of the work are separated in the Appendices.

\section{Data}\label{sec:Data}
To train and validate our models we created a dataset consisting of real \xmm observations (section \ref{sec:real_xmm}) and a separate dataset of simulated \xmm observations (section \ref{sec:sim_xmm}).

\subsection{Real XMM-Newton Dataset}\label{sec:real_xmm}
\xmm observations are in the form of \emph{eventlists} that record the time photons of a certain energy hit a specific CCD pixel. We need to transform these event lists into images, so in order to limit the scope of this research we focus on the EPIC-pn detector in the (extended) full-frame mode and build images using events with energy in band [0.5,2.0] keV.

We use the entire \xmm Science Archive (XSA), filtering out observations with less than 20~ks exposure times, bad time intervals and events. We split the event lists in 10ks intervals for each observation, i.e. for a 40~ks observation, we generate 4x10~ks images, 2x20~ks images, 1x30~ks and 1x40~ks image. The images with multiple exposure times enables us to train super-resolution and de-noising models using the same exposure-time for different observations. It also enables us to train a de-noising model with pairs of low and high exposure images. The exact implementation details of generating the real \xmm dataset are described in \autoref{app:real_dataset_gen}.

The final dataset contains 5554 unique EPIC pn exposures giving rise to the same number of full exposure images and almost 24000 sub-images after splitting the eventfiles into sub-images with multiples of 10ks exposure times. We only use 20ks, 50ks and 100ks sub-images in our final datasets (\autoref{tab:real_data_dist}). 

\begin{table}
\caption{The distribution of all the sub-images extracted from the total 5554 EPIC pn exposures as a function of their exposure time and their corresponding train, validate and test splits.}
\label{tab:real_data_dist}
\begin{tabular}{|c|c|c|c|c|}
\hline
Exposure time (ks) & Number of images & Train & Validate & Test \\ \hline
20                 & 5022             & 3489  & 775      & 758  \\ \hline
50                 & 834              & 583   & 123      & 128  \\ \hline
100                & 109              & 81    & 9        & 19   \\ \hline
\end{tabular}
\end{table}

\subsection{Simulated XMM-Newton Dataset}
\label{sec:sim_xmm}

The real \xmm dataset cannot be used to train a super-resolution model. To train a super-resolution model we need a data set consisting of low resolution input images and their high resolution counterparts as our targets. 

The creation of such a dataset, consisting of high and low resolution pairs, is often achieved digitally, through down-sampling high resolution images \citep[e.g.][]{wang2018esrgan,lugmayr2020srflow,Ledig_2017_CVPR,chen2022real} or optically, through aligning images taken with different optical zoom scales \citep[e.g.][]{Ledig_2017_CVPR, chen2022real,zhang2019zoom}. 
In this study we want to achieve higher than \xmm resolution images both spatially (smaller pixel size) and with smaller PSF size. Thus, the down-sampling approach is not an option as it would only decrease the pixel size and not the PSF.

Zooming is also not an option, since the \xmm telescope (a glancing reflector) cannot zoom. It would be possible to combine low resolution \xmm images with  higher spatial resolution images taken with another X-ray telescope such as \textit{Chandra}, however, we would be limited by the number of fields that have been observed by both \xmm and \textit{Chandra} and therefore there would not be enough data to train the model. Additionally, we note that most X-ray sources are variable, hence ideally we would need simultaneous observations, making the available data even more limited, and the different telescopes have different properties that need to be taken into consideration. 

An appropriate training dataset can be achieved through the use of simulations, where we can artificially increase the resolution (both the angular resolution and the sensor resolution) whilst maintaining the required observational properties of real \xmm\ images.

\subsubsection{Simulating \xmm\ Images}
For our simulations we use the SIXTE X-Ray simulation software package \citep{dauser2019sixte}. This is a X-ray simulation software package provided by ECAP/Remeis observatory\footnote{\url{https://github.com/thdauser/sixte}}. We create custom configuration files to resemble the \xmm\ EPIC-pn detector and individual events in the [0.5,2] keV energy band. This configuration provides realistic images with all important instrumental properties: vignetting, position-dependent PSF, background noise. We use a recent calibration file to replicate the vignetting properties of the \xmm\footnote{\url{https://xmmweb.esac.esa.int/docs/documents/CAL-SRN-0321-1-2.pdf}}.
For a detailed description of the simulation configuration see appendix \ref{app:simulation}.

We create two sets of simulated images: one with the actual \xmm\ PSF and another one with a rescaled PSF with twice the resolution (i.e. FWHM and HEW are two times smaller). As this study is a proof of concept we chose to be somewhat conservative and only increase the resolution twice. We keep the pixel size of the images in the second set at half  the size of the original (i.e. keeping the PSF sampling the same), hence they contain 4 times more pixels for the same field-of-view.

\begin{figure*}
    \centering
    \includegraphics[width=\linewidth, trim={0 3.cm 0 0}]{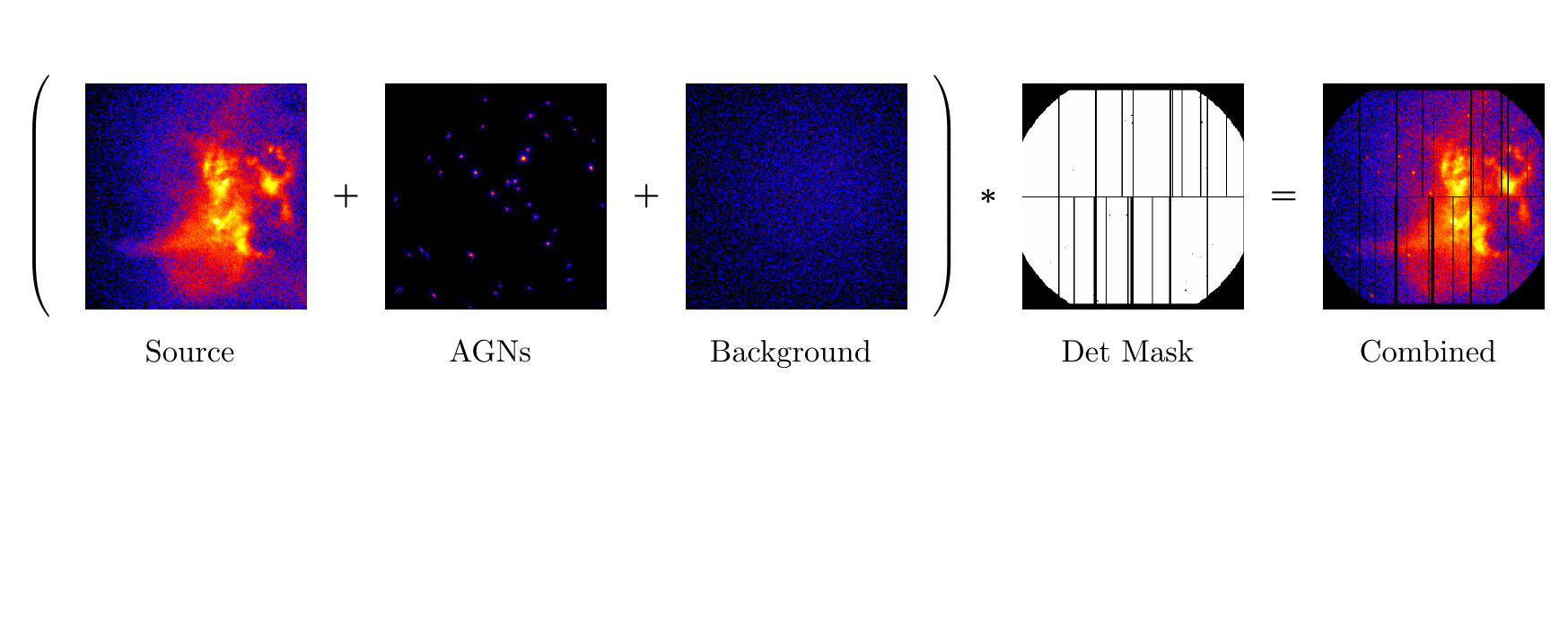}
    \caption{Different components that make up a simulated \xmm observation. The simulated extended source, AGNs and background are added together and then multiplied by the detector mask to create a simulated \xmm observation. The images are logarithmic scaled for visualisation.}
    \label{fig:sim_comb}
\end{figure*}

We focus on simulated observations of complex fields, like observations of galaxies or clusters of galaxies which provide images with three main components: the extended emission from the galaxy or the cluster, ``contaminant'' point sources which are mainly Active Galactic Nuclei (AGN) and the background \autoref{fig:sim_comb}). We will simulate these three components separately and describe them in the next sub-sections.

\subsubsection{Extended Source Component}
To simulate extended sources, we use the IllustrisTNG\footnote{\url{https://www.tng-project.org/}} suite of a large scale, cosmological magnetohydrodynamical simulations of galaxy formation. The three simulations we are using are IllustrisTNG 50-1 \citep{nelson2019first,pillepich2019first}, IllustrisTNG 100-1 and IllustrisTNG 300-1 \citep{springel2018first,marinacci2018first,nelson2018first,naiman2018first,pillepich2018first} at a redshift of 0.01. These are simulated at different scales (cubic volumes of $\sim$ 50, 100, and 300 Mpc side lengths) and mass resolutions that enable the study of different types of sources - supernova remnants at the smaller scale and galaxy clusters at the larger scale. The simulations include full baryonic physics. In each simulation, we select the top 400 subhalos based on the $M_{gas}$. We then project the subhalo from the x,y, and z axes on two different scales. We project at two different scales for a close-up of the source and a projection that is four times further away (tng50-1: 100 kpc and 400 kpc, tng100-1: 400 kpc and 1.6 Mpc, tng300-1: 1 Mpc and 4 Mpc) in order to capture different spatial information from the same source. From these projections, we calculate the X-ray photon intensity in the [0.5,2] keV energy range at redshift 0.01. X-ray dim subhalos are manually removed from our dataset. This results in:
\begin{itemize}
    \item TNG50-1: 1632 images
    \item TNG100-1: 2165 images
    \item TNG300-1: 2374 images
\end{itemize}

To convert the intensity maps from the TNG simulations to photon flux maps in $[0.5, 2.0]$ keV, we need to assume a spectral model. Galaxy clusters exhibit thermal spectrum, however, the subhalos from different TNG scales are not necessarily galaxy clusters. Therefore, for simplicity, we decided to use an absorbed power law. We use \texttt{XSPEC} \citep{arnaud1996xspec} to do this, assuming $N_H = 0.04\times 10^{22}$ cm$^{-2}$, and photon index $\Gamma=2$. The flux of the extended sources are further modified to reflect real extended sources that \xmm observed: we set the central part of the source (a box at 5$\%$ of the image width/height at the center of the image) to have a flux value randomly sampled uniformly to be between 5 and 50 times the standard deviation of background noise ($\sigma_B$) at the boresight. 

Additionally, to increase our training sample size, we artificially augment the data by apply a random zoom of scale $[1,2]$, and a random x, y perturbation offset using a standard distribution with a standard deviation of $5\%$ of the image height/width.

After the various augmentations are applied we are left with a total of 30855 augmented inputs that are used as the extended source component of the simulated \xmm observations.

\subsubsection{Point-Source Component}
The point source component of the observations are mostly AGNs. Based on measurements of AGN number counts $N (<S)$ as function of flux $S$ \citep{gilli2007synthesis}, we compute the expected number of AGNs in each observation corrected for the \xmm FOV \footnote{The expected number of AGNs was determined using the following tool: \url{http://www.bo.astro.it/~gilli/counts.html}}. \cite{gilli2007synthesis} do not publish their uncertainties, therefore we include an additional Poisson uncertainty $\pm\sqrt{N_\mathrm{AGN}}$.

We simulate AGN absorption at different Galactic latitudes by shifting the $\log(N)/\log(S)$ distribution down by a certain factor, i.e. dividing the flux $S$. We set the absorption factor to 100 at the Galactic plane ($b=0\deg$) and to 1 at the Galactic poles $|b|=90\deg$. We draw random absorption factors from log distribution from 1 to 100, such that we favour more extragalactic fields. For example, a field with absorption factor of 20 will have much less AGNs than a field with no absorption (absorption factor of 1).

\subsubsection{Background Component}
The background of the EPIC data has different components. The first component is due to the astrophysical sky background, from thermal low energy emission, unresolved cosmological sources and solar wind charge exchange. There is also particle induced background and finally electronic noise. The sky background is position dependent, while the quiescent particle background (QPB) is time-variable and correlates with the solar cycle. In general, in the $[0.5,2]$~keV energy band, the sky background level is $\sim2$ times higher than the QPB.

We decided not to simulate the individual background components but use the available Blank Sky event files\footnote{\url{https://xmm-tools.cosmos.esa.int/external/xmm_calibration/background/bs_repository/blanksky_all.html}} \texttt{pn\_t\_ff\_g} \citep{carter2007xmm} which contain all components as they are produced using real \xmm observations. The background component is simulated using the spectrum extracted from these Blank Sky event files.

\subsubsection{Co-Added Images}
Simulating the extended, point source and background components separately enables us to create many combinations of images. For example, if we look at uniquely simulated sources at 100ks, we have 30855 simulations of extended sources, 25000 simulations of point sources and 25000 simulations of background noise. This amounts to $30855 \times 25000 \times 25000 \approx 2 \times 10^{13}$ possible unique simulated pn images. We generate noisy images to use as inputs to both of our networks, paired with noiseless counterpart images as our DN network targets, and high-resolution, noiseless counterpart images as our SR network targets. Although many of these will be visually similar, the different combinations help reduce overfitting of the model. This compensates for the lack of traditional image augmentations such as spatial and colour transforms that would change the properties of the observation.

\subsection{Data Pre-Processing}\label{subsec:datapreprocessing}
To help accelerate the optimization of the model and more efficient convergence, the data need to be pre-processed. 

We transform the data from counts to counts/s by dividing the image by the exposure time. This enables us to use training data with different exposure times whilst maintaining the input pixel intensity distribution. 

Bright sources can have large pixel values that are orders of magnitudes higher than other pixels in a particular observation. This big difference can make training a deep learning model very unstable and therefore we clip pixel values to 200 times the mean background rate $\mu_B = 1.1168 \times 10^{-5}$ counts/s for the denoising data set and 50$\mu_B$ for the (2x) super resolution dataset\footnote{The clipping value for 2x is four times smaller because the pixel density is four times larger, meaning that the pixel counts on one pixel on the 1x resolution scale will be distributed over four pixels in the 4x resolution scale.}. This can lead to the loss of detail in bright regions, however, the majority of the extended features have X-ray counts below 200 times the mean background. 

The image is then normalized to $[0, 1]$. Even on normalised images, fainter features would not be visible to the human eye when visualizing them without a suitable data scaling (or stretch). Many interesting structures have pixel counts a few times above the background noise ($\sigma_b$), while bright parts of the image, such as centres of point-like sources, can have pixel counts in the hundreds of $\sigma_b$. 

\begin{figure}
    \centering
    \includegraphics[width=0.11\textwidth]{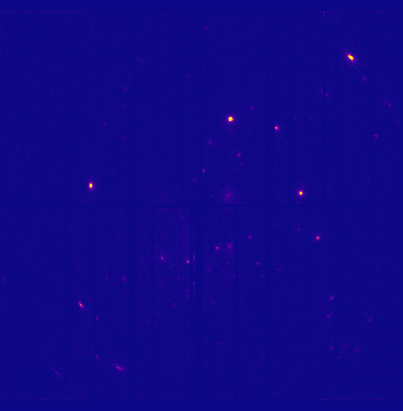}
    \includegraphics[width=0.11\textwidth]{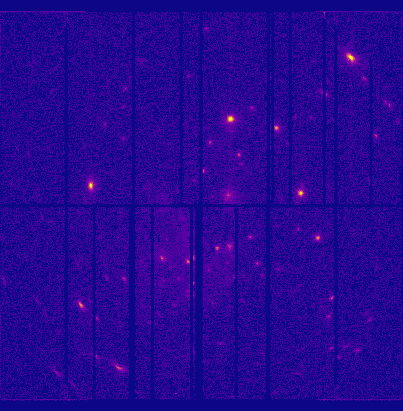}
     \includegraphics[width=0.11\textwidth]{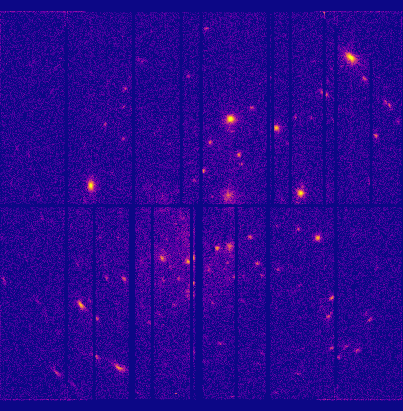}
    \includegraphics[width=0.11\textwidth]{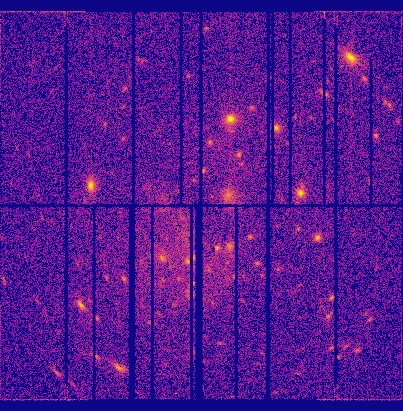}
    \caption{Examples of the different scaling applied to galaxy M101 (obs id: 0824450501). From \textit{left} to \textit{right}: linear, sqrt, asinh and log.}
    \label{fig:data_scaling_example}
\end{figure}

The pixel intensity distribution can also affect the training of the model. For example, an L1 loss, would put more weight on features with higher pixel values and bias the results. Our main focus is to enhance the visual clarity of faint details, and for this reason we explore several different data scaling functions. We compare linear, square root (sqrt), logarithmic (log) and hyperbolic arcsine (asinh) stretch functions. Each of these highlights different levels of the normalised pixel values, as shown in \autoref{fig:data_scaling_example}, with asinh lying between the sqrt and log stretch. 

\subsection{Train, Validation and Test split}
The final images are split into train, validation and test sets where only the training dataset is used to update the weights of the network, the validation data is used to monitor the performance of the network and the test data is reserved for final evaluation of network and is not seen by the network during training. 

For the simulated dataset, the splits are made in a way that all the spatial augmentations done during the simulations are always in the same set. Note that a specific source can appear multiple times across the sets but with different projections and distances. The choice to not split based on the sub-halos themselves was made because rare source structures could then be over-represented in one of the splits. Since different projections and distances of the same source look very different, this should not be an issue.

For each component (extended sources, point sources, background) of the simulated dataset, we split the distribution to have 80\%, 10\% and 10\%, train, validation and test subsets respectively. The number of sub-components in each subset is shown in \autoref{tab:sim_splits}.

Since the real dataset has a smaller number of images we need a larger percentage of the images to validate and test the results in comparison to the simulated dataset. The train, val, test split distribution chosen for real \xmm images is thus 70\%, 15\% and 15\% respectively. 

\begin{table}
\centering
\caption{The train, validation and test splits of the simulated sub-components.}
\begin{tabular}{|c|c|c|c|}
\hline
Component        & Train & Validation & Test \\ \hline
Extended Sources & 24678 & 3090       & 3087 \\ \hline
AGNs             & 20000 & 2500       & 2500 \\ \hline
Background       & 20000 & 2500       & 2500 \\ \hline
\end{tabular}

\label{tab:sim_splits}
\end{table}

\section{Method}\label{sec:Model}
\subsection{De-Noising and Super-Resolution Model}
For the de-noising model, the input image is at the default \xmm resolution with 20ks exposure time and the target image is similarly at the default \xmm resolution but with 50ks exposure time. We choose this combination of exposure times to replicate more realistic observations and to ensure our results are trustworthy. Having an exposure time of for example 100ks would force the model to make more uncertain predictions based on the input image. At 50ks exposure we also have more real world data to train and validate on.

The input to the SR model is the simulated \xmm image at the normal resolution with an exposure time of 20ks and background noise. For the target image we use the simulation image with 2x resolution, an exposure time of 100ks and without background noise. Omitting the background noise from the label image allows the model to concentrate on the source.

Originally, for our super-resolution problem, we took an approach based on a GAN architecture \citep{goodfellow2014}. GANs use a generator network to generate realistic images and a discriminator network to ensure that the generated images are visually indistinguishable from the high-resolution target images. We initially chose the ESR-GAN model \citep{wang2018esrgan} for its proven success as a super-resolution model. It consists of a stacked Residual-in-Residual Dense Block (RRDB, see \autoref{sec:RRDB}) generator and a deep CNN discriminator. However, like all GAN based algorithms, ESR-GAN suffers from hallucinations of non-existent features in the model output.
These hallucinations are caused by the discriminatory network that forces the generator output to be visually similar to the images in the training dataset. However, when for example, a part of the input image does not have sufficient information to generate a high resolution counterpart the model starts to hallucinate detailed features in order to generate a visually similar output image.
This can have catastrophic consequences in astronomy. 

For more robust reconstructions, and at the expense of generating images that are less visually similar to the target, we choose to omit the adversarial component. Our main model is therefore based on only the RRDB generator used in ESR-GAN. As we only use a generator model, our architecture is no longer a GAN. This architecture generates more reliable outputs, however is only able to generate sharp reconstructions in areas where the model learned to generate the features with high confidence, such as high signal-to-noise point sources. Note that the model does not output the confidence level of the generated features. Aspects of low confidence such as the background noise will result in blurry reconstructions and the output image is unlikely to visually resemble the target. 
The areas of low and high confidence are different for each image since they are dependent on the features present in the input image.

\begin{figure*}
    \centering
    \includegraphics[width=\linewidth]{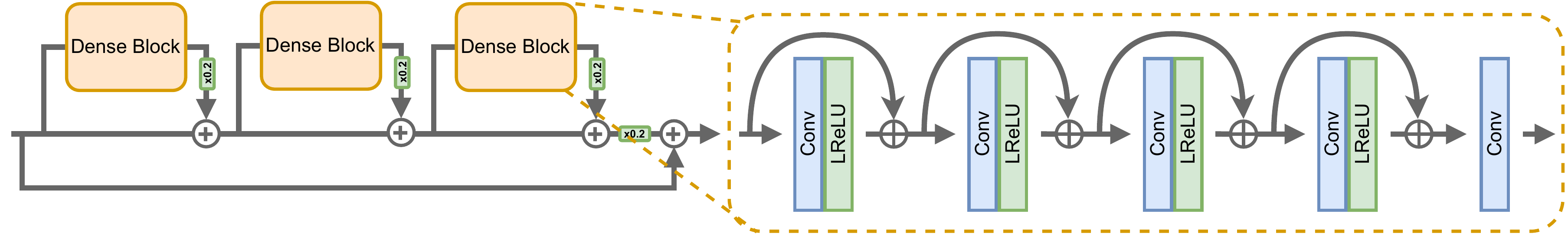}
    \caption{RRDB Basic Block, with $\oplus$ resembling concatenation. Adapted from \citep{wang2018esrgan}.}
    \label{fig:rrdb_block}
\end{figure*}

\subsection{Model Architectures}\label{sec:RRDB}
\begin{figure*}
    \centering
    \includegraphics[width=\linewidth]{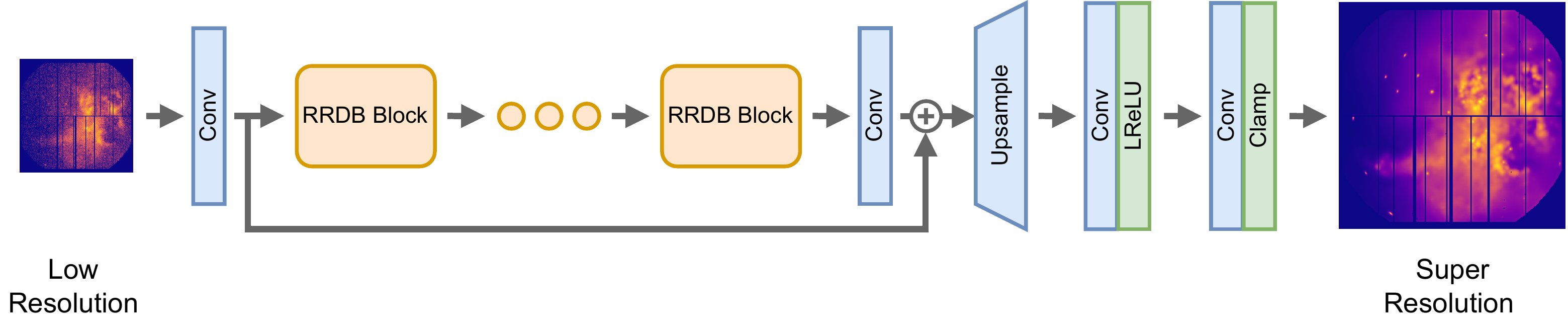}
    \caption{RRDB Super-Resolution Model Architecture. The network takes a low resolution image and undergoes a convolutional layer followed by a series of RRDB blocks, another convolutional layer with skip connections, an upsampling layer, and finally 2 more convolutional layers to return a higher resolution mapping.}
    \label{fig:sr_model_architecture}
\end{figure*}
 
The main feature of our architectures is the use of the RRDB block (\autoref{fig:rrdb_block}). This block is inspired by the DenseNet architecture \citep{iandola2014} and connects all layers within the residual block with each other. 
The RRDB block consists out of three Dense Blocks, within which contain 4 consecutive convolution layers each followed by Leaky ReLU activations and an additional convolutional layer. The concatenated output of every previous layer is fed into the next convolution layer. Thus the number of input channels  in every consecutive convolution layer increases linearly:
\begin{equation}
    N_{i+1} = N_{i} + N_{0} 
\end{equation}
Where $N_{i+1}$ is the number of input channels in the next layer, $N_{i}$ is the number input channels in the current layer and $N_{0}$ is the number of input channels in the first layer.

For our task of super-resolution, we base our architecture on the original ESR-GAN generator (\autoref{fig:sr_model_architecture}), however we replace the nearest-neighbour interpolation upsampling layer with pixel shuffle upsampling \cite{shi2016real}. Pixel shuffle has more connections and does not interpolate the upsampled image. This should improve quantitative details on smaller scales. 

\begin{figure*}
    \centering
    \includegraphics[width=\linewidth]{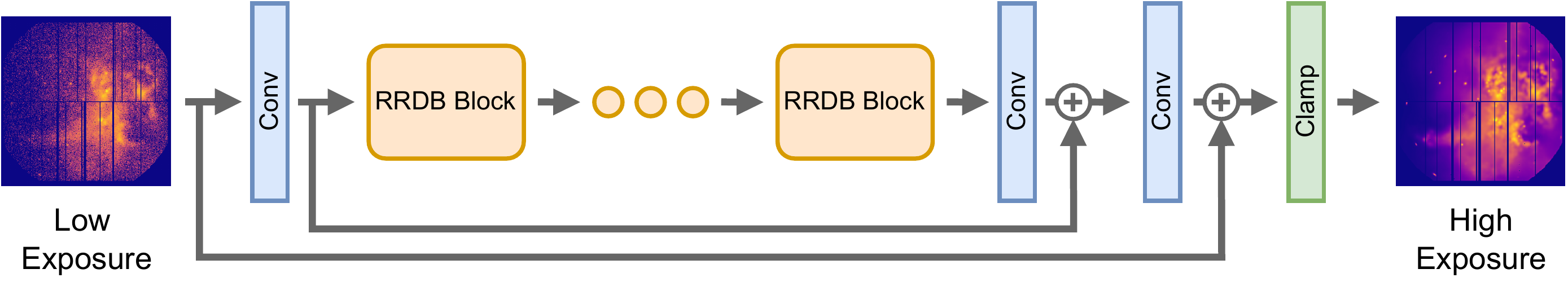}
    \caption{RRDB de-noise model architecture. The network takes a noisy low exposure image and undergoes a convolutional layer followed by a series of RRDB blocks, another 2 convolutional layer, with skip connections to output a higher SNR mapping.}
    \label{fig:dn_model_architecture}
\end{figure*}

For the denoising model, we use the same architecture however we remove the upsampling layer and the last convolutional layer (\autoref{fig:dn_model_architecture}). Additionally more skip connections are introduced as inspired by \cite{zhang2020residual}. This helps to learn smaller features in the image and improves training speed since the model does not have to process all the small features. Instead, it only learns the features to suppress to create the de-noised output image. 

\subsubsection{Weight Initialisation}
Since our output images have values between 0.0 and 1.0, we need to ensure that the first pass through the model results in values in this interval. If this does not happen, when for example the values are all negative, everything will be clipped to 0. This will result in no usable gradients for back-propagation, i.e. the model will not train. Therefore we skew the initial weights in the last convolution layer to be slightly more positive.

In general, the weights are initialized using a random normal distribution where the standard deviation is based on the size of the convolution layer:

\begin{equation}
    std = \frac{1}{\sqrt{\rm layer\, size}}
\end{equation}

In order to prevent the initial forward passes from being outside the image range, we initialize the weights in the last convolution layer to be uniformly distributed from $[-std, std + 0.01 \times std]$, this ensures the weights are slightly more positive. An alternative solution would be to explore  different final activation layers but this is beyond the scope of this work.  

\subsection{Loss Functions}\label{sec:loss_functions}
The loss function determines how good a prediction of the model is with respect to the reference image. In preliminary testing, we observe a substantial difference in the visual appearance of the generated images and the target images. However, we require a more quantitative measurement of the reconstruction than a simple visual comparison. Different loss functions optimise the model for different attributes of the output. We consider L1, Poisson, Peak Signal to Noise Ratio (PSNR), Structural Similarity Index \citep[SSIM,][]{wang2004SSIM}, and Multi-scale Structural Similarity Index \citep[MS-SSIM,][]{wang2003multiscale} loss functions. 

The L1 loss minimizes the mean absolute difference between pixel values of the generated and target images. This is the simplest loss function. We do not include the mean square error loss (L2) as it is sensitive to outliers and extreme values which can lead to bad performance on e.g. observations of AGNs. We include the Poisson loss, which measures the likelihood of the generated pixel values assuming that the target comes from a Poisson distribution conditioned on the input. It's relevant here because our data is count data and follows a Poisson distribution. The PSNR is a measure of the ratio between the maximum signal and the distorting noise. It is one of the basic metrics in denoising models. A higher PSNR value equates to better denoising. Lastly SSIM and MS-SSIM are perceptual metrics that incorporate the idea that spatially close pixels have strong inter-dependencies. These losses therefore measure the similarity of structure in images on a single scale and a combination of different scales respectively. The SSIM and MS-SSIM have parameters that needed to be fine-tuned for our problem. For SSIM we empirically found the following parameters to work well: $window$ $size = 13$, $\sigma = 2.5$, $K_1 = 0.01$ and $K_2 = 0.05$.  For MS-SSIM we used the same parameters as for SSIM with the weight for each scale being $[0.0448, 0.2856, 0.3001, 0.2363, 0.1333]$.

To meaningfully combine loss functions, we need to normalize them since the values of different loss functions can differ by orders of magnitude. 
We normalize the loss based on trial runs of the model trained with Poisson loss with the various data scaling functions and an untrained model. A model trained with a loss function different from the Poisson loss will generate different output images. These will have different loss values. However, the difference between a trained and untrained model with any of our loss functions will be huge. Therefore, the final loss metrics should be approximately on the same scale.

We aim to have the normalized loss value of the untrained model at 1 and the trained model at 0. We calculated the normalization with the following formula:
\begin{equation}\label{eq:lnorm}
    L_{normalized} = \alpha \ L_{unnormalized} + \beta
\end{equation}
With:
\begin{align}\label{eq:slope}
    \alpha &= \frac{y_2 - y_1}{x_2 - x_1} \\
    \beta &= y_1 - a\ x_1
\end{align}
Where $y_1$ is the target loss value for the untrained model (in our case $y_1 = 1$), $y_2$ is the target loss value for the trained model (in our case  $y_2 = 0$), $x_1$ is the measured loss of the untrained model and $x_2$ is the measured loss of the trained model.
We can now combine different loss functions by adding the normalized loss functions together since the loss functions are now on the same scale.

\begin{figure}
    \centering
    \includegraphics[width=0.24\linewidth]{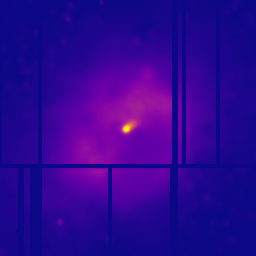}
    \includegraphics[width=0.24\linewidth]{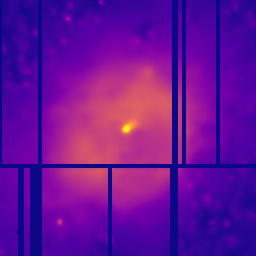}
    \includegraphics[width=0.24\linewidth]{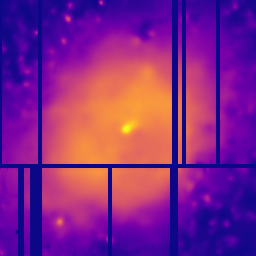}
    \includegraphics[width=0.24\linewidth]{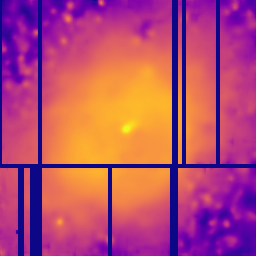}
    \includegraphics[width=0.24\linewidth]{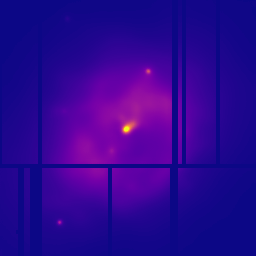}
    \includegraphics[width=0.24\linewidth]{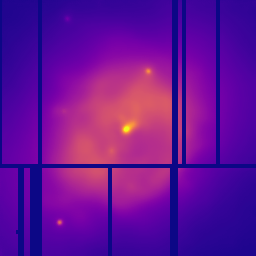}
    \includegraphics[width=0.24\linewidth]{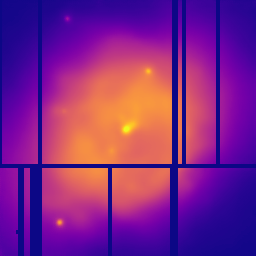}
    \includegraphics[width=0.24\linewidth]{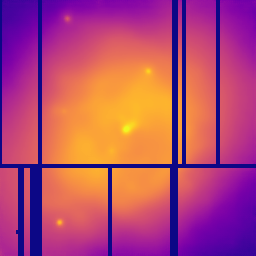}
    \caption{A model trained at linear data scale (top row) and a model trained at the sqrt data scale (bottom row). The display data scales from left to right: linear, sqrt, asinh, and log. At the asinh and log data scale, the \textit{blobs} generated by the linear trained model are not present with the sqrt trained model.}
    \label{fig:lin_scaling_prob}
\end{figure}
 
\subsection{Evaluation Metrics}\label{sec:imagemetrics} 
To evaluate our models, in addition to the loss metrics discussed in \autoref{sec:loss_functions}, we make use of the Feature Similarity Index \citep[FSIM,][]{zhang2011fsim} and the Haar wavelet-based Perceptual Similarity Index \citep[HaarPSI,][]{reisenhofer2018haar} metrics.

In a similar fashion to SSIM and MS-SSIM, FSIM is a metric that aims to mimic human vision. The human visual system perceives images through salient low-level features, and FSIM uses 2 kinds of these features to determine image quality - the phase congruency (PC) and the gradient magnitude (GM). Rather than the areas with sharp changes in contrast, PC highlights features as areas where the order in the phase component of the Fourier transform is high. Thus PC is an illumination and contrast invariant measure of feature significance. However since contrast is also an important aspect of human vision, FSIM also incorporates gradient magnitude to encode contrast information. 

HaarPSI uses coefficients obtained from a discrete wavelet transform to construct local similarity maps between two images. The Haar wavelet is used, being the simplest and  most efficient to compute. Next an non-linearity is applied in the form of a logistic function to highlight the relative importance of those areas. 

\section{Model Optimisation}\label{sec:model_tuning}
Hyper-parameters influence the training of a model and its performance. To tune for the optimal configuration we perform a parameters search where we trained many models with different hyper-parameters to gain insight into the influence of each hyper-parameter on the model performance. There are two categories of hyper-parameters: the model hyper-parameters (\autoref{sec:modelparameters}) and the data hyper-parameters (\autoref{sec:dataparameters}). The model hyper-parameters are tuned first and fixed before the data hyper-parameters are tuned. 

We use a grid-search approach to hyper parameter tuning, which can be computationally expensive and therefore we only train on a $25\%$ subset of the simulated dataset where the inputs are further cropped to 128x128 pixels around the boresight. We train the models for 50 epochs on this reduced dataset. Although this is slightly different from the final model training, we argue that it gives enough insight into the model performance to make informed choices on the hyper-parameters used in the final model.

\subsection{Model Hyper-Parameter Tuning}\label{sec:modelparameters}
The model hyper-parameter-search aims to optimise parameters based on the model's learning ability. For this sweep, we use a Poisson loss with square root data-scaling since this resulted in desirable results in initial testing. 
We train models with a range of combination of hyper-parameters (180 models) and monitor the loss of the validation data. For exact details see \autoref{app:modelparameters}. The final model hyper-parameters are shown in \autoref{tab:final_hyper_params}.

\begin{table}
\centering
\caption{Final model hyper-parameters.}
\begin{tabular}{|l|l|}
\hline
Hyper-parameter & Value                      \\ \hline
RRDB convolutional filters & 32                      \\ \hline
RRDB blocks & 4                       \\ \hline
Batch size                & 1                       \\ \hline
Learning rate             & 0.0001                  \\ \hline
Data scaling              & square root                   \\ \hline
Loss function            & PSNR and MS\_SSIM \\ \hline
\end{tabular}
\label{tab:final_hyper_params}
\end{table}

\begin{table*}
\centering
\begin{tabular}{|l|l|l|l|l|}
\hline
         Metric & Input                           & Simulated Data                 & Real Data & \multicolumn{1}{c|}{\begin{tabular}[c]{@{}l@{}}Fine-Tuned\\ (\textit{XMM-DeNoise})\end{tabular}} \\ \hline
L1       & 0.006528  & 0.005202                       & 0.004628  & 0.004408                                                        \\ \hline
PSNR     & 39.349  & 41.728                         & 42.227    & 42.693                                                          \\ \hline
Poisson  & 0.07616 & 0.04782                        & 0.04856   & 0.04778                                                         \\ \hline
SSIM     & 0.9484                          & 0.9359 & 0.9512    & 0.9567                                                          \\ \hline
MS\_SSIM & 0.9922                          & 0.9910  & 0.9930     & 0.9939                                                          \\ \hline
FSIM     & 0.9688                          & 0.9577 & 0.9745    & 0.9783                                                          \\ \hline
HaarPsi  & 0.8879  & 0.9006                         & 0.9139    & 0.9253                                                          \\ \hline
\end{tabular}

\caption{Various de-noising models to test the influence of the training data used when applied on the real test set. The input column refers to the direct comparison between the real input data and target image. We show the results for models trained on simulated data, real data and simulated then fine-tuned to real data compared to the target. The rows correspond to the different metric scores when applied to the real data test set.}
\label{fig:denoise_metrics}
\end{table*}

\subsection{Data Hyper-Parameter Tuning}\label{sec:dataparameters}
Having determined the model hyper-parameters we tune the hyper-parameters that influence the visual properties of the generated images: the loss function and data scaling. Here, we fix the model hyper-parameters to the optimal values determined in \autoref{sec:modelparameters} however the batch size used is set to 4 to increase the training speed. We train a model for all possible combination of loss functions (\autoref{sec:loss_functions}) and data scalings (\autoref{subsec:datapreprocessing}, 128 models). 

To determine the optimal data hyper-parameters we visually compare the generated images and their image quality metrics. Since we cannot consistently inspect the thousands of generated images, we first select the best performing models based on the evaluation metrics and before deciding the final data hyper-parameters based on both a quantitative and qualitative visual inspection. For the exact details of this process see \autoref{app:dataparameters}.

We correlate each hyper-parameter with the combined metric score and find that the sqrt data scaling performs the best. Fixing the image scaling to sqrt, we then determine the optimal loss function.

Based the models performance on the image quality metrics and visual inspections of the generated images we chose to train the final model with the PSNR combined with MS\_SSIM loss function. Since the PSNR ($L_{\text{psnr}}$) and MS\_SSIM ($L_{\text{ms\_ssim}}$) losses are at different scales we needed to normalize them in order to meaningfully combine them, as described in \autoref{sec:loss_functions}. The final normalized combined loss ($L_c$) is defined as: 
\begin{equation*}\label{eq:loss_combined}
    L_c = 5.43 - 0.0609 \ L_{\text{psnr}} - 1.51 \ L_{\text{ms\_ssim}}
\end{equation*}
The final SR model was trained on the full simulated training dataset. For the DN models we trained on the full simulated training dataset, the full real training dataset and a combination of the two using transfer learning. We selected the best preforming model out of these three as our final DN model.
We train the final models for 50 epochs using an Adam optimiser \citep{kingma2014adam}. After the training is complete we select the model from the epoch which achieved the best validation loss as our final model.

\subsection{Transfer Learning}
For the de-noising model we have access to real data. Whilst the simulated dataset contains more images, the real data encodes the domain that we are interested in. With the smaller dataset of real images, the performance of training a model on real data alone could be limited. Instead we make use of transfer learning \citep{tan2018survey} by taking the model trained on the larger simulated dataset and fine-tuning the weights to optimise for the real data. Fine-tuning is done by further training the model using the real data for another 50 epochs. We again select the model from the epoch which performed best on the validation loss as our final model.

\begin{figure*}
    \centering
    \includegraphics[width=0.24\linewidth]{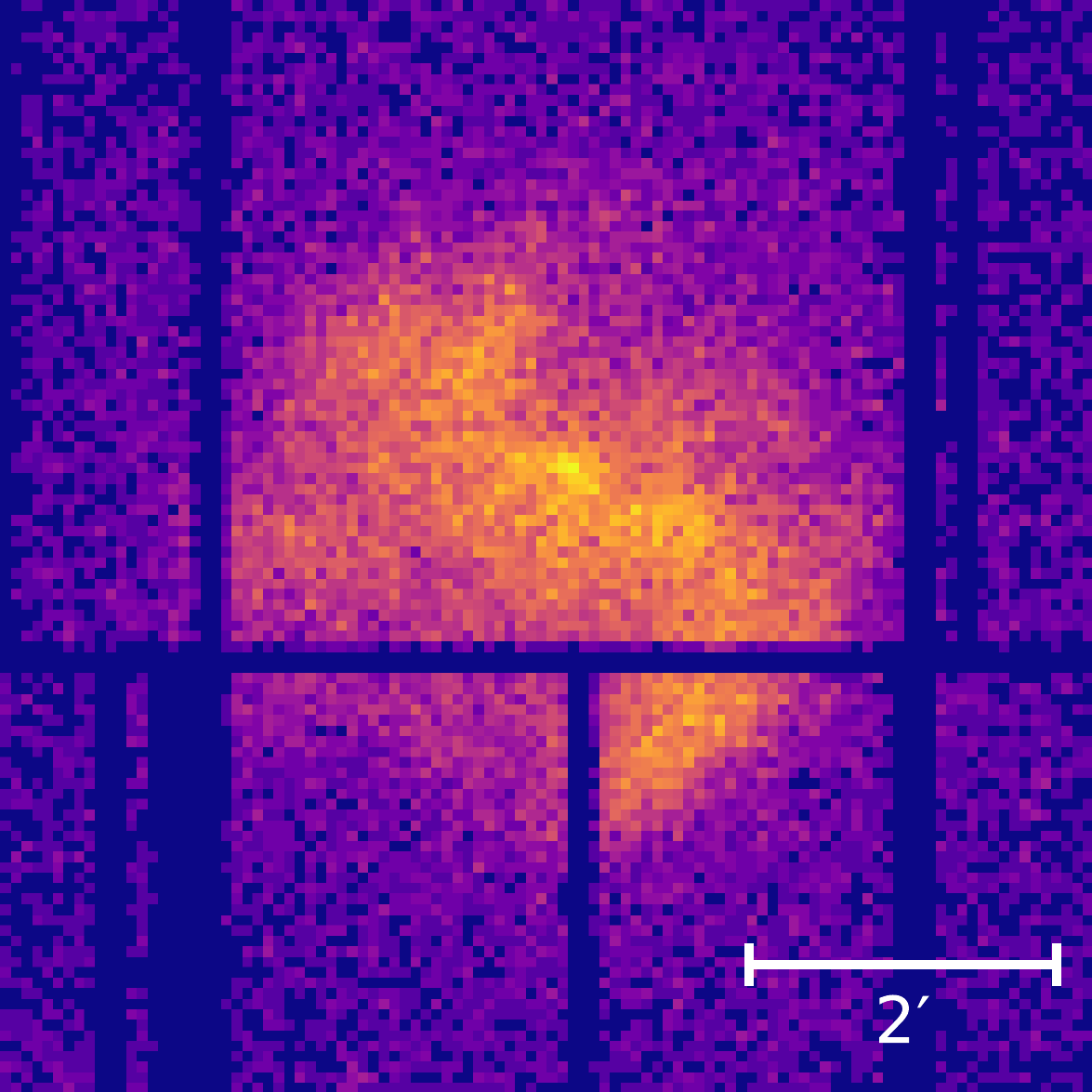}
    \includegraphics[width=0.24\linewidth]{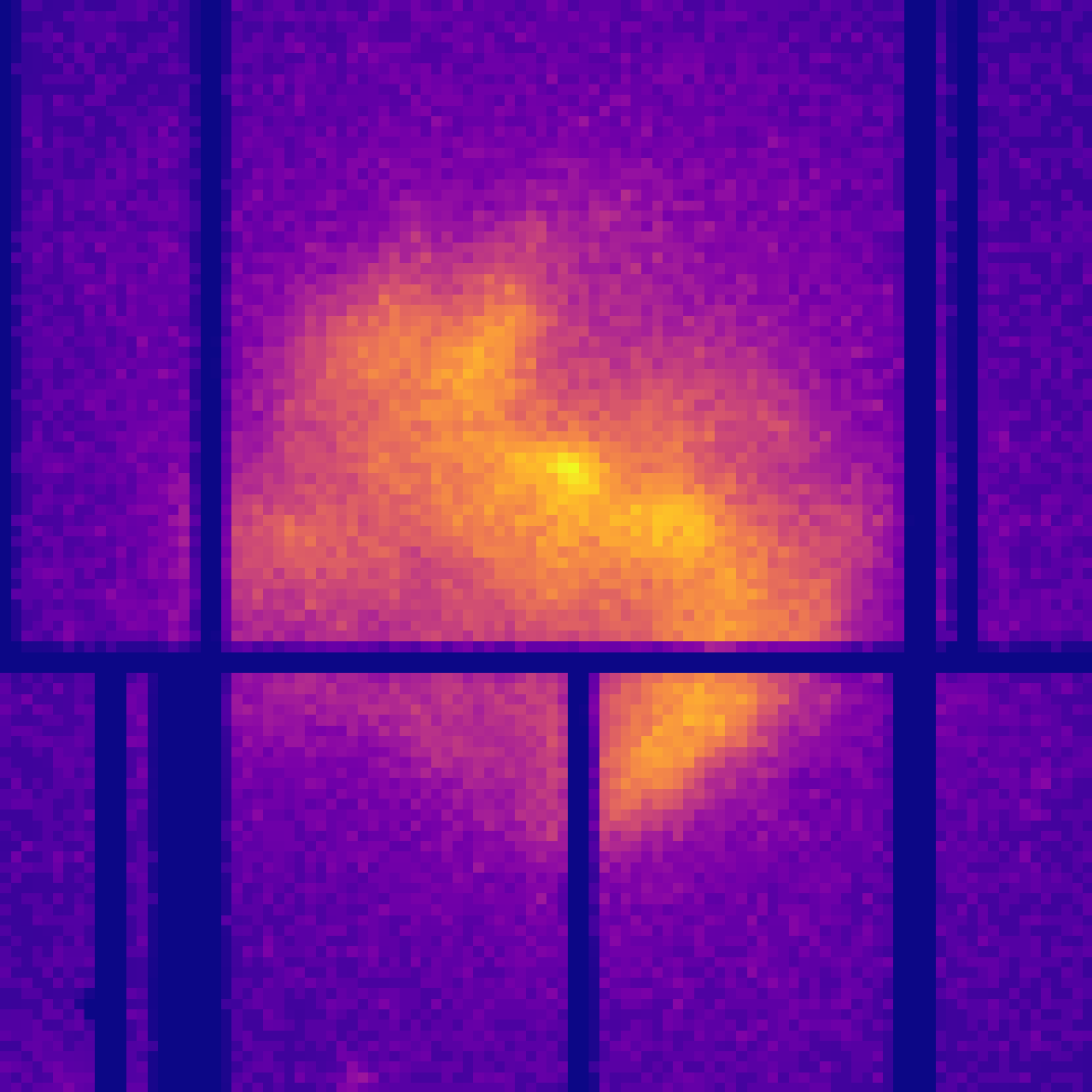}
    \includegraphics[width=0.24\linewidth]{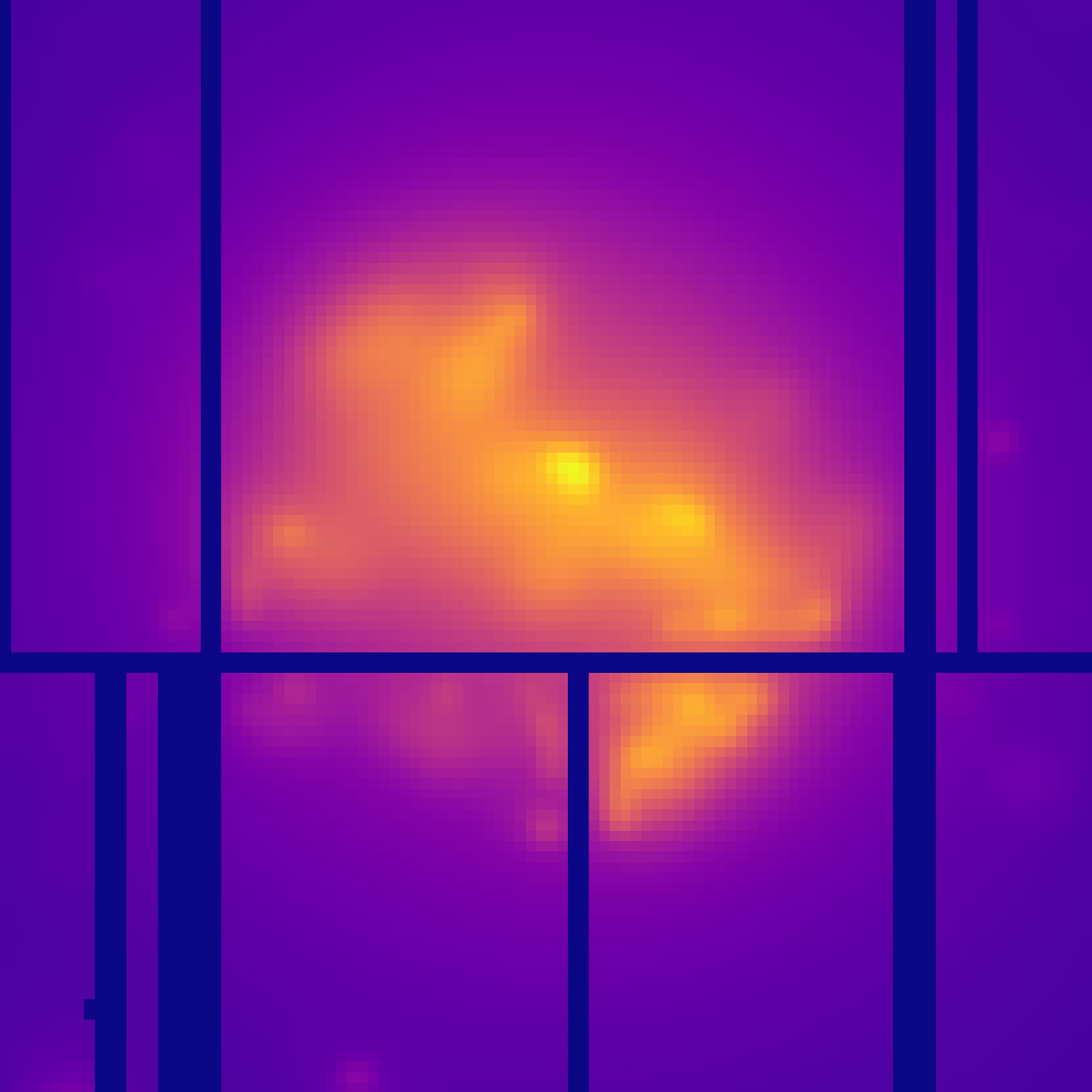}
    \includegraphics[width=0.24\linewidth]{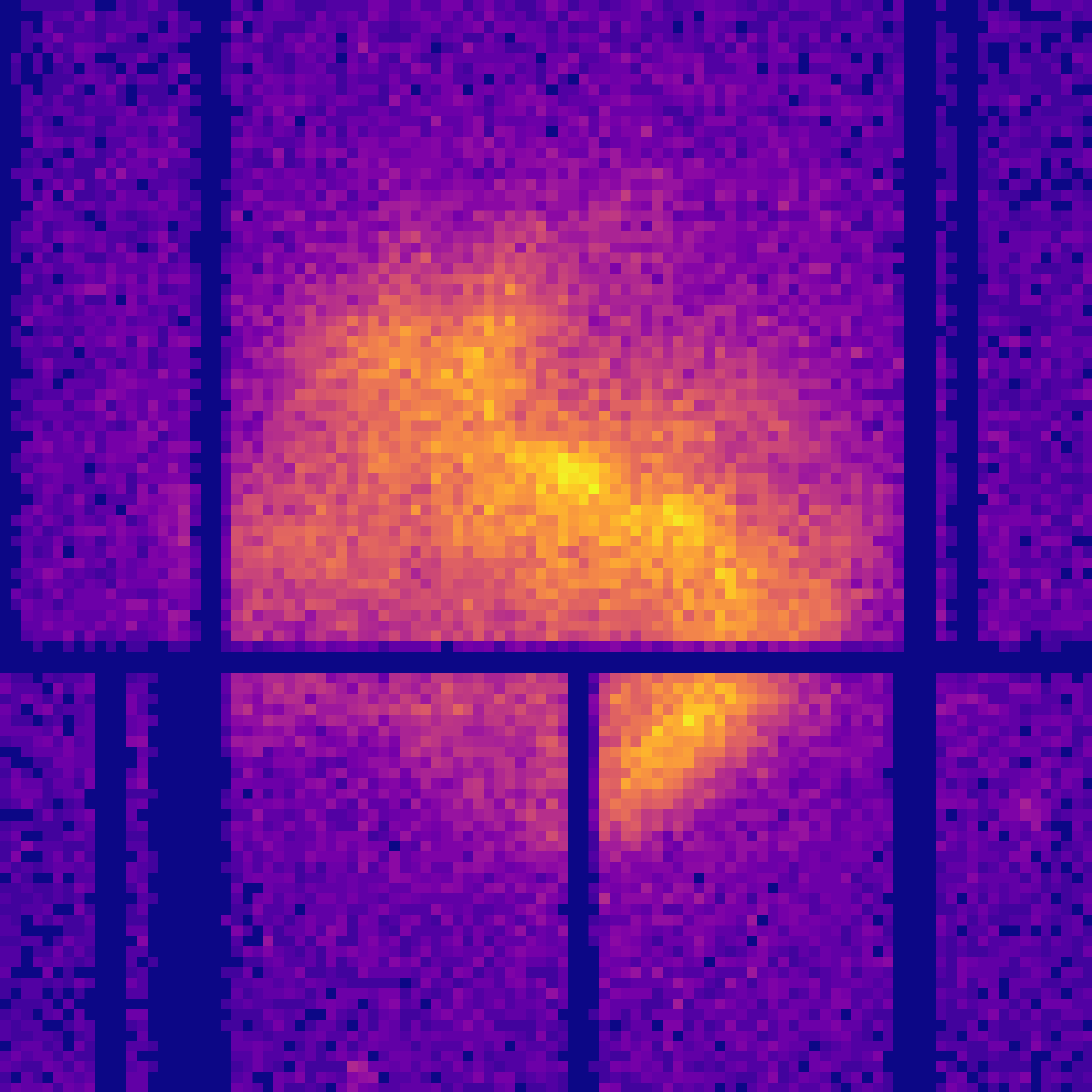}
    
    \includegraphics[width=0.24\linewidth]{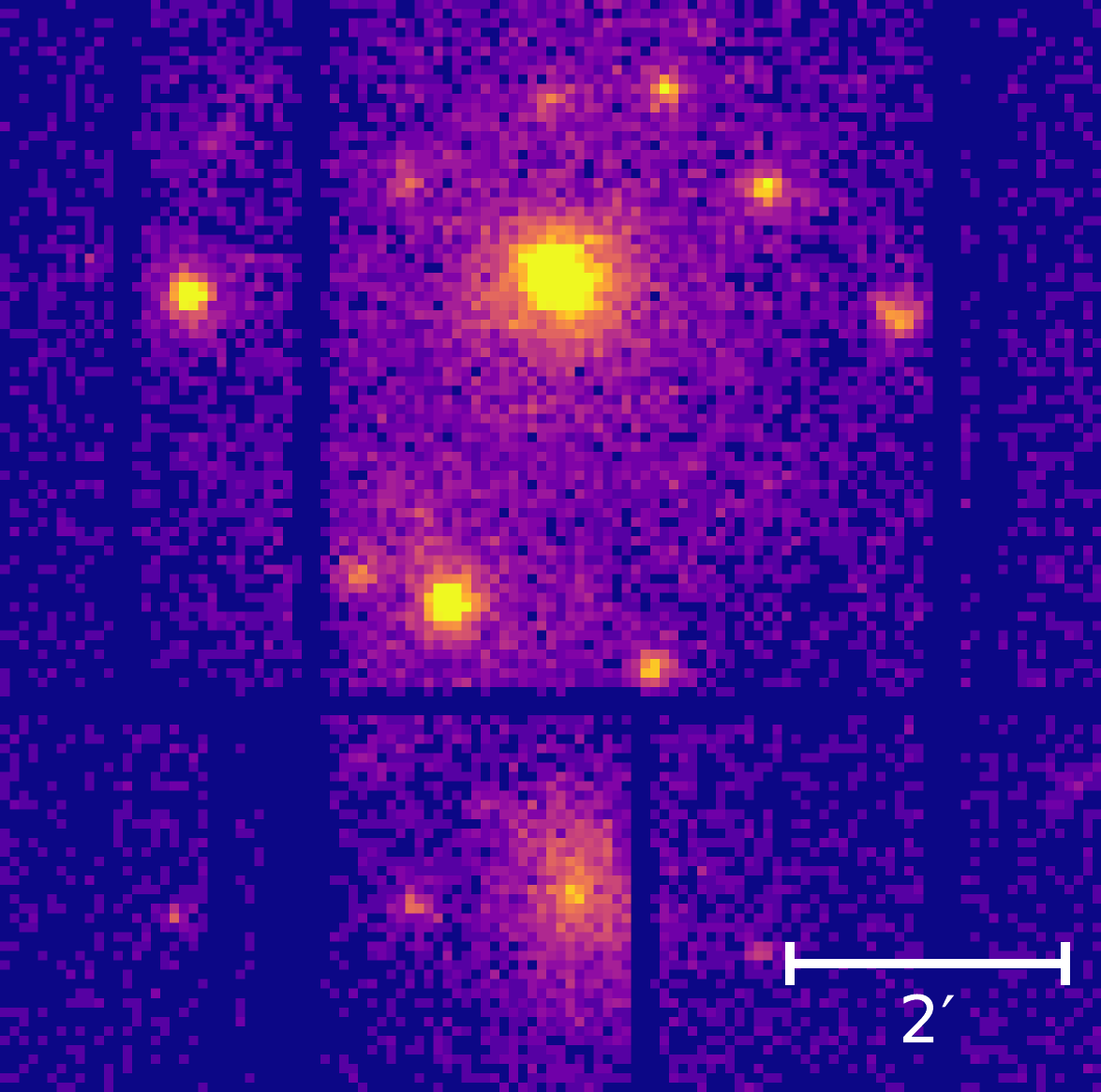}
    \includegraphics[width=0.24\linewidth]{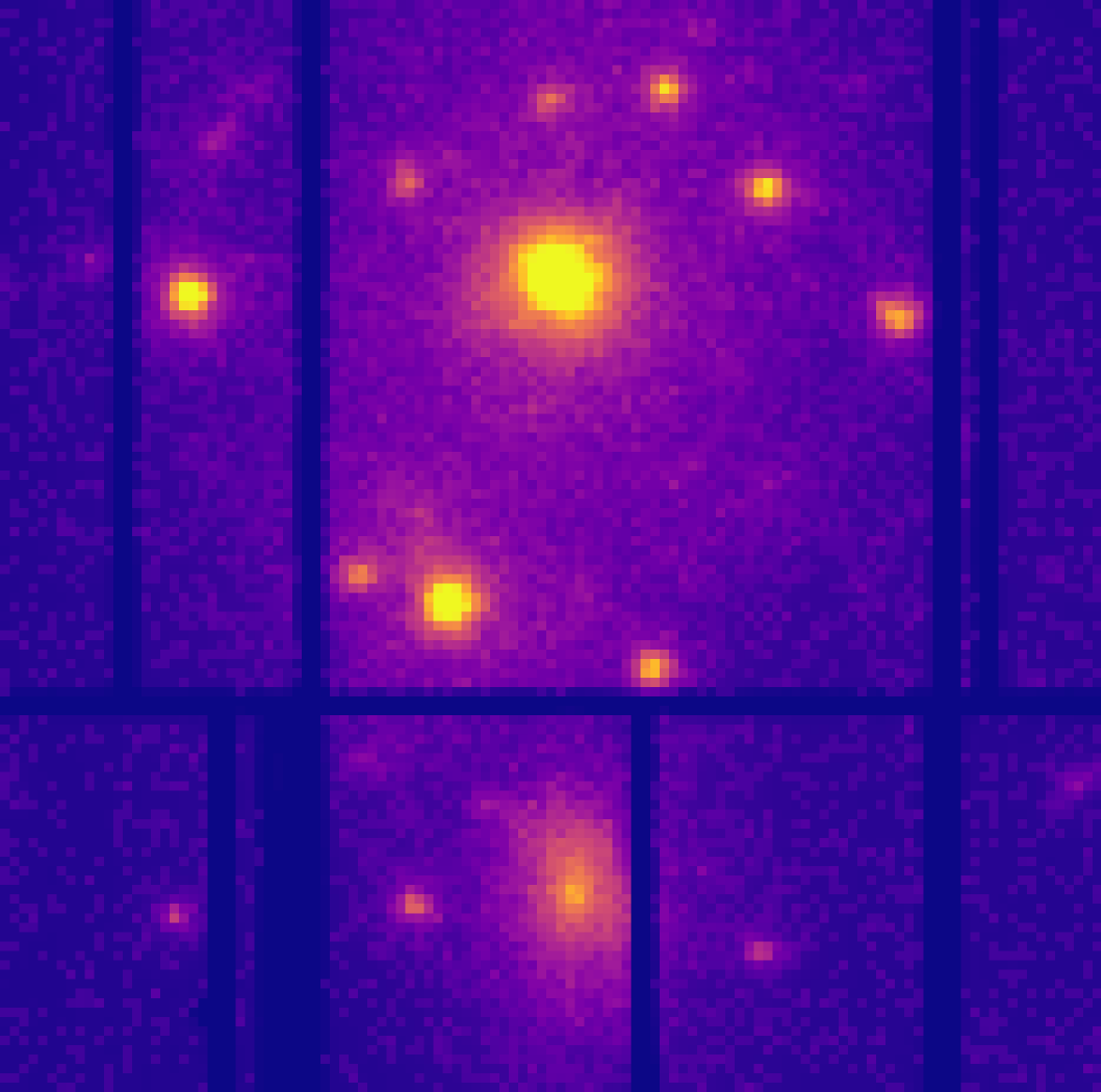}
    \includegraphics[width=0.24\linewidth]{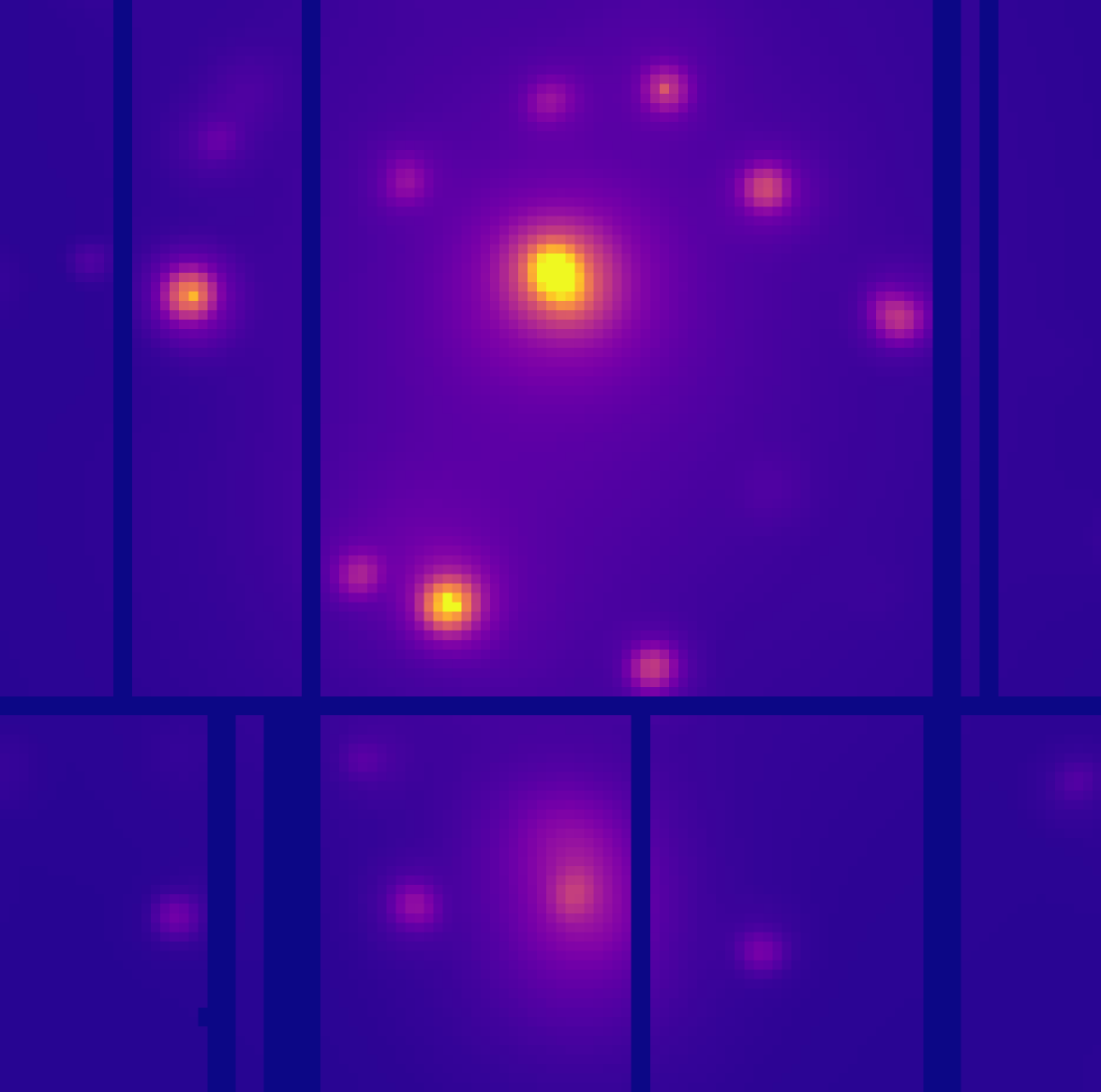}
    \includegraphics[width=0.24\linewidth]{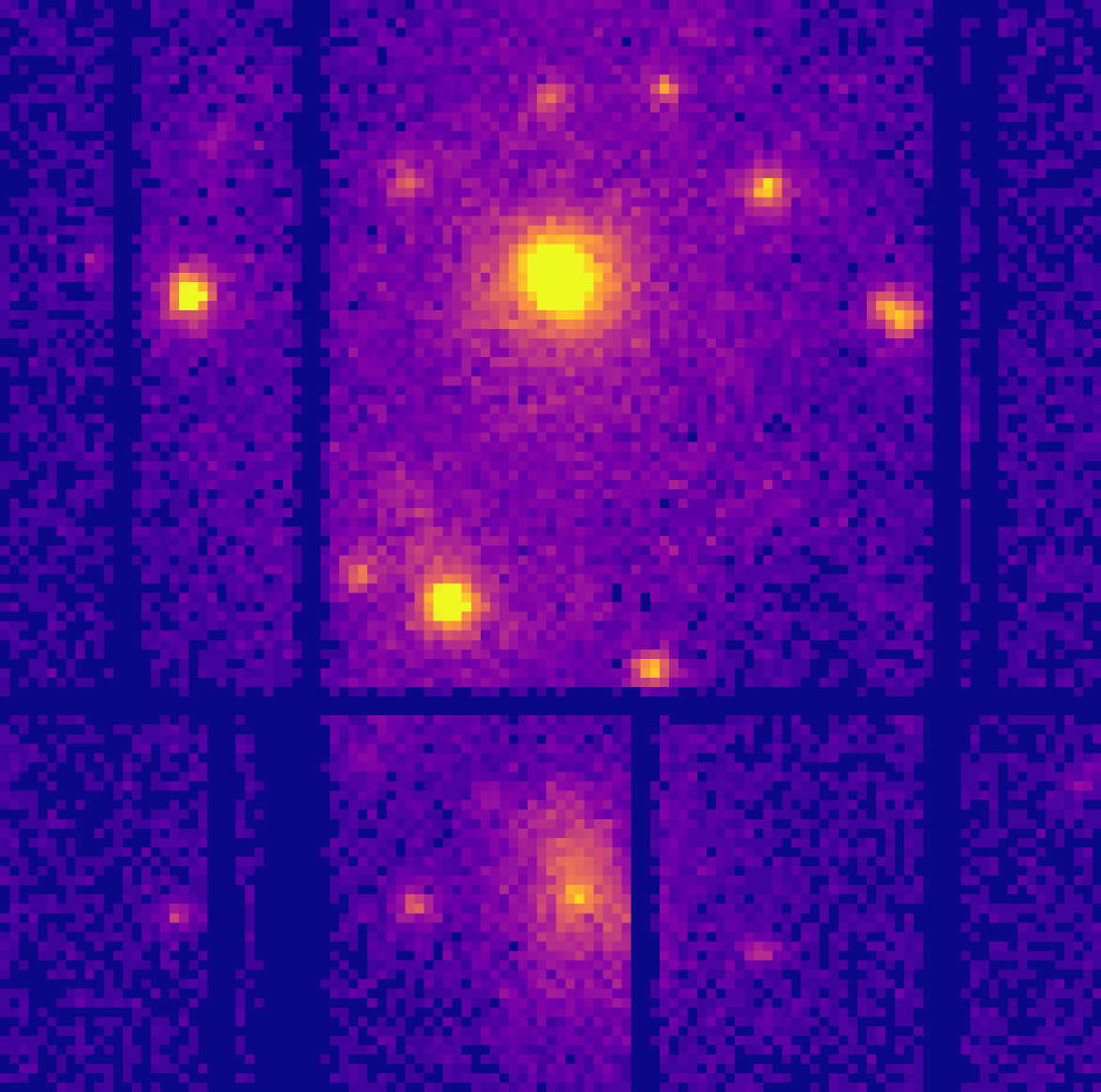}
    
    \caption{De-noised and wavelet transformed examples, W49B (top) and M51 (bottom), from the real \xmm dataset. Cropped to the central source and scaled with the square root function. From \textit{left} to \textit{right}: Input image at 1x resolution with 20ks exposure, generated de-noised image for 50ks, wavelet transformed image and the target image at 50ks.}
    \label{fig:dn_real_example}
\end{figure*}

\section{Results}\label{sec:Results}
\subsection{XMM-Denoise}\label{sec:res:denoising}
In \autoref{fig:denoise_metrics} we quantify the performance of the best de-noising models trained either on simulated data, real data or simulated and later fine-tuned to real data when applied on a real data based test sample. The models trained on the real dataset generally score better than those trained on the simulated data. This is expected as the test set was based on real data and certain features present in the real data will not be present in the simulated data. The model that performed the best overall is the model that was first trained on simulated data and then fine-tuned on real data. We, therefore, select this as our final DN model, named \textit{XMM-DeNoise}.

\subsubsection{Wavelet Comparison}
We qualitatively compare our \textit{XMM-Denoise} model to the non machine learning based wavelet transform. The use of wavelet based de-noising methods has been shown to optimize the detection of AGNs, galaxy clusters and other features in X-ray images of different telescopes (e.g. \citealt{valtchanov01,faccioli2018xxl,xu2018new,zhang2020high}. Our implementation is based on \citet{faccioli2018xxl}.

\noindent In \autoref{fig:dn_real_example} de-noised examples generated by \textit{XMM-DeNoise} are shown compared to wavelet transformed image and the target image. The images are cropped to highlight the details. We can see that our de-noised images are, compared to the more smoothed wavelet transformed images, visually much closer to the target images. Note that for the wavelet technique the goal is not to mimic the higher exposure time image but to de-noise the images substantially. Certain features, such at the shock waves in W49B (top row) are better defined because of this. However, this also comes at the risk of having more artifacts or filtering out too much information --- the wavelet transform will filter out regions with constant gradient, e.g. flat background. For example, in the M51 images (bottom row) we can see that the wavelet transformations filtered out the extended features of the source in the center left of the image. And using radially symmetric wavelet function will predominantly produce spherical morphologies.

\begin{figure*}
    \centering
    \includegraphics[width=0.24\linewidth]{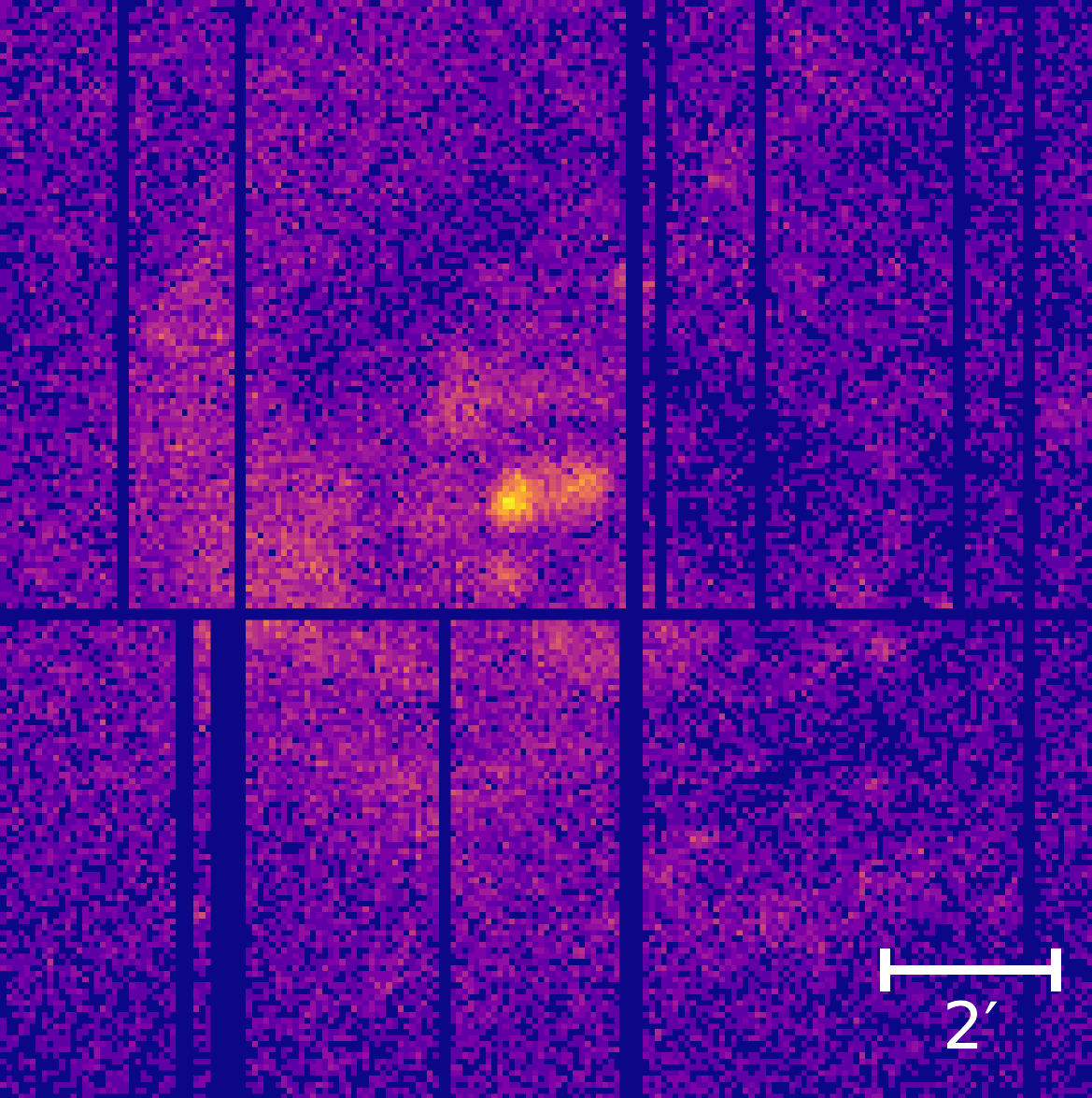}
    \includegraphics[width=0.24\linewidth]{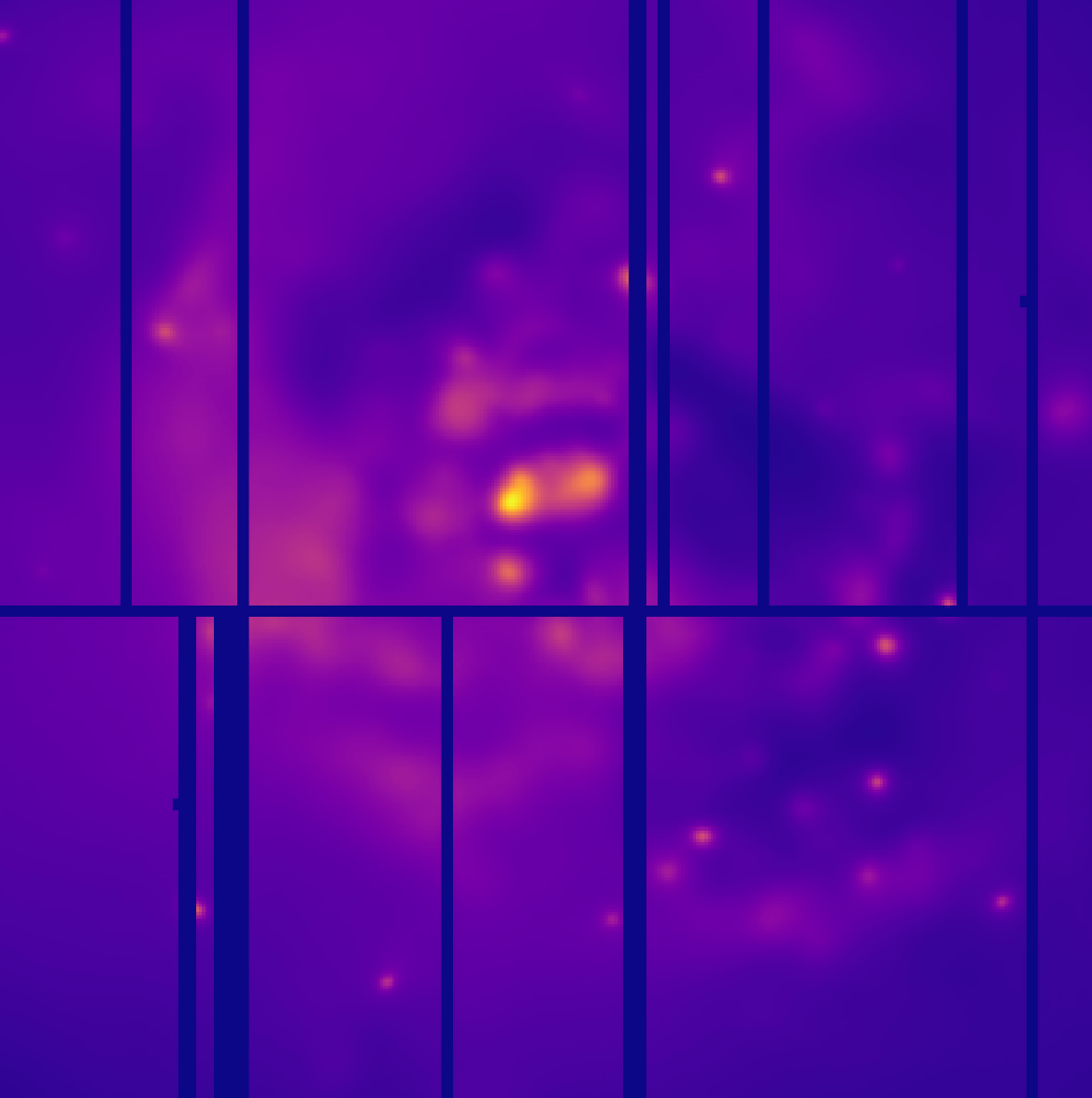}
    \includegraphics[width=0.24\linewidth]{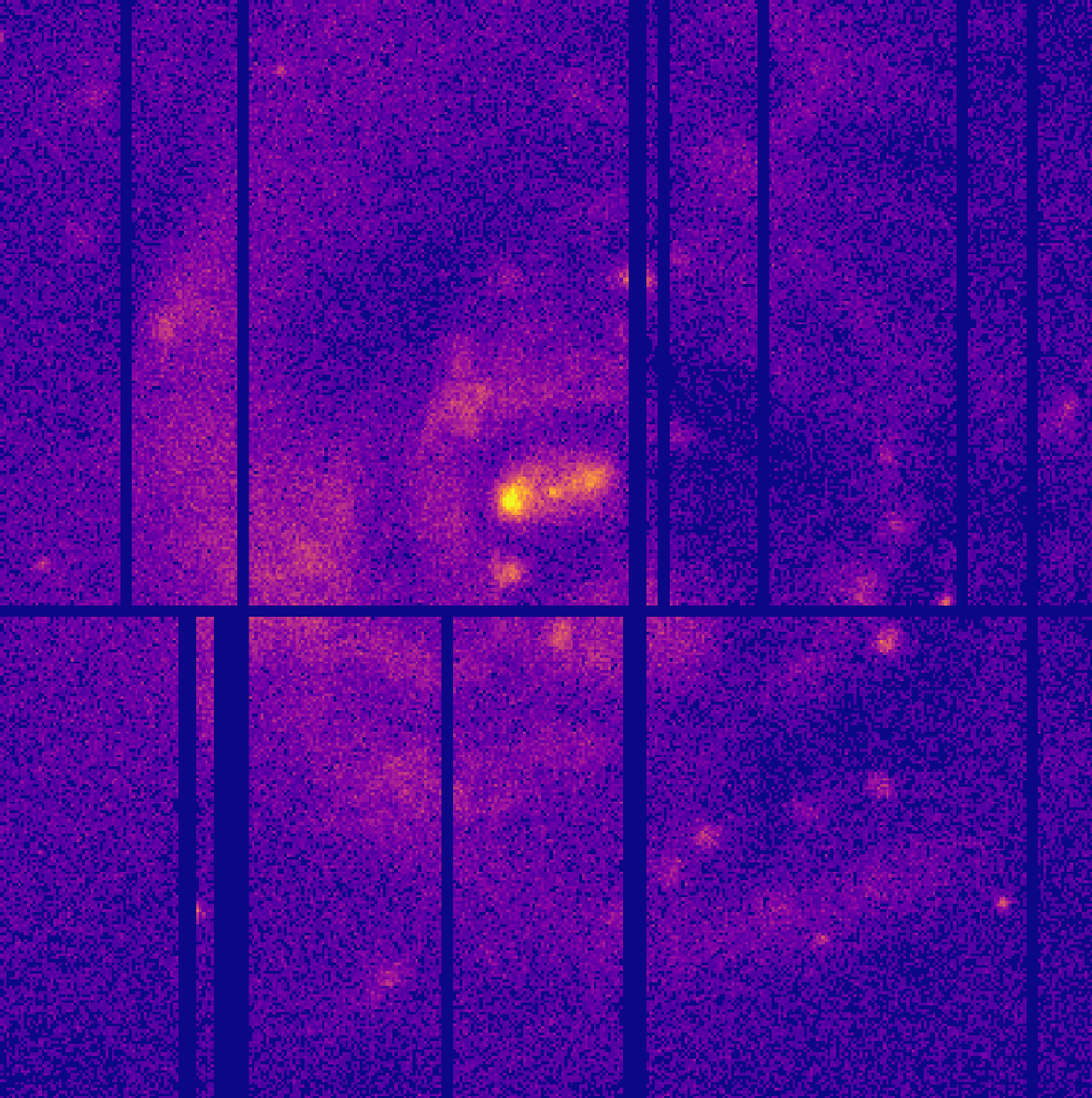}

    \includegraphics[width=0.24\linewidth]{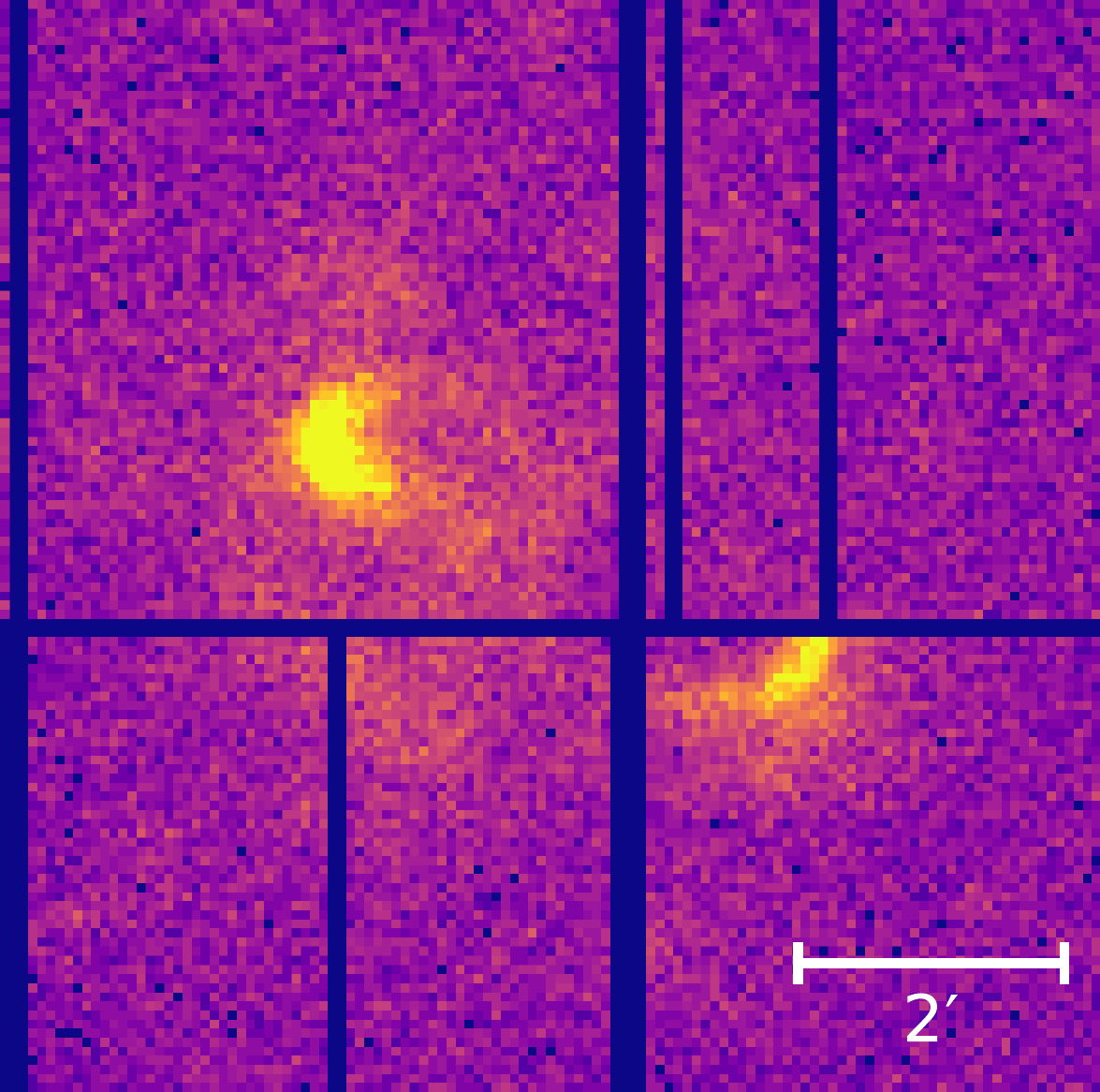}
    \includegraphics[width=0.24\linewidth]{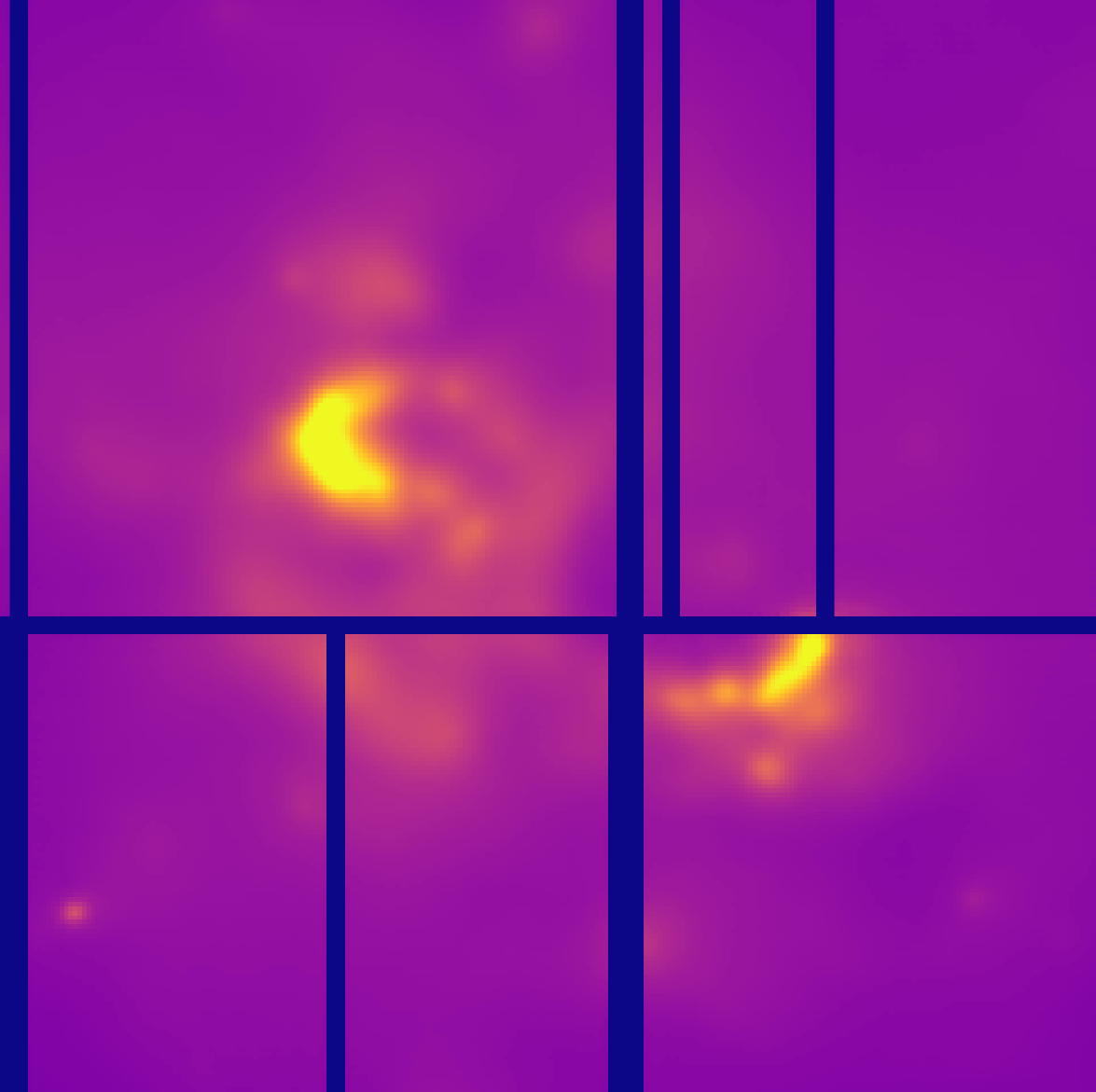}
    \includegraphics[width=0.24\linewidth]{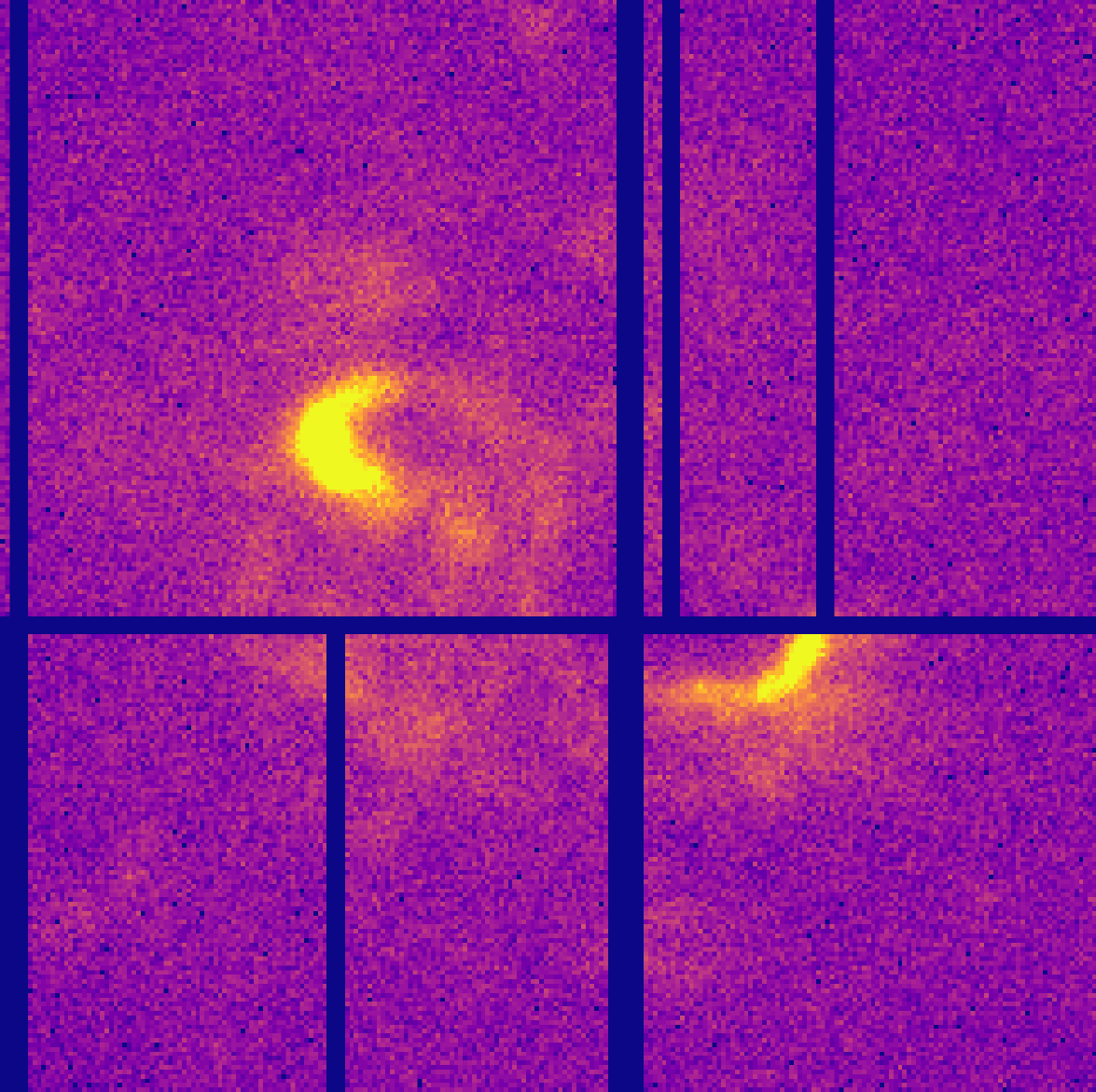}

    \includegraphics[width=0.24\linewidth]{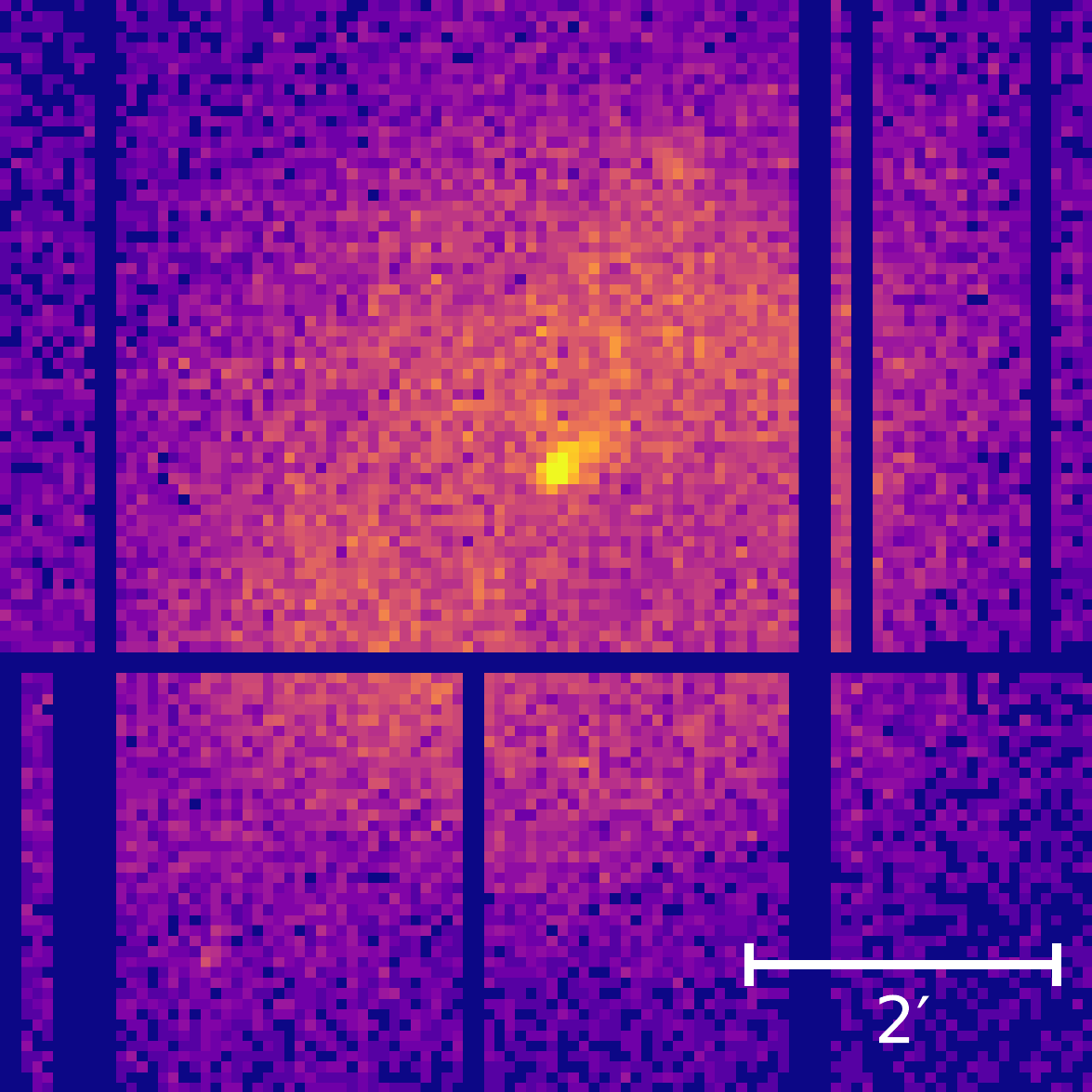}
    \includegraphics[width=0.24\linewidth]{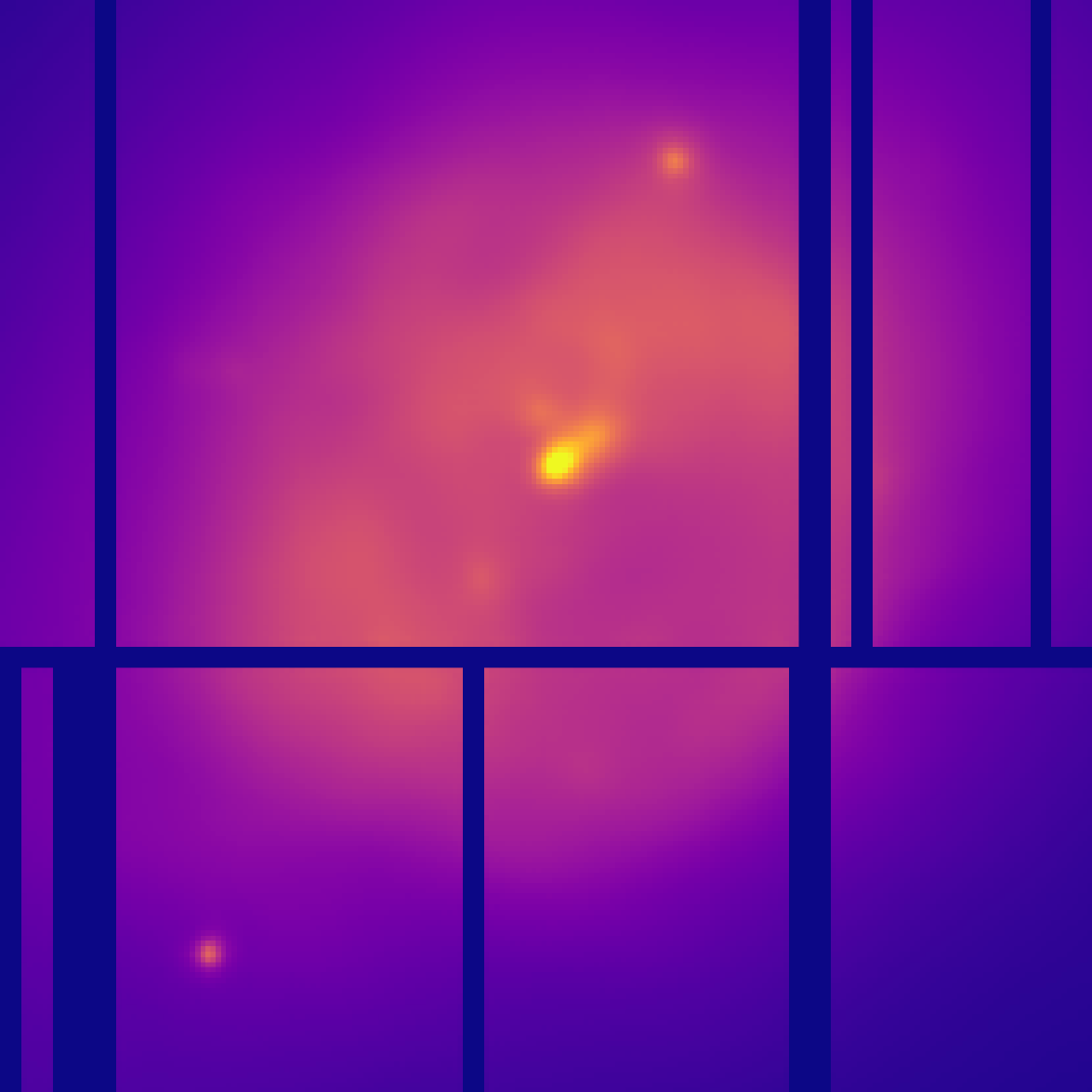}
    \includegraphics[width=0.24\linewidth]{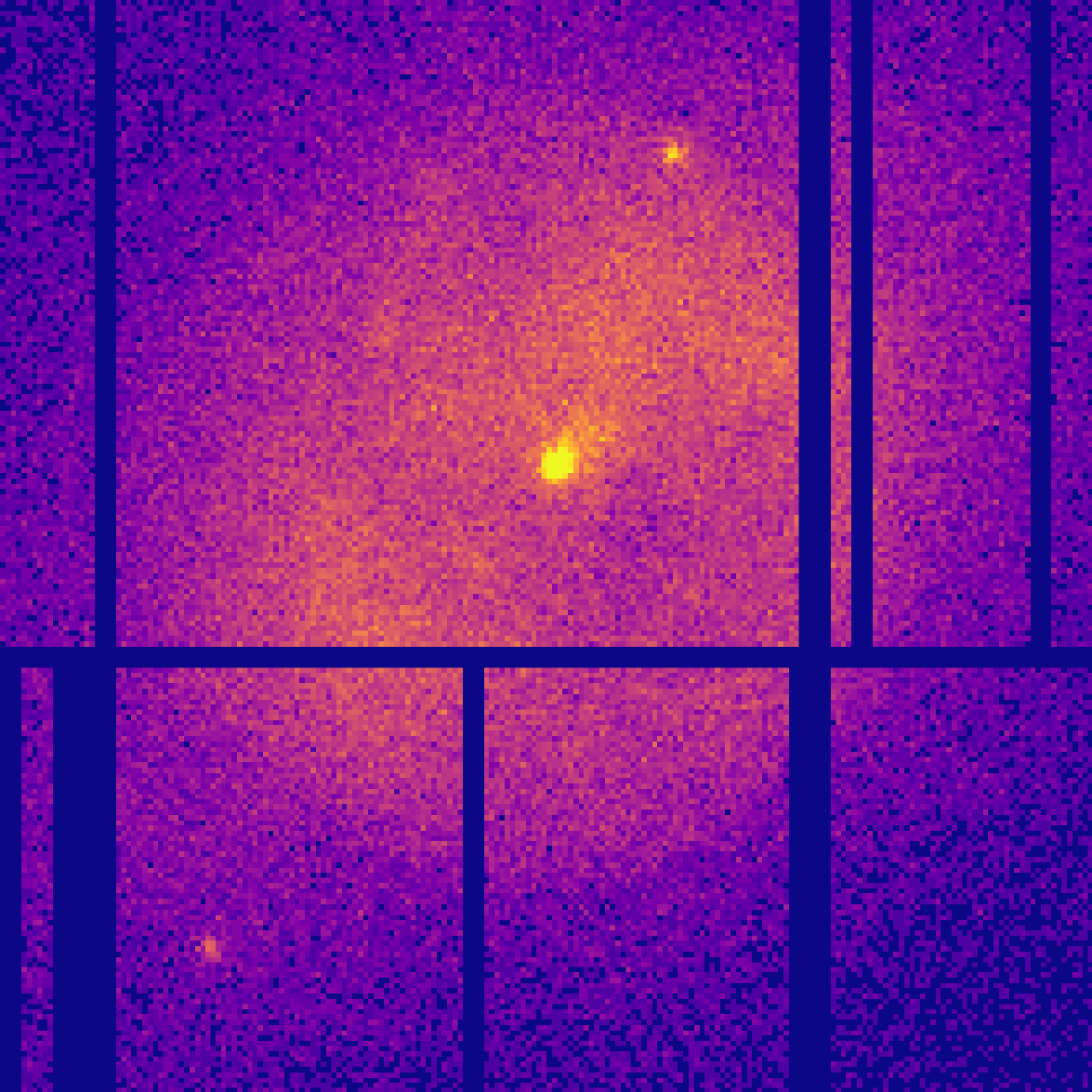}

    \caption{Super-resolution examples from the simulated \xmm dataset. Cropped to the central source and scaled with the square root function. Each row from \textit{top} to \textit{bottom}: TNG50 Subhalo 382215, TNG100 Subhalo 41583 and TNG300 Subhalo 296363. From \textit{left} to \textit{right}: Input image at 1x resolution with 20ks exposure, generated super-resolution image for 100ks and the label image at 100ks without background noise.}
    \label{fig:sr_sim}
\end{figure*}

\subsection{XMM-SuperRes}
\autoref{fig:sr_sim} shows a select few examples of generated super-resolution images. The generated images tend to contain more defined structures and more AGN. The performance of the \textit{XMM-SuperRes} model based on the simulated test set are shown in \autoref{tab:sr_sim}. The metrics are calculated compared to the target image using the unscaled (linear) data. To be able to do a comparison between the input and the target images, we need to match their resolutions. We use a naive method, namely nearest-neighbour upsampling. Our model improved the input image on all metrics.

\subsubsection{Brightness Analysis}
\begin{figure}
    \centering
    \includegraphics[width=0.32\linewidth]{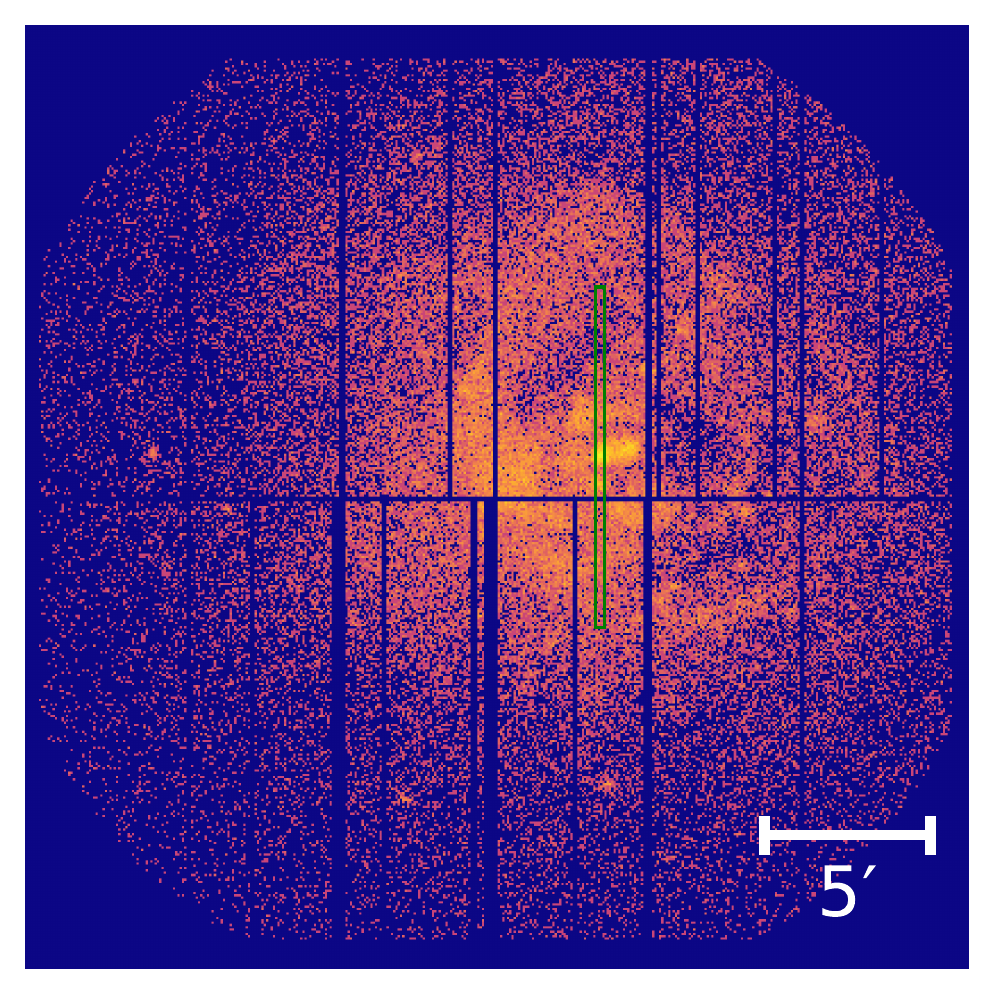}
    \includegraphics[width=0.32\linewidth]{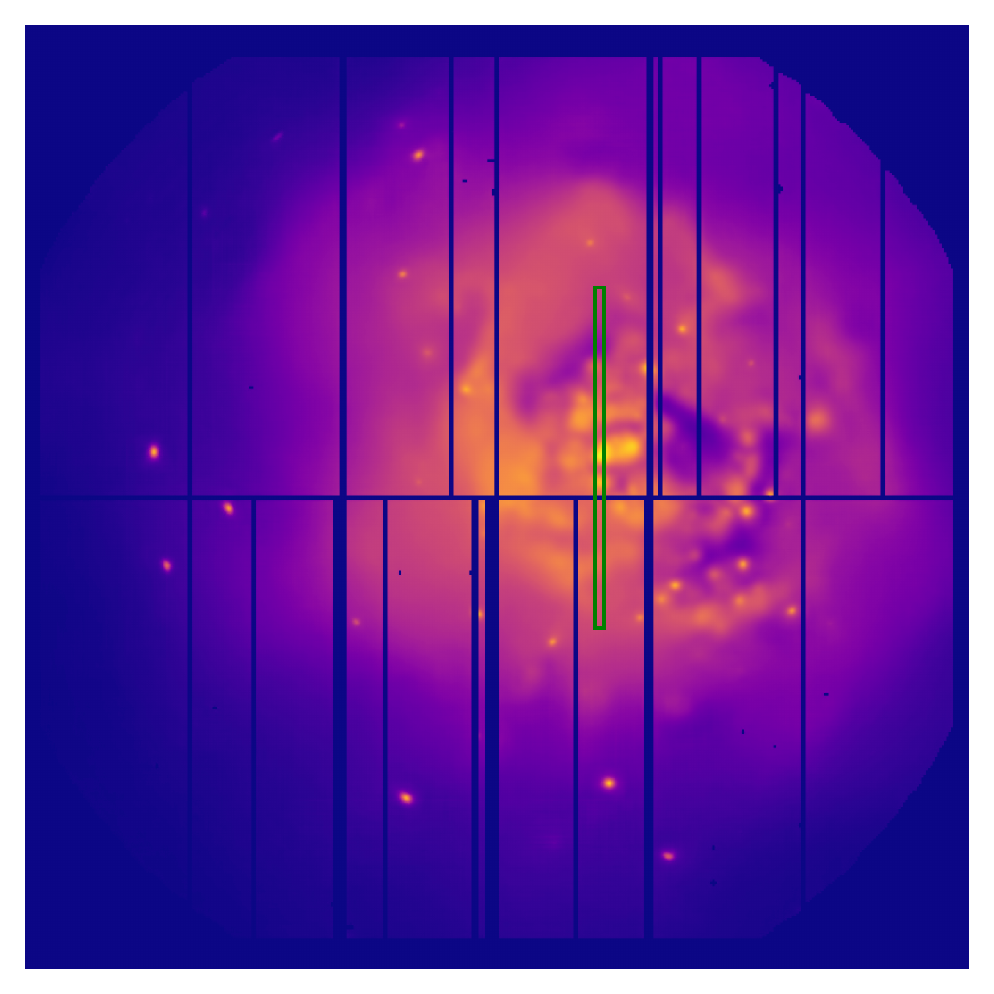}
    \includegraphics[width=0.32\linewidth]{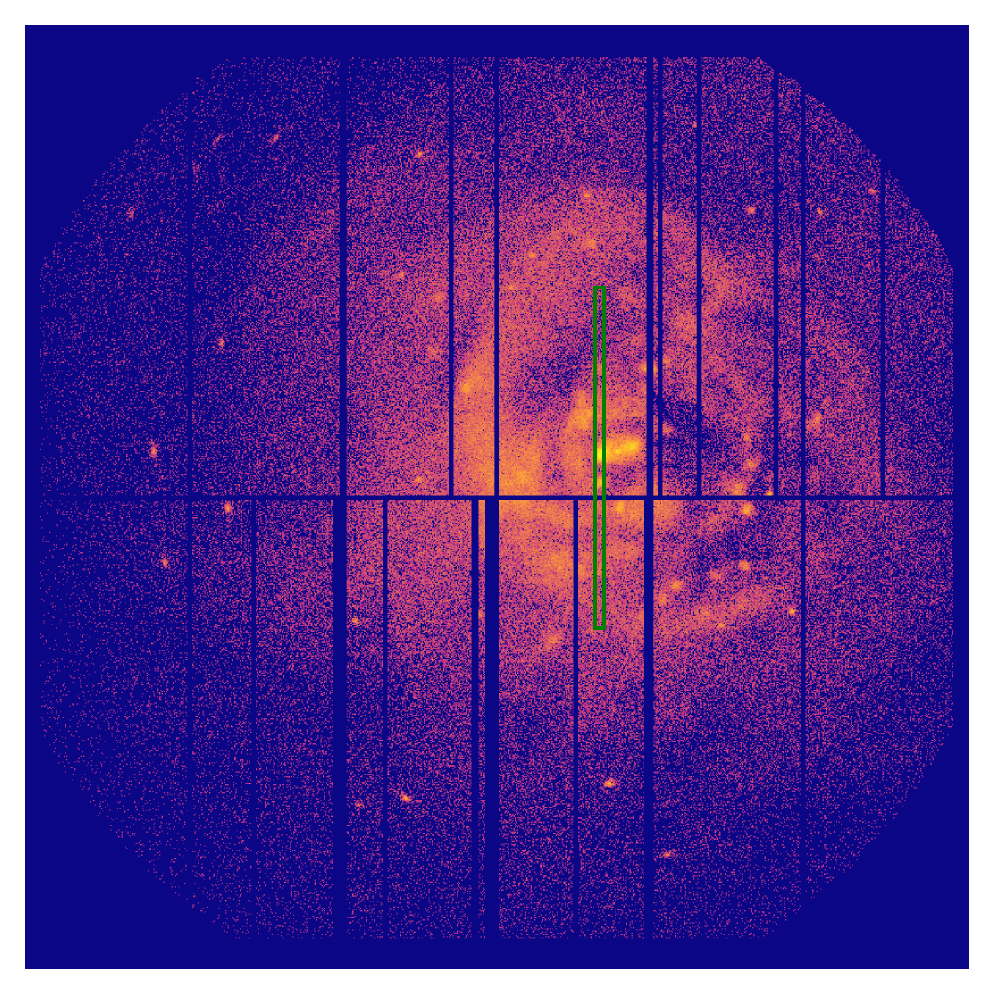}
    \caption{Strip plot regions TNG50 Subhalo 382215, scaled with a logaritmic funtion. From left to right: input image at 1x resolution at 20ks, \textit{XMM-SuperRes} generated image with 2x resolution at 100ks and the target image with 2x resolution at 100ks without background noise. The green regions indicate the regions that we will analyse on brightness.}
    \label{fig:strip_reg}
\end{figure}

\begin{figure}
    \centering
    \includegraphics[width=\linewidth]{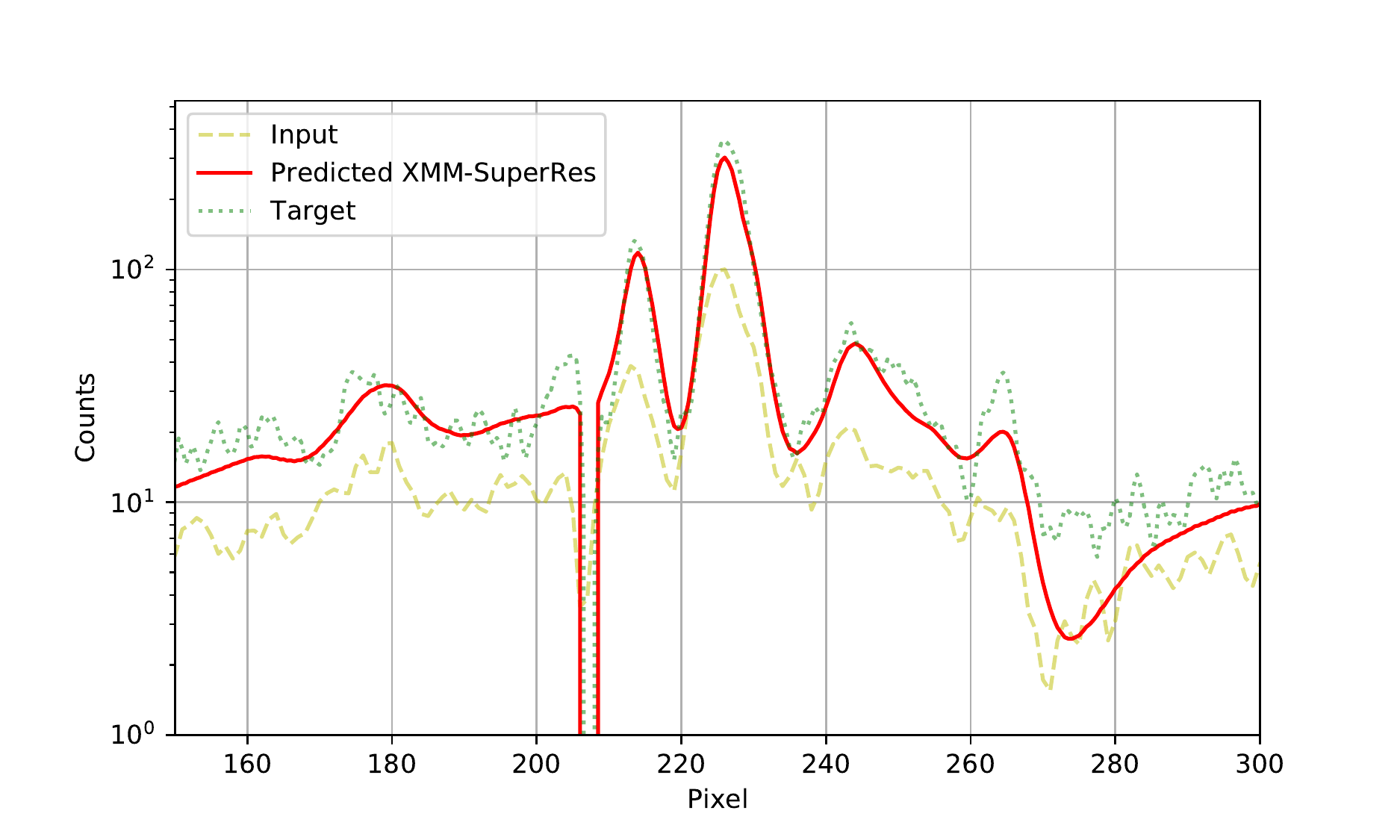}
    \caption{Count plots of the input image, \textit{XMM-SuperRes} generated image and the target image of TNG50 Subhalo 382215 corresponding to the vertical cutout regions in \autoref{fig:strip_reg}.}
    \label{fig:ver_cutout}
\end{figure}

To analyze the performance SR models in detail, we take vertical segments through the boresight of the input and generated images (\autoref{fig:strip_reg}) and plot the pixel value distribution summed along the minor axis (\autoref{fig:ver_cutout}). The generated images are smoother than the input and target images; therefore, we smooth the input and target images using 1d convolution with a Gaussian kernel of size $5$ and $\sigma = 1.0$ for a fairer comparison. Since the input image is at 20ks and the target image is at 100ks, the input image will have lower counts per region. The sudden drop in count values corresponds with the chip-gaps where the count is zero. The predicted \textit{XMM-SuperRes} image (red) bares better resemblance to the target brightness in comparison to the input image.

\begin{table}
\centering
\begin{tabular}{|l|l|l|}
\hline
Metric   & Input                           & \begin{tabular}[c]{@{}l@{}}Predicted\\ (\textit{XMM-SuperRes})\end{tabular} \\ \hline
L1       & 0.01096 & 0.006508                                   \\ \hline
PSNR     & 33.525  & 38.034                                     \\ \hline
Poisson  & 0.08285 & 0.04997                                    \\ \hline
SSIM     & 0.8248  & 0.907                                      \\ \hline
MS\_SSIM & 0.9499  & 0.9846                                     \\ \hline
FSIM     & 0.8657  & 0.8688                                     \\ \hline
HaarPsi  & 0.5312  & 0.697                                      \\ \hline
\end{tabular}
\caption{Super Resolution model metrics based on the simulated data test-set. The input column refers to the direct comparison between the simulated input and target image. The rows correspond to the different metric scores when applied to the simulated data test set compared to the target.}
\label{tab:sr_sim}
\end{table}

\subsubsection{Chandra Comparison}
For the super-resolution model, it is not possible to learn the domain mapping for real \xmm observations. However, we can probe its performance with real data by comparing SR-generated images with their Chandra counterparts. 

As a qualitative measure of our results we compare our SR \xmm generated images with Chandra observations of the same source. The \textit{Chandra} images have a higher resolution of 0.5~arcsec HEW compared to the 17~arcsec of the \xmm EPIC-pn. We use the full exposure time of the \textit{Chandra} images. We do however stress that the properties of the two telescopes are not equivalent. The PSF of the two instruments are not the same and \xmm is more sensitive than \textit{Chandra}, so these images can not be considered as ground-truth. Nonetheless we present a few examples to cover a variety of fields and source morpholigies.

\begin{figure*}
    \includegraphics[width=0.19\linewidth]{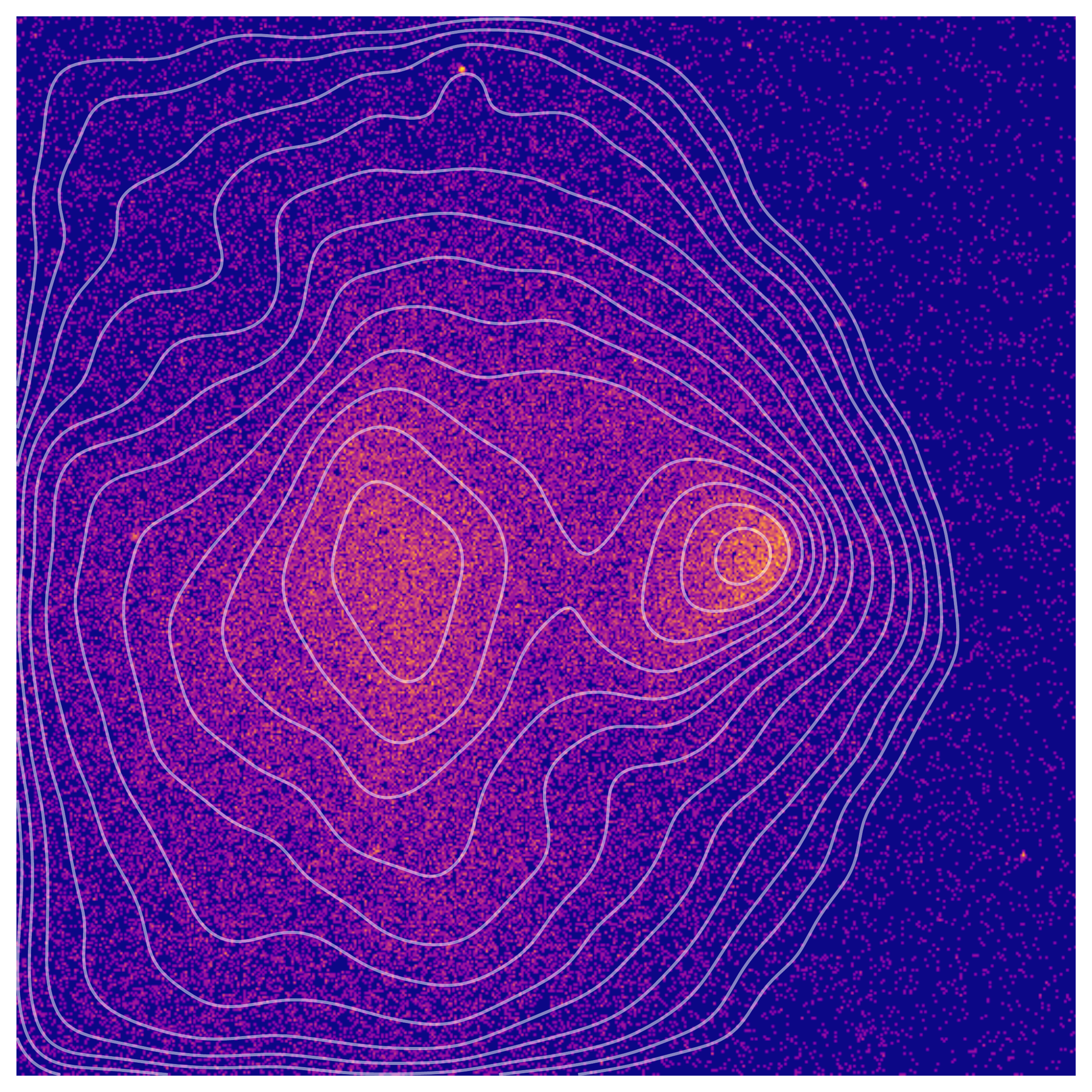}
     \includegraphics[width=0.19\linewidth]{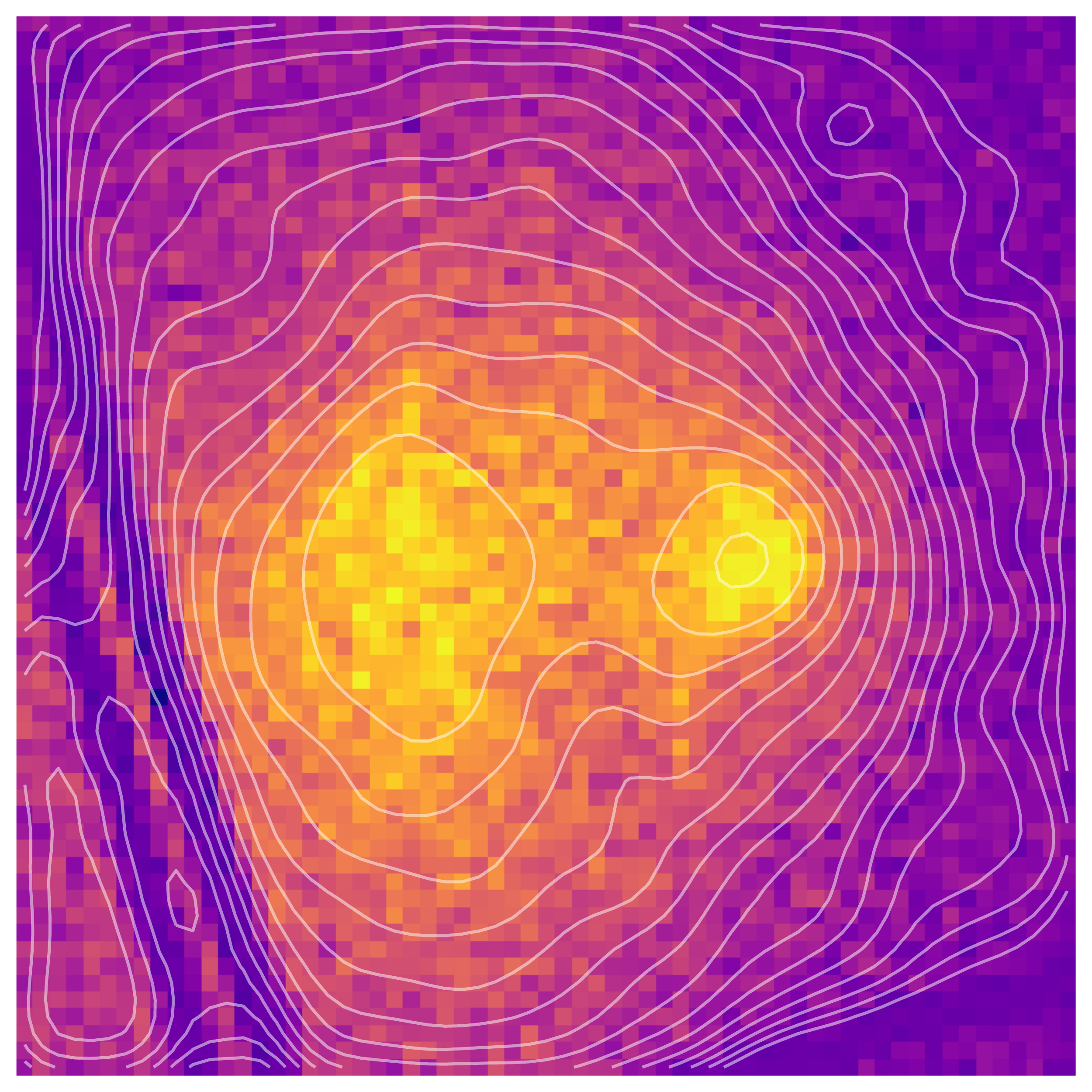}
    \includegraphics[width=0.19\linewidth]{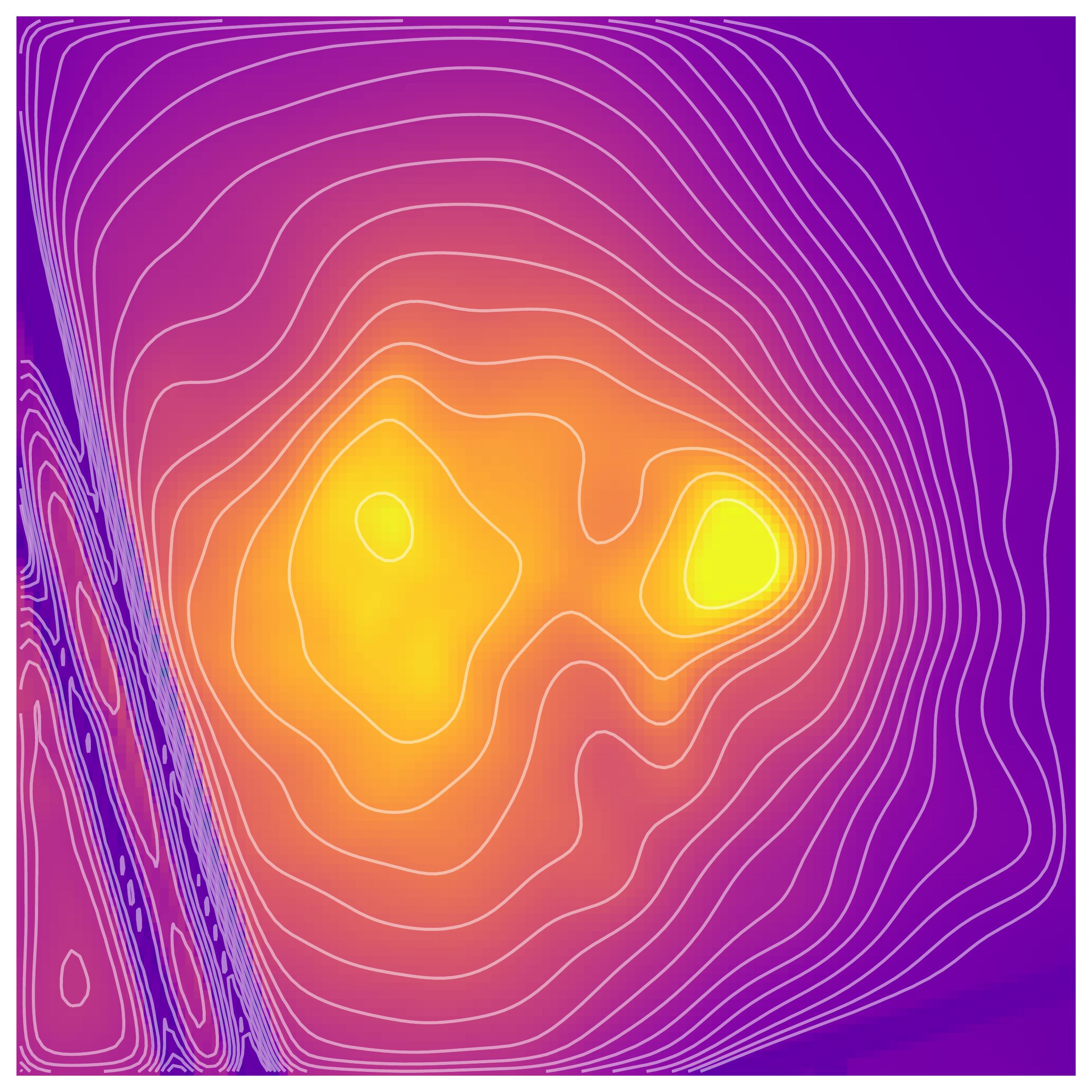}
     \includegraphics[width=0.19\linewidth]{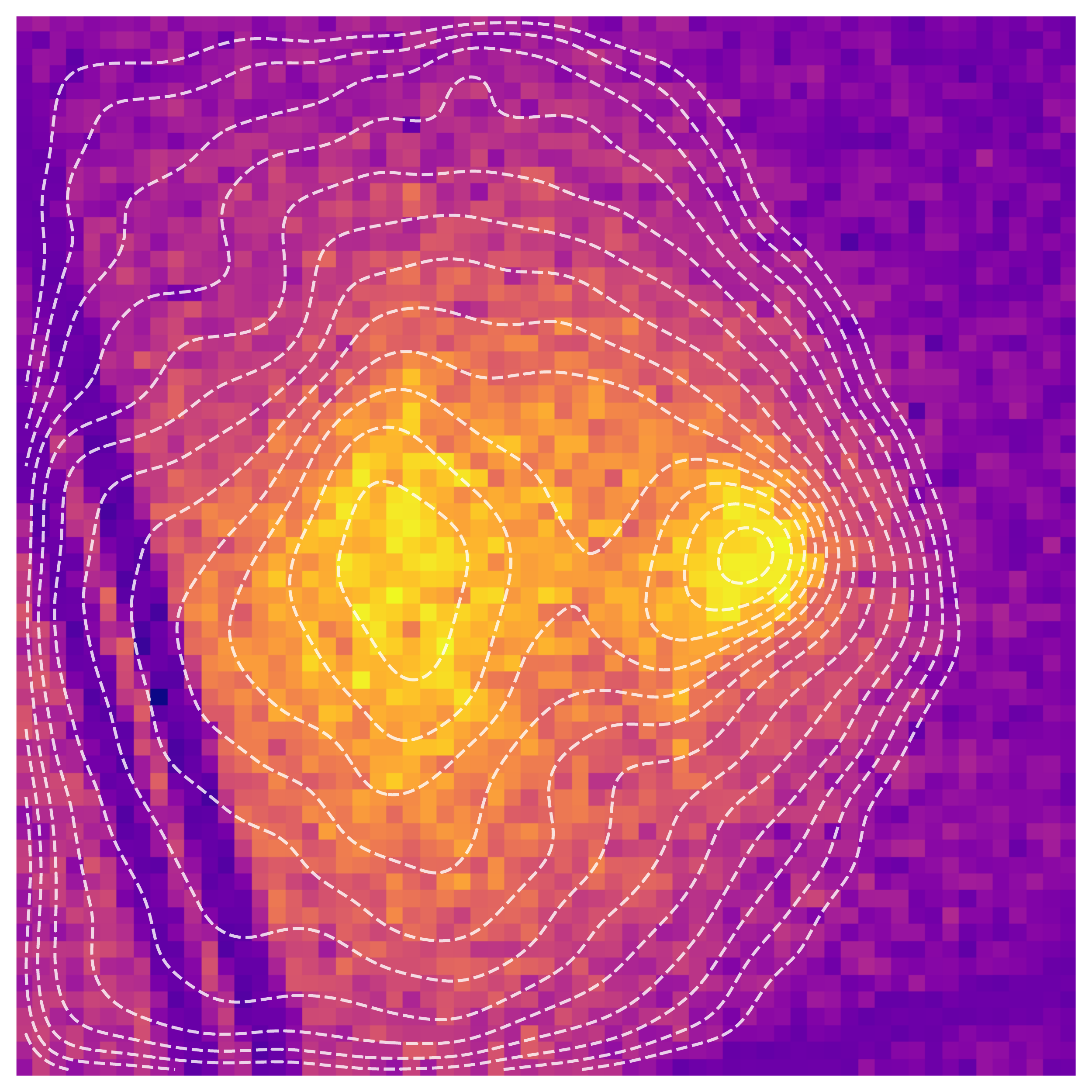}
    \includegraphics[width=0.19\linewidth]{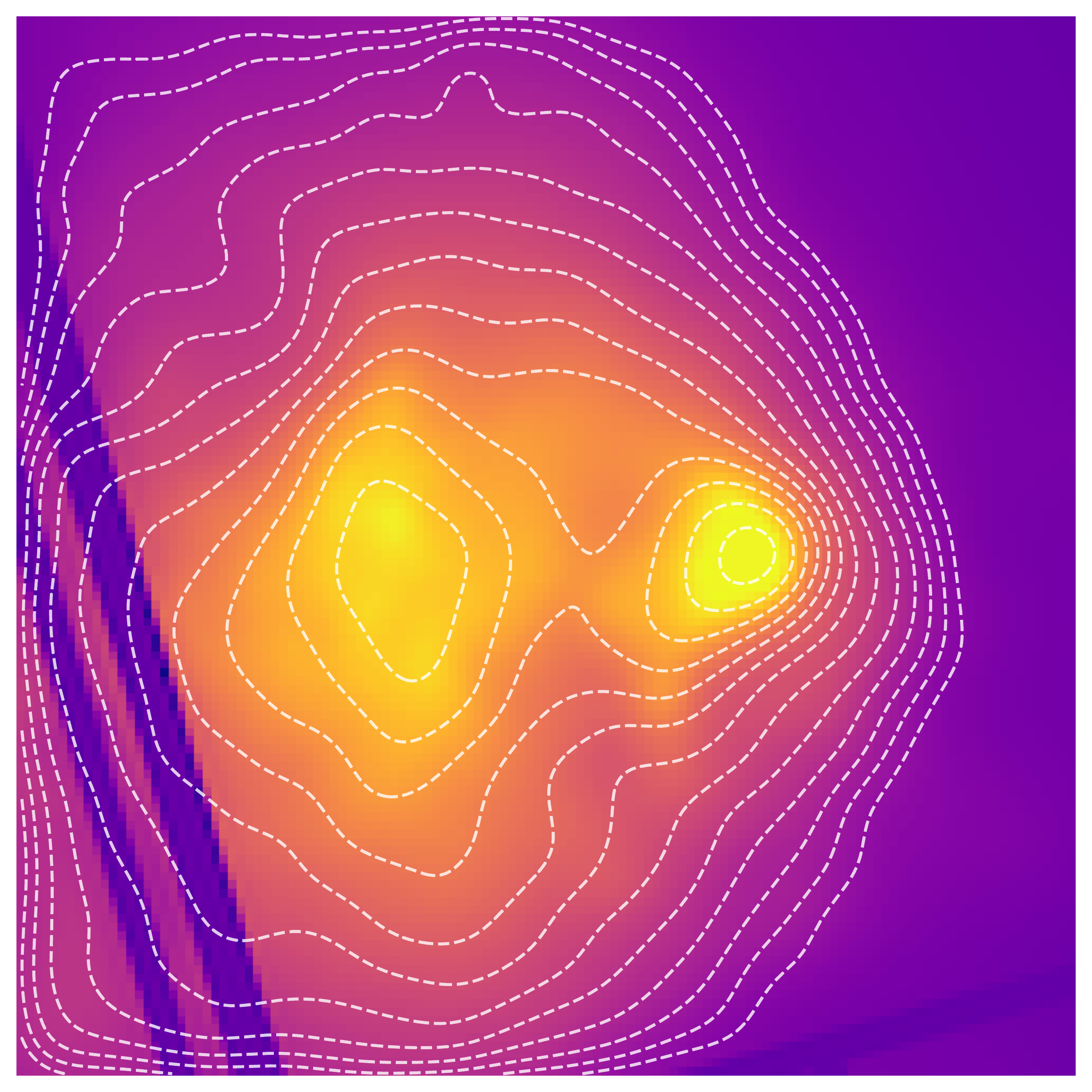}\caption{Images of the two colliding clusters of galaxies Bullet Cluster (1E 0657-56) with contours highlighted in white. Cropped to a frame size of $4.2`$. From \textit{left} to \textit{right}: \textit{Chandra} at 88ks exposure, \xmm at 20ks exposure, generated \xmm SR, \xmm at 20ks exposure overlayed with the \textit{Chandra} contours and generated \xmm SR overlayed with the \textit{Chandra} contours.} 
\label{fig:bc_contour}
\end{figure*}

Our first case study is the Bullet cluster (\autoref{fig:bc_contour}) a well know system of two interacting galaxy clusters. The cavity between the two X-ray components is enhanced in both the \textit{Chandra} and the generated SR image in comparison to the input \xmm image. Looking at the real \xmm image and the SR-generated one, with white contours from \textit{Chandra} overlayed, we can see that the cavity between the two clusters is much better defined in the SR and DN image compared to the original \xmm image, the \textit{Chandra} image also clearly contains this feature. 

\begin{figure*}
    \includegraphics[width=0.19\linewidth]{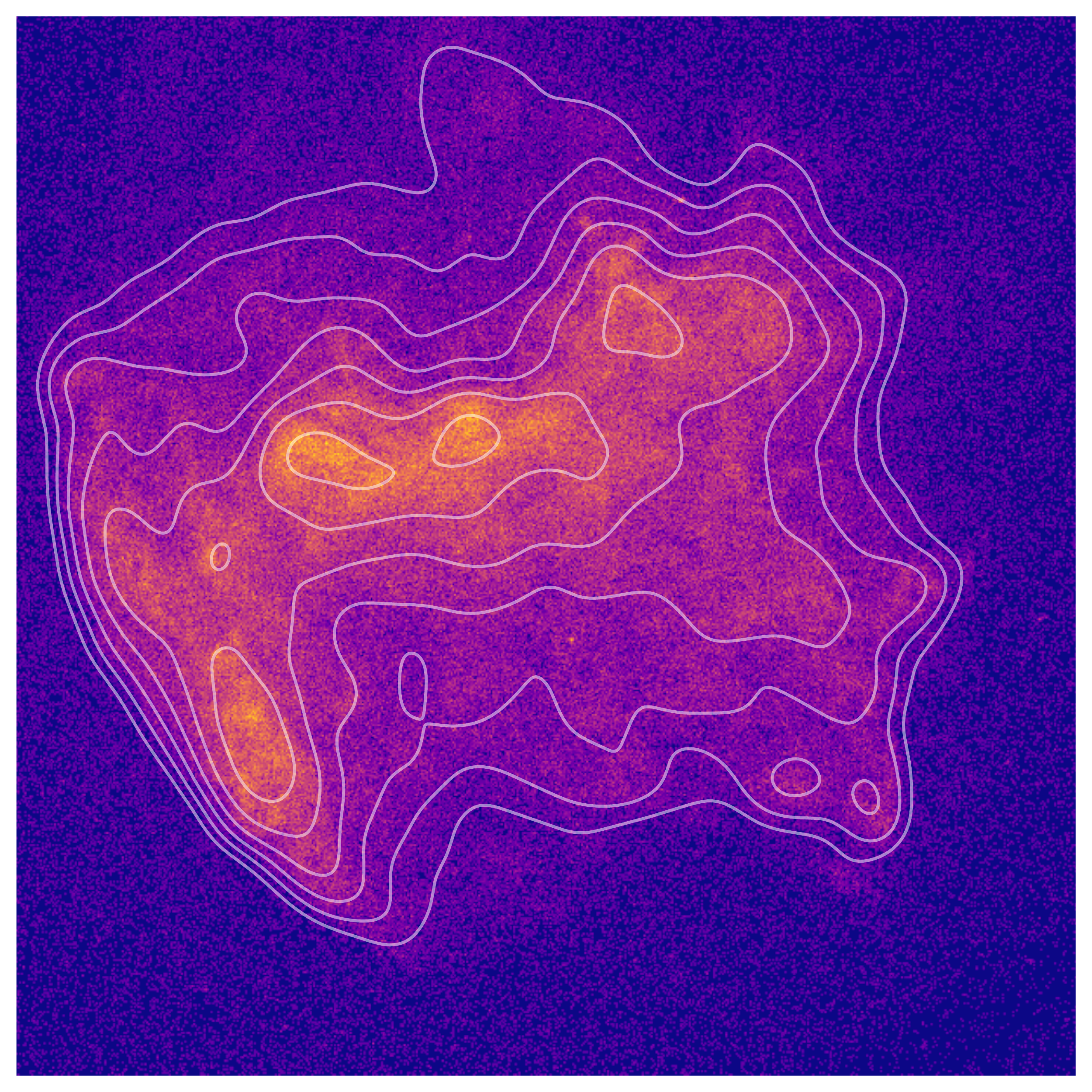}
     \includegraphics[width=0.19\linewidth]{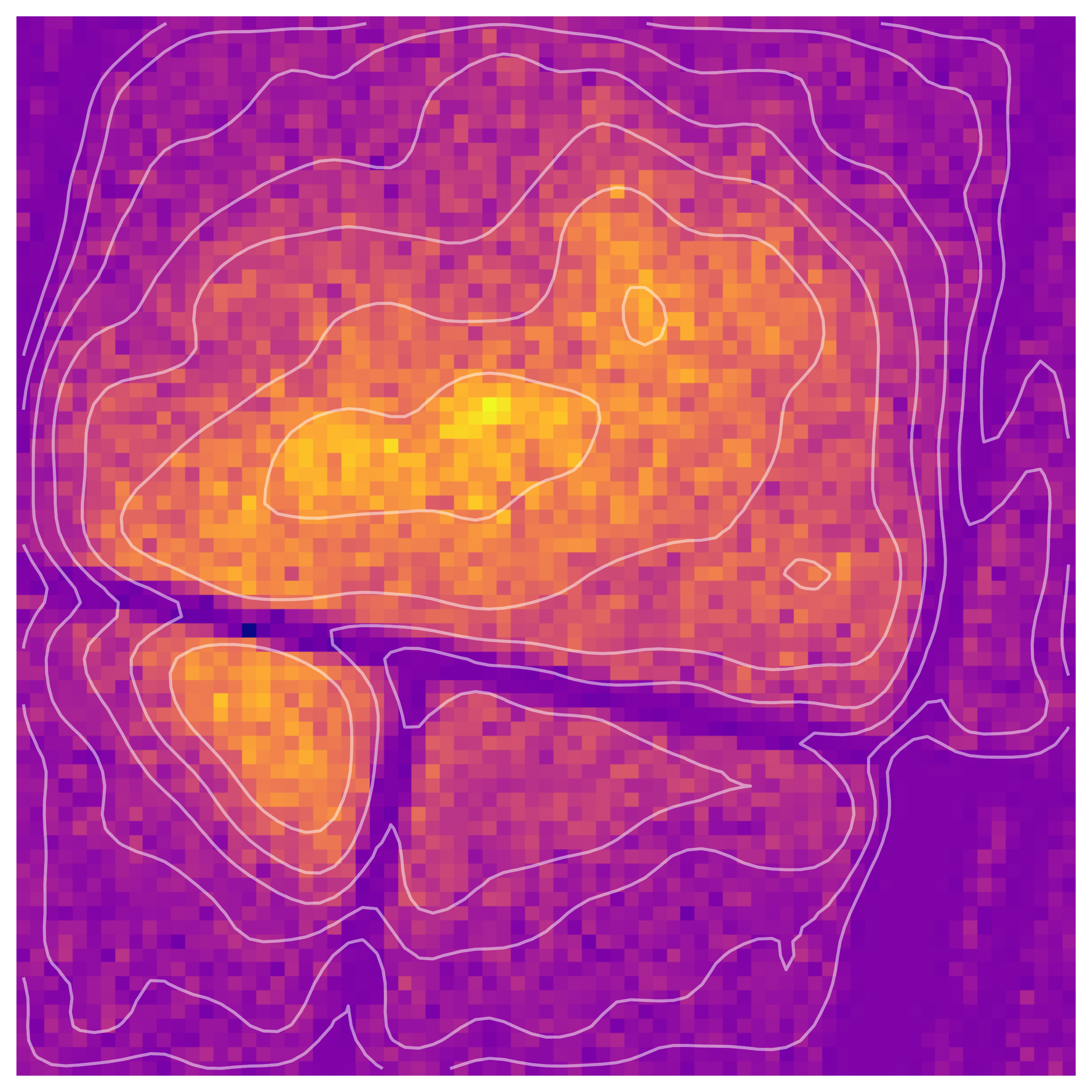}
     \includegraphics[width=0.19\linewidth]{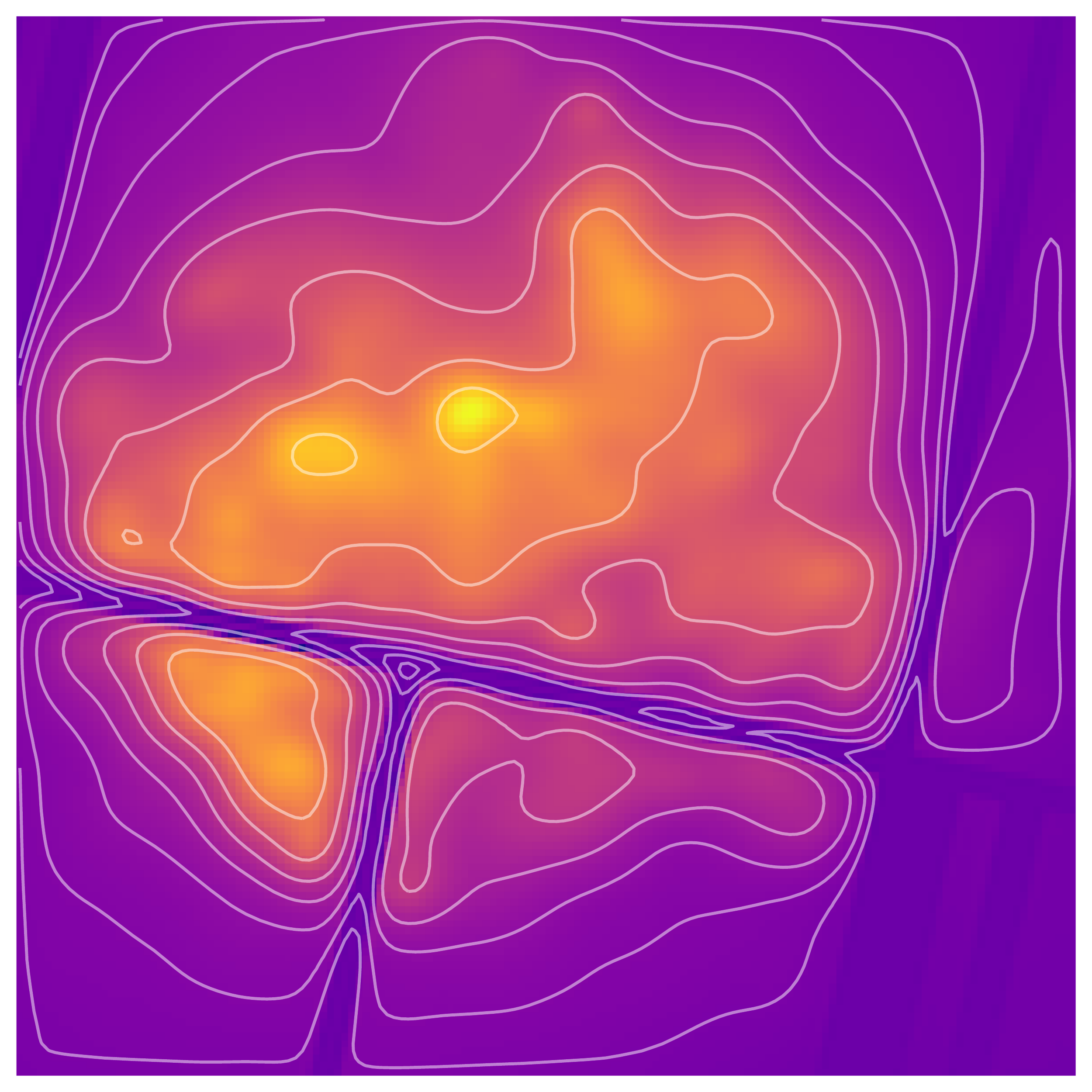}
     \includegraphics[width=0.19\linewidth]{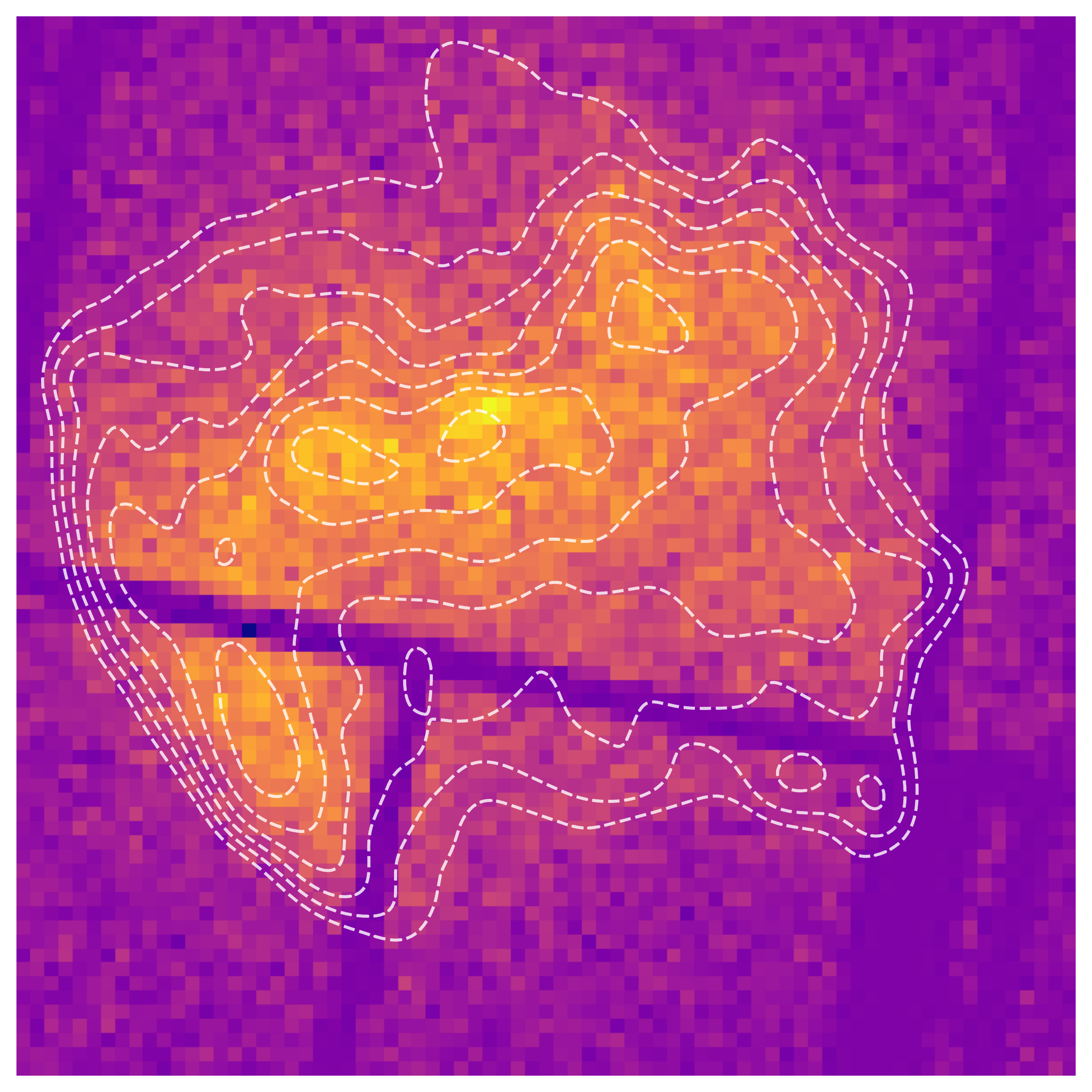}
     \includegraphics[width=0.19\linewidth]{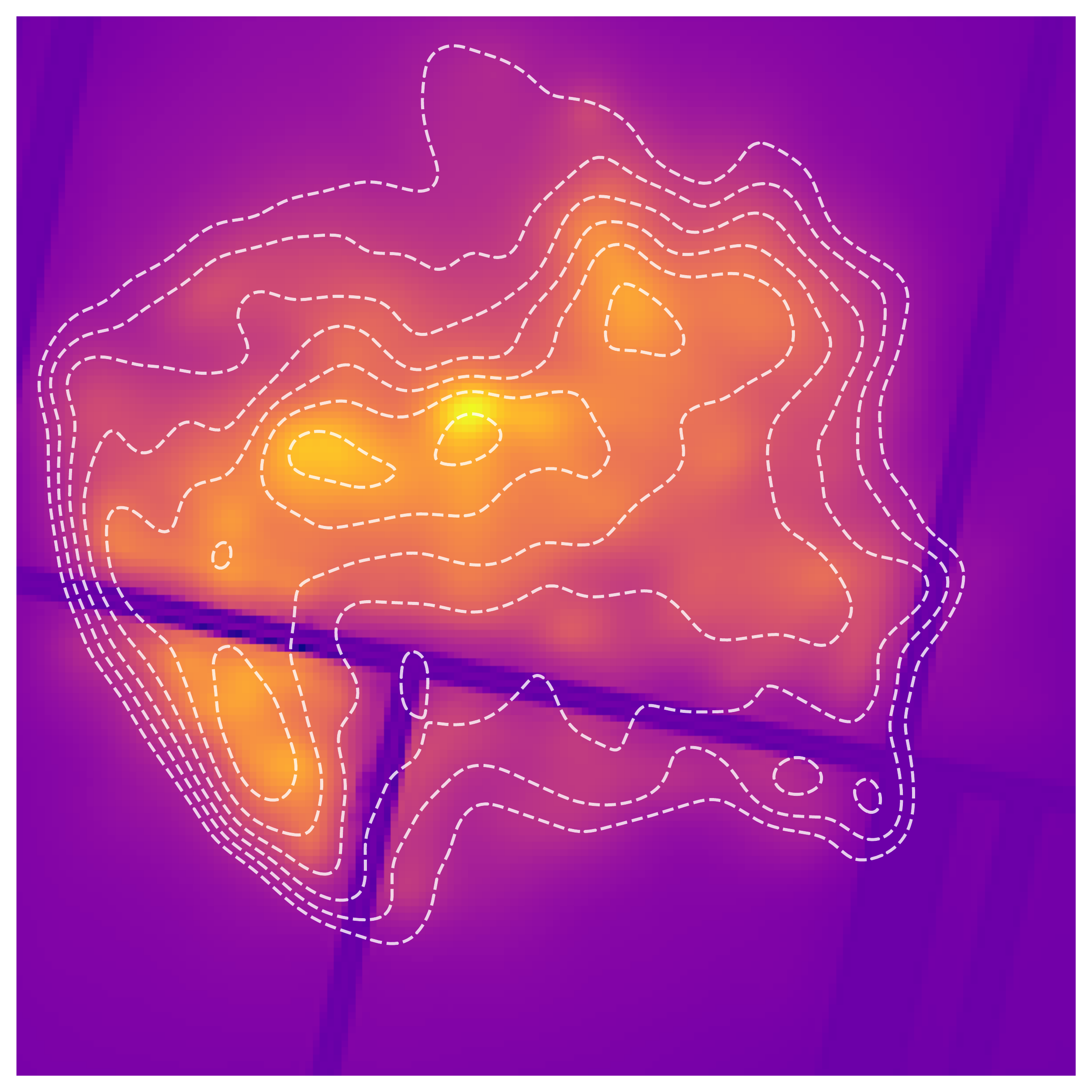}
\caption{Images of the supernova remnant W49B (SNR G043.3-00.2), with contours highlighted in white. Cropped to a frame size of $5`$. From \textit{left} to \textit{right}: \textit{Chandra} at 158ks exposure, \xmm at 20ks exposure, generated \xmm SR, \xmm at 20ks exposure overlayed with the \textit{Chandra} contours and generated \xmm SR overlayed with the \textit{Chandra} contours.} \label{fig:w49b_contour}
\end{figure*}

Our next case is supernova remnant W49B (\autoref{fig:w49b_contour}). Here, again, we see more pronounced features in the SR image in comparison to the input. The extended features on the top of the image seen in the \textit{Chandra} contour lines are better defined in the generated SR image compared with the real \xmm image.

\begin{figure*}
    \includegraphics[width=0.19\linewidth]{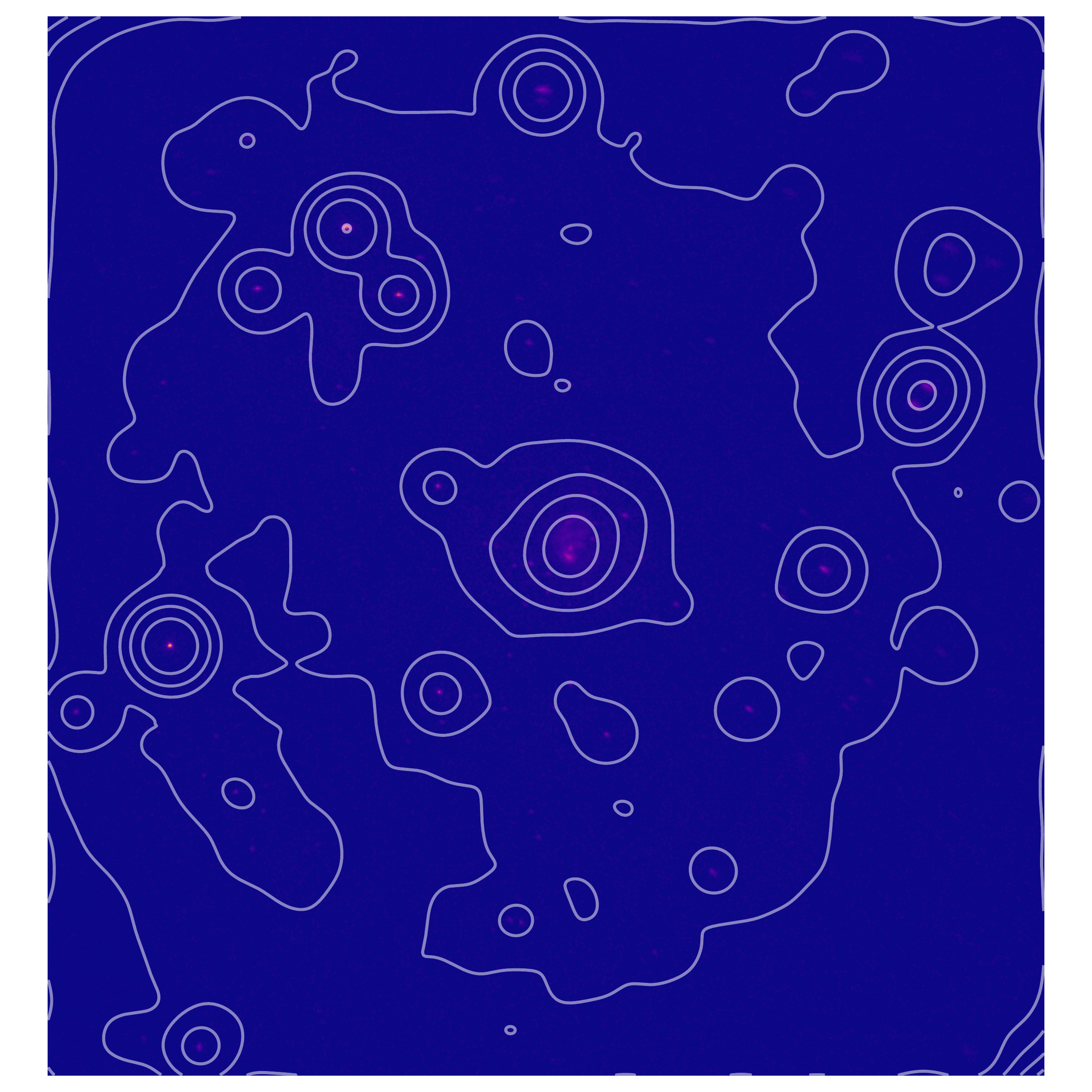}
    \includegraphics[width=0.19\linewidth]{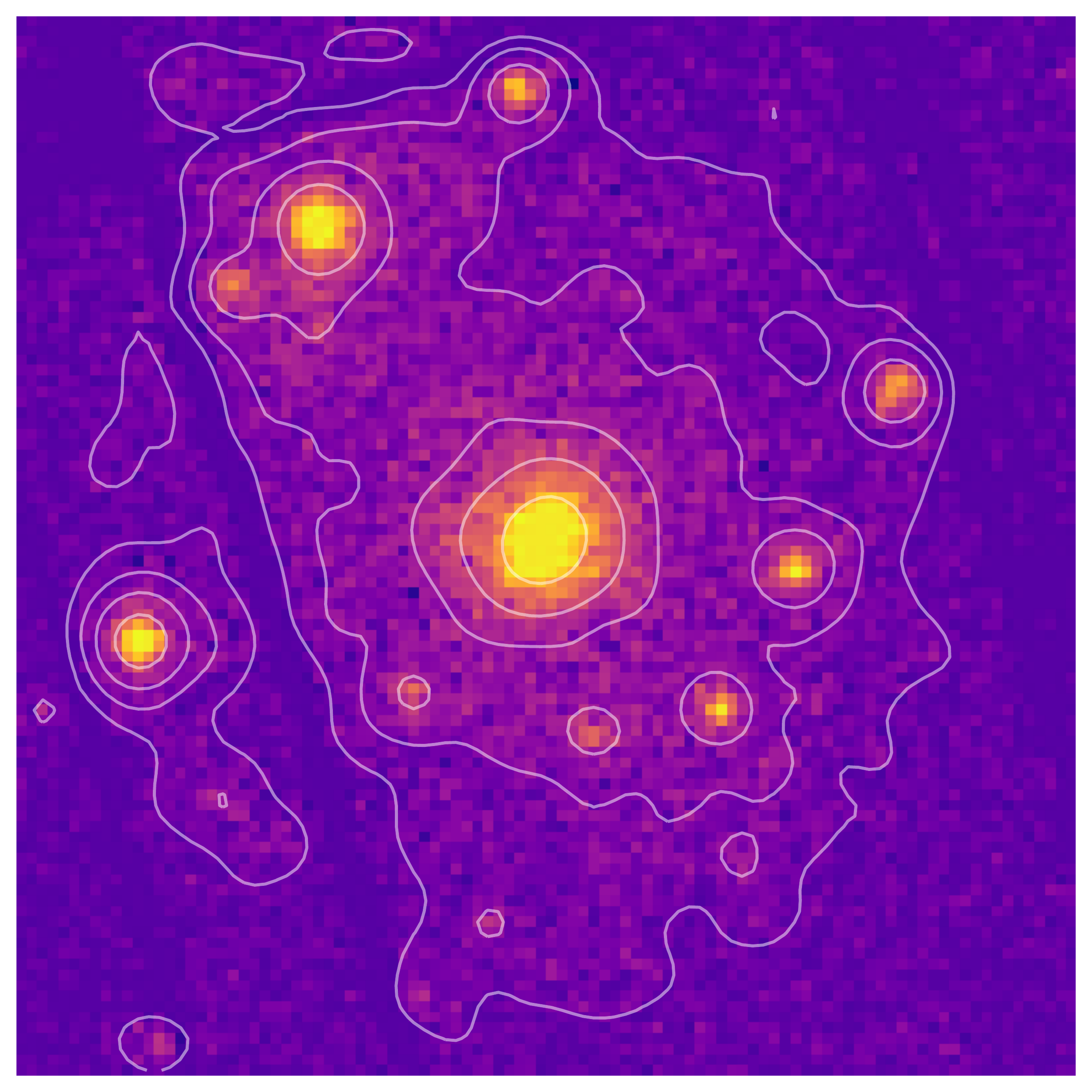}
    \includegraphics[width=0.19\linewidth]{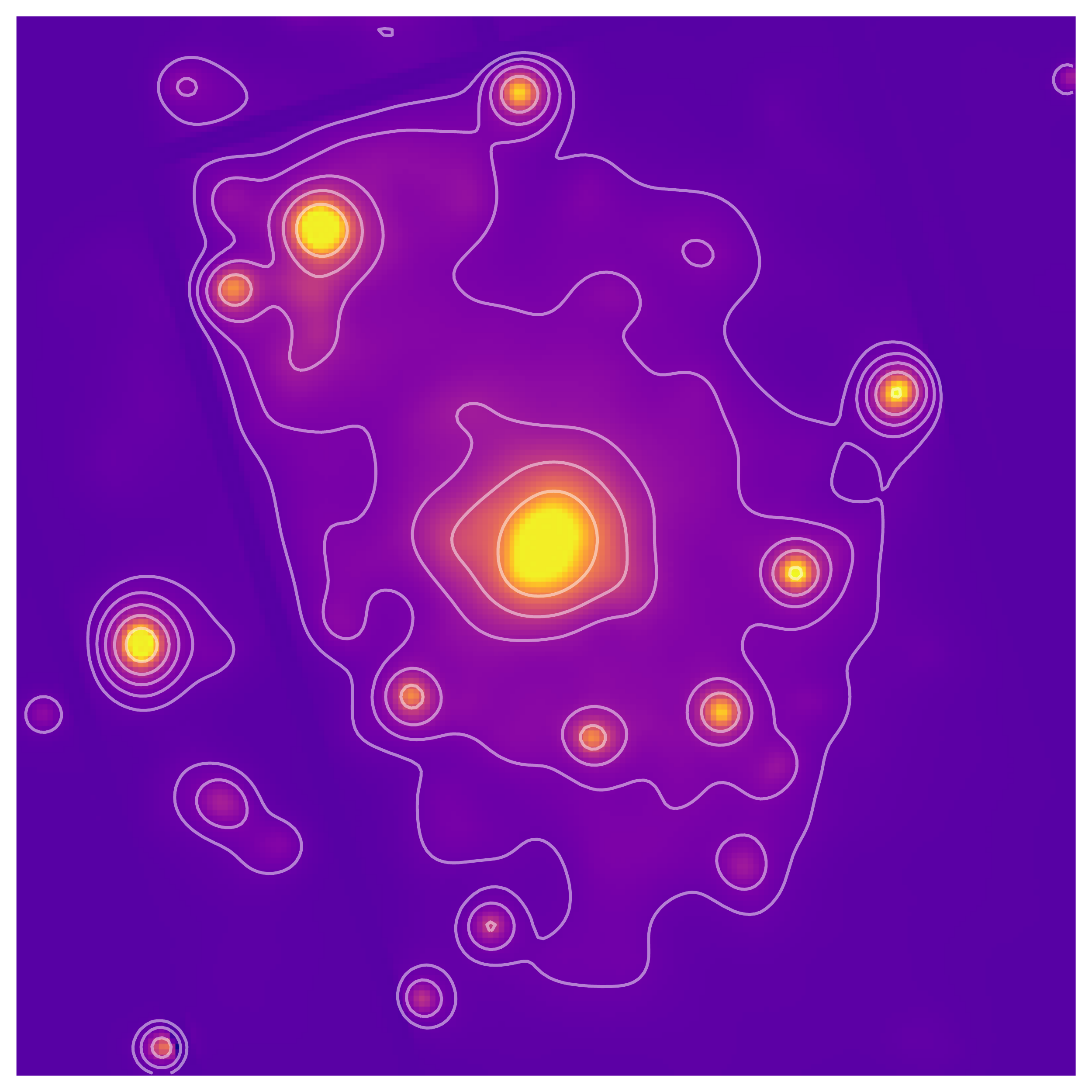}
    \includegraphics[width=0.19\linewidth]{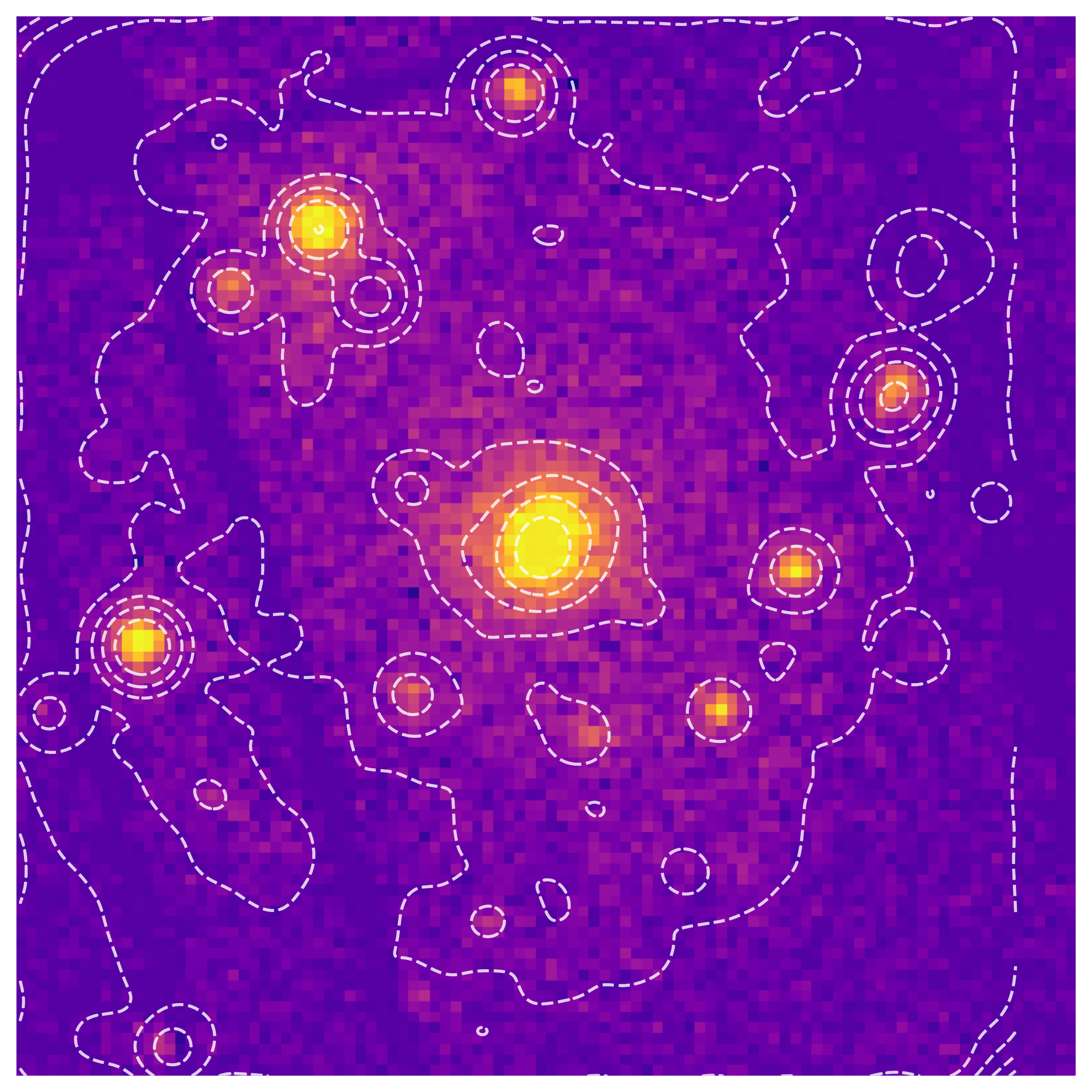}
    \includegraphics[width=0.19\linewidth]{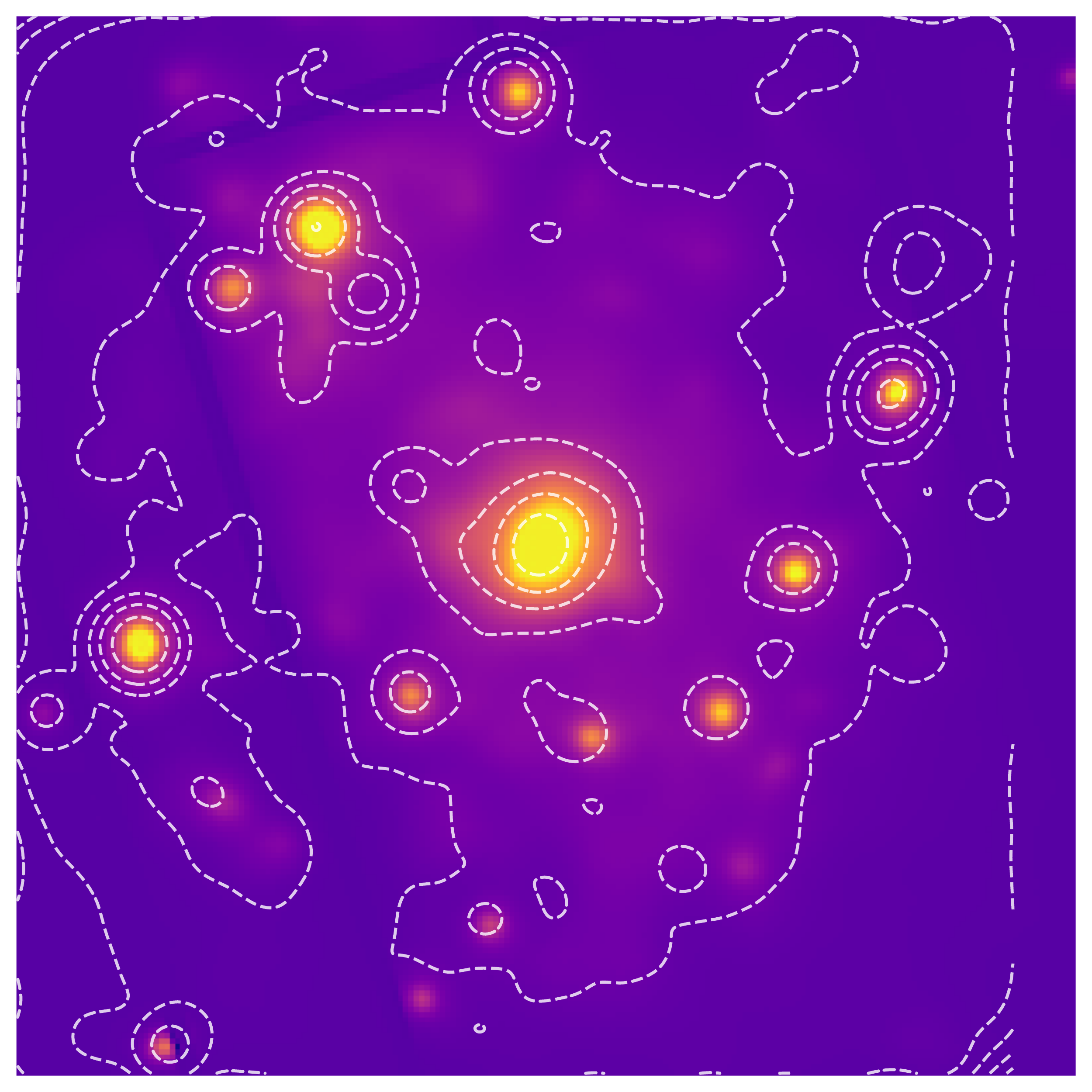}
\caption{Images of the group of galaxies M51 with with contours highlighted in white. Cropped to a frame size of $6.7`$. From \textit{left} to \textit{right}: \textit{Chandra} at 190ks exposure, \xmm at 20ks exposure, generated \xmm SR, \xmm at 20ks exposure overlayed with the \textit{Chandra} contours and generated \xmm SR overlayed with the \textit{Chandra} contours.} \label{fig:m51_contour}
\end{figure*}

Messier 51 (M\.51) is an interacting spiral galaxy with an active galactic nuclei, and it is another useful case study to see how the network performs (\autoref{fig:m51_contour}).  
In this example we see that the generated image has point sources that are better defined compared to the real \xmm image. For example, the faint source at the bottom left of the centre is clearly visible in the SR and \textit{Chandra} image but are barely in the real \xmm image. Looking at the real \xmm image and generated SR \xmm with in white the contours of \textit{Chandra} overlayed. We can also see that in the top left of the SR image an extended feature is visible that matches with the contours of \textit{Chandra}, this extended feature is harder to be seen in the real \xmm image. However, the SR image also does sometimes mis-predict features, for example on the right side the is a circular \textit{blob} visible in the \textit{Chandra} contours. In the real \xmm image it is hard to tell if there is anything present. However, the SR model predicted almost no counts in that area.

\section{Discussion}\label{sec:Discussion}
\subsection{Detector Coordinates}
Our models use the \xmm\ images in detector coordinates instead of sky coordinates. This was done to make it easier for the models to learn the image imperfections such as the chip-gaps and bad pixels, which in detector coordinates are always at the same location. The exact pipeline we used for processing real observations is described in \autoref{app:real_dataset_gen}.

\subsection{Reliability}
We have shown that our model is able to generate images with enhanced features that correspond well to features in the higher spatial resolution \textit{Chandra} observations. However, due to the nature of deep-learning-based SR and DN, our reconstructed images are susceptible to "hallucinated" features. Multiple images can be consistent as the SR/DN counterpart to a given low-resolution/noisy image, however our model only predicts one possibility. Other approaches such as flow based models \citep{kobyzev2020normalizing} allow us to actively tune the generated image to coincide with different characteristics. We explored the use of the flow based super resolution model SRFlow \citep{lugmayr2020srflow} for SR/DN of \xmm images, with promising initial results. With this approach we could tune the model to minimize artifacts, albeit with more blurry output images, or create perceptually realistic images with the compromise of an increased number of artefacts. To limit the scope of this paper, we did not continue further with this model, although it could be interesting for future research.

The goal of the metrics in \autoref{fig:denoise_metrics} and \autoref{tab:sr_sim} is to compare the models to each other and the input.
We did not calculate the error on the metrics, e.g. by retraining with different initial random weights, since it would not add any constraints on the reliability of the individually generated outputs, because of the diversity of their contents. In addition, during the hyper-parameter tuning we observed similar performance of similar models within narrow margins of the validation loss (\autoref{fig:model_param_lr_filtered}) and therefore we expect that \textit{XMM-SuperRes} and \textit{XMM-DeNoise} will have similar performance after retraining.

\subsection{AGN Deblending}
Our research primarily focused on extended sources and less on AGNs. However, the models perform well at enhancing faint AGNs and deblending them. Deblending allows us to resolve two sources when the spatially separation is smaller than the telescope resolution. Future research could focus on using the \textit{XMM-SuperRes} model for this purpose or even training a new SR and DN model specifically for deblending.

\subsection{Limitations}
\subsubsection{Clipping}
We clip extreme pixel values of the data used in this work to improve the training stability of our models, however this destroys information associated with bright sources. Since our main objective is to improve the resolution and de-noise to enhance the visibility of extended structures, losing detail in bright sources is an acceptable compromise. Some sources (not within our data sample) will have interesting features above our chosen clipping limit, and any user should take precaution when applying our model to sources of interest with count rates above our threshold. Whilst it is possible to retrain our models with a higher clipping threshold, we note that such action will affect other variables such as the data scaling function and loss. 

\subsubsection{Data Sample}
The current models are limited to the energy range $[0.5,2]$~keV which corresponds to the energy range of the majority of extended source emission. We would advise against applying our model to images in different energy bands as whilst the PSF is only marginally dependent on the energy range, the vignetting and noise properties on the other hand are known to be energy-dependent. 
Also, our model inputs are 20~ks exposure time observations from the \xmm EPIC-pn sensor since it has the largest effective area and, therefore, good spectral and spatial resolutions.  
We argue that the domain chosen for this research is sufficient to show the effectiveness of our proof-of-concept method. 
A future extension to this work could look into expanding the energy range, incorporating the MOS detectors, incorporating spectral information and increasing the flexibility of the input image exposure time. 

\subsubsection{Simulations}
Our simulations do not contain telescope properties such as out-of-time events. These phenomena tend to be caused by extremely bright sources. These are not the sources that we are interested in this research and therefore it is more efficient for us to negate out-of-time events in our simulator. 

\section{Conclusions}\label{sec:Conclusions}
We have developed deep-learning-based super-resolution (SR) and denoising (DN) models to enhance \textit{XMM-Newton} X-ray EPIC images. As a proof of concept, we only considered the EPIC-pn detector and images with photon events with energies in [0.5,2] keV. We increase the resolution of the observations and de-noise to improve the SNR and enhance features that are challenging to locate in the original images. 

To train the SR and DN models, we simulated EPIC-pn images with twice the nominal spatial resolution and images with larger exposure times. We explored the influence of the model architecture parameters, data pre-processing, and loss functions on the model's performance. To enhance the image quality, we proposed using a combined loss function consisting of both PSRN and MS\_SSIM. To address the problem of the high dynamical range of pixel values present in X-ray images, we implemented data-scaling with different stretch functions. We showed that using suitable data-scaling, our models generated fewer artifacts in low surface brightness areas in extended sources while preserving the details we are interested in.

Our SR and DN model (\textit{XMM-SuperRes}) generates enhanced SR and DN images with twice the spatial resolution and an improved image quality metrics as quantified by the PSNR. The network-produced images have the desired properties, such as a smaller PSF, and all the tested image quality metrics were improved when the model was applied on the test dataset. Specifically, with the simulated datasets, it improved the PSNR by $21.5\%$ and reduced the L1 by $40.3\%$. We have validated the performance of the model by applying it on real data and visually comparing with observations taken by NASA's \textit{Chandra} telescope, which has much higher spatial resolution.

We find that the model produce images that are able to enhance features with obvious counterparts in the \textit{Chandra} observations. Nevertheless, due to the nature of the reconstruction, some of the generated SR features may be spurious; hence whilst this model may be able to find and uncover interesting details, further detailed analysis and ideally follow-up observations at higher spatial resolution will be needed for their confirmation.

Our denoising model (\textit{\textit{XMM-DeNoise}}) based on  \textit{XMM-SuperRes}, generates images with 2.5 times higher exposure without increasing the resolution. This enabled us to train and validate the model with real \xmm observations. We found that training the denoising model on simulated data first and fine-tuning it on real data resulted in the best results for most image quality metrics. \textit{XMM-DeNoise} similarly improves the quality of real \xmm  images on all measured global quality metrics. Specifically, it improved the PSNR by $8.15\%$ and reduced the L1 by $38.4\%$.

In conclusion, we have demonstrated the feasibility of using deep-learning models to improve the spatial resolution and denoising of \xmm EPIC-pn X-ray astronomy images to increase their scientific value. The \textit{XMM-SuperRes} and \textit{XMM-DeNoise} models developed in this paper could be used as a proof-of-concept to create more elaborated methods. Such as creating a model that can output a range of possible SR and DN images emphasising on different characteristics (e.g. shock fronts in supernova remnants or deblending of point sources) to more directly tackle the ill-posed nature of the problem. The next steps for future work after this pilot study are obvious: training the models with the other \xmm instruments, incorporating a set of different energy ranges and exposure times and also extend it to other current and future X-ray telescopes.

\section*{Acknowledgements}
We would like to thank the anonymous referee for the valuable and constructive comments that helped improve the paper. We also thank the \xmm Project Scientist, Norbert Schartel, for many helpful discussions during the initiation and the development of the study. We appreciate the valuable assistance from the SIXTE team. Based on observations obtained with \xmm, an ESA science mission with instruments and contributions directly funded by ESA Member States and NASA. The IllustrisTNG simulations were undertaken with compute time awarded by the Gauss Centre for Supercomputing (GCS) under GCS Large-Scale Projects GCS-ILLU and GCS-DWAR on the GCS share of the supercomputer Hazel Hen at the High Performance Computing Center Stuttgart (HLRS), as well as on the machines of the Max Planck Computing and Data Facility (MPCDF) in Garching, Germany. SFS acknowledges support from the ESA traineeship program and further support from ESA through the Science Faculty of the European Space Astronomy Centre (ESAC) --- Funding reference ESA-SCI-SC-LE-033. ML acknowledges a Machine Learning and Cosmology research fellowship from the University of Nottingham. 

\section*{Data Availability}
More detail about our data generation process is provided in \autoref{app:real_dataset_gen} and \autoref{app:simulation}. Data will be available on request. Our code is publicly available. The EPIC-pn simulator and dataset generation code is available on \url{https://github.com/SamSweere/xmm-epicpn-simulator} and the code to train and run inference on the SR and DN models is available on \url{https://github.com/SamSweere/xmm-superres-denoise}, as well as more implementation details in the SFS's Master's thesis which is included in the GitHub folder. 


\bibliographystyle{mnras}
\bibliography{sources} 

\begin{thebibliography}{}
\makeatletter
\relax
\def\mn@urlcharsother{\let\do\@makeother \do\$\do\&\do\#\do\^\do\_\do\%\do\~}
\def\mn@doi{\begingroup\mn@urlcharsother \@ifnextchar [ {\mn@doi@}
  {\mn@doi@[]}}
\def\mn@doi@[#1]#2{\def\@tempa{#1}\ifx\@tempa\@empty \href
  {http://dx.doi.org/#2} {doi:#2}\else \href {http://dx.doi.org/#2} {#1}\fi
  \endgroup}
\def\mn@eprint#1#2{\mn@eprint@#1:#2::\@nil}
\def\mn@eprint@arXiv#1{\href {http://arxiv.org/abs/#1} {{\tt arXiv:#1}}}
\def\mn@eprint@dblp#1{\href {http://dblp.uni-trier.de/rec/bibtex/#1.xml}
  {dblp:#1}}
\def\mn@eprint@#1:#2:#3:#4\@nil{\def\@tempa {#1}\def\@tempb {#2}\def\@tempc
  {#3}\ifx \@tempc \@empty \let \@tempc \@tempb \let \@tempb \@tempa \fi \ifx
  \@tempb \@empty \def\@tempb {arXiv}\fi \@ifundefined
  {mn@eprint@\@tempb}{\@tempb:\@tempc}{\expandafter \expandafter \csname
  mn@eprint@\@tempb\endcsname \expandafter{\@tempc}}}

\bibitem[\protect\citeauthoryear{Arnaud}{Arnaud}{1996}]{arnaud1996xspec}
Arnaud K.,  1996, in Astronomical Data Analysis Software and Systems V. p.~17

\bibitem[\protect\citeauthoryear{Bourdin, Slezak, Bijaoui  \& Arnaud}{Bourdin
  et~al.}{2001}]{bourdin2001}
Bourdin H.,  Slezak E.,  Bijaoui A.,   Arnaud M.,  2001, arXiv preprint
  astro-ph/0106138

\bibitem[\protect\citeauthoryear{Carter \& Read}{Carter \&
  Read}{2007}]{carter2007xmm}
Carter J.,  Read A.,  2007, \aa, 464, 1155

\bibitem[\protect\citeauthoryear{Chen, Chen, Xu  \& Koltun}{Chen
  et~al.}{2018}]{chen2018learning}
Chen C.,  Chen Q.,  Xu J.,   Koltun V.,  2018, in Proceedings of the IEEE
  Conference on Computer Vision and Pattern Recognition. pp 3291--3300

\bibitem[\protect\citeauthoryear{Chen, He, Qing, Wu, Ren, Sheriff  \& Zhu}{Chen
  et~al.}{2022}]{chen2022real}
Chen H.,  He X.,  Qing L.,  Wu Y.,  Ren C.,  Sheriff R.~E.,   Zhu C.,  2022,
  Information Fusion, 79, 124

\bibitem[\protect\citeauthoryear{Dauser et~al.,}{Dauser
  et~al.}{2019}]{dauser2019sixte}
Dauser T.,  et~al., 2019, \aap, 630, A66

\bibitem[\protect\citeauthoryear{Dong, Loy, He  \& Tang}{Dong
  et~al.}{2014}]{dong2014learning}
Dong C.,  Loy C.~C.,  He K.,   Tang X.,  2014, in European conference on
  computer vision. pp 184--199

\bibitem[\protect\citeauthoryear{Faccioli et~al.,}{Faccioli
  et~al.}{2018}]{faccioli2018xxl}
Faccioli L.,  et~al., 2018, Astronomy \& Astrophysics, 620, A9

\bibitem[\protect\citeauthoryear{{Feng}, {Chen}, {Zhang}, {Lu}  \& {Li}}{{Feng}
  et~al.}{2003}]{Feng2003}
{Feng} H.,  {Chen} Y.,  {Zhang} S.~N.,  {Lu} F.~J.,   {Li} T.~P.,  2003,
  \mn@doi [\aap] {10.1051/0004-6361:20030324}, \href
  {https://ui.adsabs.harvard.edu/abs/2003A&A...402.1151F} {402, 1151}

\bibitem[\protect\citeauthoryear{Gilli, Comastri  \& Hasinger}{Gilli
  et~al.}{2007}]{gilli2007synthesis}
Gilli R.,  Comastri A.,   Hasinger G.,  2007, \aap, 463, 79

\bibitem[\protect\citeauthoryear{Goodfellow, Pouget-Abadie, Mirza, Xu,
  Warde-Farley, Ozair, Courville  \& Bengio}{Goodfellow
  et~al.}{2014}]{goodfellow2014}
Goodfellow I.,  Pouget-Abadie J.,  Mirza M.,  Xu B.,  Warde-Farley D.,  Ozair
  S.,  Courville A.,   Bengio Y.,  2014, Advances in neural information
  processing systems, 27

\bibitem[\protect\citeauthoryear{Iandola, Moskewicz, Karayev, Girshick, Darrell
   \& Keutzer}{Iandola et~al.}{2014}]{iandola2014}
Iandola F.,  Moskewicz M.,  Karayev S.,  Girshick R.,  Darrell T.,   Keutzer
  K.,  2014, arXiv preprint arXiv:1404.1869

\bibitem[\protect\citeauthoryear{Jain \& Seung}{Jain \&
  Seung}{2008}]{jain2008natural}
Jain V.,  Seung S.,  2008, Advances in neural information processing systems,
  21

\bibitem[\protect\citeauthoryear{Jansen et~al.,}{Jansen
  et~al.}{2001}]{jansen2001}
Jansen F.,  et~al., 2001, \aap, 365, L1

\bibitem[\protect\citeauthoryear{Johnson, Alahi  \& Fei-Fei}{Johnson
  et~al.}{2016}]{johnson2016perceptual}
Johnson J.,  Alahi A.,   Fei-Fei L.,  2016, in European conference on computer
  vision. pp 694--711

\bibitem[\protect\citeauthoryear{Kingma \& Ba}{Kingma \&
  Ba}{2014}]{kingma2014adam}
Kingma D.~P.,  Ba J.,  2014, arXiv preprint arXiv:1412.6980

\bibitem[\protect\citeauthoryear{Kobyzev, Prince  \& Brubaker}{Kobyzev
  et~al.}{2020}]{kobyzev2020normalizing}
Kobyzev I.,  Prince S.~J.,   Brubaker M.~A.,  2020, IEEE transactions on
  pattern analysis and machine intelligence, 43, 3964

\bibitem[\protect\citeauthoryear{Lauritsen, Dickinson, Bromley, Serjeant, Lim,
  Gao  \& Wang}{Lauritsen et~al.}{2021}]{lauritsen2021super}
Lauritsen L.,  Dickinson H.,  Bromley J.,  Serjeant S.,  Lim C.-F.,  Gao Z.-K.,
    Wang W.-H.,  2021, arXiv preprint arXiv:2102.06222

\bibitem[\protect\citeauthoryear{LeCun, Boser, Denker, Henderson, Howard,
  Hubbard  \& Jackel}{LeCun et~al.}{1989}]{LeCun1989}
LeCun Y.,  Boser B.,  Denker J.~S.,  Henderson D.,  Howard R.~E.,  Hubbard W.,
   Jackel L.~D.,  1989, Neural computation, 1, 541

\bibitem[\protect\citeauthoryear{LeCun, Bottou, Bengio  \& Haffner}{LeCun
  et~al.}{1998}]{LeCun1998}
LeCun Y.,  Bottou L.,  Bengio Y.,   Haffner P.,  1998, Proceedings of the IEEE,
  86, 2278

\bibitem[\protect\citeauthoryear{Ledig et~al.,}{Ledig
  et~al.}{2017}]{Ledig_2017_CVPR}
Ledig C.,  et~al., 2017, in Proceedings of the IEEE Conference on Computer
  Vision and Pattern Recognition (CVPR).

\bibitem[\protect\citeauthoryear{Li, Peng, Bhanu, Zhang  \& He}{Li
  et~al.}{2018}]{li2018}
Li Z.,  Peng Q.,  Bhanu B.,  Zhang Q.,   He H.,  2018, \apss, 363, 1

\bibitem[\protect\citeauthoryear{Lugmayr, Danelljan, Van~Gool  \&
  Timofte}{Lugmayr et~al.}{2020}]{lugmayr2020srflow}
Lugmayr A.,  Danelljan M.,  Van~Gool L.,   Timofte R.,  2020, in European
  Conference on Computer Vision. pp 715--732

\bibitem[\protect\citeauthoryear{Marinacci et~al.,}{Marinacci
  et~al.}{2018}]{marinacci2018first}
Marinacci F.,  et~al., 2018, \mnras, 480, 5113

\bibitem[\protect\citeauthoryear{Naiman et~al.,}{Naiman
  et~al.}{2018}]{naiman2018first}
Naiman J.~P.,  et~al., 2018, \mnras, 477, 1206

\bibitem[\protect\citeauthoryear{Nelson et~al.,}{Nelson
  et~al.}{2018}]{nelson2018first}
Nelson D.,  et~al., 2018, \mnras, 475, 624

\bibitem[\protect\citeauthoryear{Nelson et~al.,}{Nelson
  et~al.}{2019}]{nelson2019first}
Nelson D.,  et~al., 2019, \mnras, 490, 3234

\bibitem[\protect\citeauthoryear{Pillepich et~al.,}{Pillepich
  et~al.}{2018}]{pillepich2018first}
Pillepich A.,  et~al., 2018, \mnras, 475, 648

\bibitem[\protect\citeauthoryear{Pillepich et~al.,}{Pillepich
  et~al.}{2019}]{pillepich2019first}
Pillepich A.,  et~al., 2019, \mnras, 490, 3196

\bibitem[\protect\citeauthoryear{Puschmann \& Kneer}{Puschmann \&
  Kneer}{2005}]{puschmann2005}
Puschmann K.~G.,  Kneer F.,  2005, \aap, 436, 373

\bibitem[\protect\citeauthoryear{Reisenhofer, Bosse, Kutyniok  \&
  Wiegand}{Reisenhofer et~al.}{2018}]{reisenhofer2018haar}
Reisenhofer R.,  Bosse S.,  Kutyniok G.,   Wiegand T.,  2018, Signal
  Processing: Image Communication, 61, 33

\bibitem[\protect\citeauthoryear{Sanders \& Fabian}{Sanders \&
  Fabian}{2001}]{Sanders2001}
Sanders J.,  Fabian A.,  2001, \mnras, 325, 178

\bibitem[\protect\citeauthoryear{{Sanders}, {Fabian}, {Russell}, {Walker}  \&
  {Blundell}}{{Sanders} et~al.}{2016}]{sanders2016}
{Sanders} J.~S.,  {Fabian} A.~C.,  {Russell} H.~R.,  {Walker} S.~A.,
  {Blundell} K.~M.,  2016, \mn@doi [\mnras] {10.1093/mnras/stw1119}, \href
  {https://ui.adsabs.harvard.edu/abs/2016MNRAS.460.1898S} {460, 1898}

\bibitem[\protect\citeauthoryear{Santos-Lleo, Schartel, Tananbaum, Tucker  \&
  Weisskopf}{Santos-Lleo et~al.}{2009}]{santos2009}
Santos-Lleo M.,  Schartel N.,  Tananbaum H.,  Tucker W.,   Weisskopf M.~C.,
  2009, \nat, 462, 997

\bibitem[\protect\citeauthoryear{Schawinski, Zhang, Zhang, Fowler  \&
  Santhanam}{Schawinski et~al.}{2017}]{Schawinski_2017}
Schawinski K.,  Zhang C.,  Zhang H.,  Fowler L.,   Santhanam G.~K.,  2017,
  \mn@doi [\mnras] {10.1093/mnrasl/slx008}, p. slx008

\bibitem[\protect\citeauthoryear{Shi, Caballero, Husz{\'a}r, Totz, Aitken,
  Bishop, Rueckert  \& Wang}{Shi et~al.}{2016}]{shi2016real}
Shi W.,  Caballero J.,  Husz{\'a}r F.,  Totz J.,  Aitken A.~P.,  Bishop R.,
  Rueckert D.,   Wang Z.,  2016, in Proceedings of the IEEE conference on
  computer vision and pattern recognition. pp 1874--1883

\bibitem[\protect\citeauthoryear{Siu \& Hung}{Siu \& Hung}{2012}]{siu2012}
Siu W.-C.,  Hung K.-W.,  2012, in Proceedings of The 2012 Asia Pacific Signal
  and Information Processing Association Annual Summit and Conference. pp 1--10

\bibitem[\protect\citeauthoryear{Springel et~al.,}{Springel
  et~al.}{2018}]{springel2018first}
Springel V.,  et~al., 2018, \mnras, 475, 676

\bibitem[\protect\citeauthoryear{Starck, Pantin  \& Murtagh}{Starck
  et~al.}{2002}]{starck2002deconvolution}
Starck J.-L.,  Pantin E.,   Murtagh F.,  2002, \pasp, 114, 1051

\bibitem[\protect\citeauthoryear{Str{\"u}der et~al.,}{Str{\"u}der
  et~al.}{2001}]{Struder2001}
Str{\"u}der L.,  et~al., 2001, \aap, 365, L18

\bibitem[\protect\citeauthoryear{Su et~al.,}{Su et~al.}{2020}]{su2020deep}
Su Y.,  et~al., 2020, Monthly Notices of the Royal Astronomical Society, 498,
  5620

\bibitem[\protect\citeauthoryear{Tan, Sun, Kong, Zhang, Yang  \& Liu}{Tan
  et~al.}{2018}]{tan2018survey}
Tan C.,  Sun F.,  Kong T.,  Zhang W.,  Yang C.,   Liu C.,  2018, in
  International conference on artificial neural networks. pp 270--279

\bibitem[\protect\citeauthoryear{Turner et~al.,}{Turner
  et~al.}{2001}]{Turner2001}
Turner M.~J.,  et~al., 2001, Astronomy \& Astrophysics, 365, L27

\bibitem[\protect\citeauthoryear{{Valtchanov}, {Pierre}  \&
  {Gastaud}}{{Valtchanov} et~al.}{2001}]{valtchanov01}
{Valtchanov} I.,  {Pierre} M.,   {Gastaud} R.,  2001, \mn@doi [\aap]
  {10.1051/0004-6361:20010264}, \href
  {https://ui.adsabs.harvard.edu/abs/2001A&A...370..689V} {370, 689}

\bibitem[\protect\citeauthoryear{Vojtekova, Lieu, Valtchanov, Altieri, Old,
  Chen  \& Hroch}{Vojtekova et~al.}{2020}]{vojtekova2020learning}
Vojtekova A.,  Lieu M.,  Valtchanov I.,  Altieri B.,  Old L.,  Chen Q.,   Hroch
  F.,  2020, \mnras

\bibitem[\protect\citeauthoryear{Wang, Simoncelli  \& Bovik}{Wang
  et~al.}{2003}]{wang2003multiscale}
Wang Z.,  Simoncelli E.~P.,   Bovik A.~C.,  2003, in The Thrity-Seventh
  Asilomar Conference on Signals, Systems \& Computers, 2003. pp 1398--1402

\bibitem[\protect\citeauthoryear{Wang, Yu, Wu, Gu, Liu, Dong, Qiao  \&
  Change~Loy}{Wang et~al.}{2018}]{wang2018esrgan}
Wang X.,  Yu K.,  Wu S.,  Gu J.,  Liu Y.,  Dong C.,  Qiao Y.,   Change~Loy C.,
  2018, in Proceedings of the European conference on computer vision (ECCV)
  workshops. pp~0--0

\bibitem[\protect\citeauthoryear{Wang, Chen  \& Hoi}{Wang
  et~al.}{2020}]{wang2020deep}
Wang Z.,  Chen J.,   Hoi S.~C.,  2020, IEEE transactions on pattern analysis
  and machine intelligence

\bibitem[\protect\citeauthoryear{Weisskopf, Tananbaum, Van~Speybroeck  \&
  O'Dell}{Weisskopf et~al.}{2000}]{weisskopf2000}
Weisskopf M.~C.,  Tananbaum H.~D.,  Van~Speybroeck L.~P.,   O'Dell S.~L.,
  2000, in X-Ray Optics, Instruments, and Missions III. pp 2--16

\bibitem[\protect\citeauthoryear{Wells \& Greisen}{Wells \&
  Greisen}{1979}]{wells1979fits}
Wells D.~C.,  Greisen E.~W.,  1979, in Image Processing in Astronomy. p.~445

\bibitem[\protect\citeauthoryear{Wilkins, Gallo, Costantini, Brandt  \&
  Blandford}{Wilkins et~al.}{2021}]{wilkins2021}
Wilkins D.,  Gallo L.,  Costantini E.,  Brandt W.,   Blandford R.,  2021,
  Nature, 595, 657

\bibitem[\protect\citeauthoryear{Xu, Ramos-Ceja, Pacaud, Reiprich  \& Erben}{Xu
  et~al.}{2018}]{xu2018new}
Xu W.,  Ramos-Ceja M.~E.,  Pacaud F.,  Reiprich T.~H.,   Erben T.,  2018,
  Astronomy \& Astrophysics, 619, A162

\bibitem[\protect\citeauthoryear{Yang, Zhang, Tian, Wang, Xue  \& Liao}{Yang
  et~al.}{2019}]{yang2019}
Yang W.,  Zhang X.,  Tian Y.,  Wang W.,  Xue J.-H.,   Liao Q.,  2019, IEEE
  Transactions on Multimedia, 21, 3106

\bibitem[\protect\citeauthoryear{Zhang, Zhang, Mou  \& Zhang}{Zhang
  et~al.}{2011}]{zhang2011fsim}
Zhang L.,  Zhang L.,  Mou X.,   Zhang D.,  2011, IEEE transactions on Image
  Processing, 20, 2378

\bibitem[\protect\citeauthoryear{Zhang, Chen, Ng  \& Koltun}{Zhang
  et~al.}{2019}]{zhang2019zoom}
Zhang X.,  Chen Q.,  Ng R.,   Koltun V.,  2019, in Proceedings of the IEEE/CVF
  Conference on Computer Vision and Pattern Recognition. pp 3762--3770

\bibitem[\protect\citeauthoryear{Zhang, Tian, Kong, Zhong  \& Fu}{Zhang
  et~al.}{2020a}]{zhang2020residual}
Zhang Y.,  Tian Y.,  Kong Y.,  Zhong B.,   Fu Y.,  2020a, IEEE Transactions on
  Pattern Analysis and Machine Intelligence, 43, 2480

\bibitem[\protect\citeauthoryear{Zhang, Ramos-Ceja, Pacaud  \& Reiprich}{Zhang
  et~al.}{2020b}]{zhang2020high}
Zhang C.,  Ramos-Ceja M.~E.,  Pacaud F.,   Reiprich T.~H.,  2020b, Astronomy \&
  Astrophysics, 642, A17

\bibitem[\protect\citeauthoryear{{Zhou Wang}, {Bovik}, {Sheikh}  \&
  {Simoncelli}}{{Zhou Wang} et~al.}{2004}]{wang2004SSIM}
{Zhou Wang} {Bovik} A.~C.,  {Sheikh} H.~R.,   {Simoncelli} E.~P.,  2004,
  \mn@doi [IEEE Transactions on Image Processing] {10.1109/TIP.2003.819861},
  13, 600

\bibitem[\protect\citeauthoryear{Zhou, Yang  \& Liao}{Zhou
  et~al.}{2012}]{zhou2012}
Zhou F.,  Yang W.,   Liao Q.,  2012, IEEE Transactions on Image Processing, 21,
  3312

\makeatother
\end{thebibliography}



\appendix
\begin{appendices}

\section{Real XMM-Newton Dataset Generation}\label{app:real_dataset_gen}

The real \xmm dataset was created using standard workflow.

\begin{itemize}
    \item We start off with the full \xmm archive, containing all the historical observations.
    
    \item First, we select all observations of at least $20ks$ observation time using either full-frame or extended full-frame mode. These modes use all the 12 EPIC-pn CCDs. 
    \item We use the \xmm pipeline produced products (PPS) for calibrated eventlists. Sometimes, an observation is split into different exposures and we select the one with the longest on-time.
    \item We filter the eventlist by removing intervals of high background. We use the PPS-generated light-curve and the PPS-derived threshold and we apply the following filtering expression to produce cleaned event lists:
    
    \texttt{PI > 300 \&\& RAWY > 12 \&\& PATTERN <= 4 \&\& \\ 
    \hspace*{1cm}((FLAG == 0) || ((FLAG \& 0x10000) != 0))}
    
    We keep the out of field-of-view events (flagged with \texttt{0x10000}) in the corners to control the instrumental background and compare with SIXTE simulations.
    \item We convert this eventlist into smaller eventlists of different exposure times with increments of 10~ks. The biggest exposure time depends on the exposure time of the cleaned eventlist. I.e. if we have a clean exposure of 40~ks, we will generate 4x10~ks images, 2x20~ks images, 1x30~ks, and 1x40~ks images. The images with multiple exposure times make the dataset more flexible to use. It also enables us to train a de-noising model with low and high exposure image pairs.
    \item Finally, using the cleaned event lists we create images in detector coordinates. We use the default binsize of 80 (4"/pix). The final image is saved in the FITS format \cite{wells1979fits}.
\end{itemize}

\section{Simulation Setup Details}\label{app:simulation}

\subsection{EPIC-pn Image Simulation}

The EPIC-pn sensor consists of 12 CCDs. Initially, we simulated all these separately based on the physical properties, including out-of-time events. However, this significantly increased the computation time of the simulator by a factor of 12 since the whole telescope has to be simulated separately for each sensor. A single observation would require 12 separate simulations. The benefits of simulating every sensor separately are that specific properties such as out-of-time events are simulated as well. However, the sources we are interested in are usually not extremely bright, which is the main cause of out-of-time events. The impact of not having these properties is minimal on the final image. 

Therefore we simulate all 12 sensors as one big sensor, without out-of-time events. Since we now do not have any chip gaps in-between the CCDs, we multiply the final image with the \xmm\ detector mask. The detector mask filters out all the areas in the image where no recording of events is possible, including the chip gaps, known bad pixels, and areas outside the field of view. Additionally we filter out over-exposed images that can result in undesirable effects. These are typically associated solar flares. We use the exposure map as a detector mask. The resolution of this exposure map matches the resolution of the observed/simulated images at 1x scale. The detector mask can therefore be used both on the real image and the simulated one. For the higher resolution we increase the resolution of the detector mask without interpolation (by repeating every pixel).

\subsection{Boresight Determination}

Note that the optical axis (also called boresight) is not exactly in the middle of the image but is slightly offset, in order to avoid a chip gap. The boresight position also changed over time. We used the information from the latest calibration file: XMM\_MISCDATA\_0022.CCF to determine the position of the optical axis for the simulations. This is important for the vignetting and the PSF, since these depend on the off-axis angle.

\subsection{PSF}\label{app:psf}
The PSF (point spread function) of the \xmm is not constant, this also needed to be simulated. 
In SIXTE, there are two PSF implementations: Using a single PSF for the whole image or setting separate PSFs for certain sections. These sections are radially distributed, centering around the boresight using a polar coordinate system. For every X-ray photon of specific energy entering the simulated telescope, SIXTE will then use the closest given PSF for that specific energy and location. 
The PSF distributions that SIXTE uses need to be provided as images. During development, this created a problem since providing many PSF images, which make the simulation more realistic, resulted in very high memory use, limiting the number of simulations we could run in parallel. 

We decided to use three different energy levels: 0.5, 1.0 and 2.0 KeV to optimize this. Use a $\phi$ degree interval of 4 degrees, and $\theta = 0, 210, 420, 600, 720, 900$, $1200$ arcsec. Resulting in 630 unique PSF images for every energy level. The PSF image resolution was set to 120x120; this is just big enough to cover the most stretched PSF at the edge of the sensor.
The PSF images were created using the \textit{psfgen} program, which is part of the official \xmm Science Analysis System (XMM-SAS).
To increase the simulation's spatial resolution, we have to decrease the PSF size. However, we do want to keep the same PSF distortion shape. Therefore, we decreased the size of the original PSF images by the resolution multiplier.

\section{Model Hyper-Parameter Tuning}\label{app:modelparameters}
To determine the model hyper-parameters we run a hyper-parameter-search. Where we fix the loss to Poisson and use square root data-scaling since this resulted in desirable results in initial testing. 
The model hyper-parameters we try and their ranges are:
\begin{itemize}
    \item Number of RRDB convolutional filters: $[8, 16, 32]$
    \item Number of RRDB blocks: $[2, 4, 8]$
    \item Learning rate: $[10^{-3}, 5\times10^{-3}, 10^{-4}, 5\times10^{-4}, 10^{-5}, 5\times10^{-5}]$
    \item Batch size: $[2, 4, 8]$
\end{itemize}
We train models for every possible combination of these hyper-parameters (180 models) and monitor the loss of the validation data. Several runs fail to converge. Some of these runs result in a validation loss greater or equal than 1.0, and only generate blank images. 

Filtering out failed and poorly performing runs (val/loss $\ge$ 0.434) still leaves us with a huge number of viable model hyper-parameter combinations. We therefore look at the correlation of the parameters with the loss. 

The batch size has the largest positive correlation with the final loss. This positive correlation indicates that the bigger the $batch\_size$, the worse the performance. After discarding runs that use a batch size of 8, we find that the next biggest correlation comes from the learning rate. It was clear that many of the failed runs were a result of high learning rates that make the training unstable but it's also known that small learning rates risk getting stuck at local minima. We therefore discard the extreme learning rates $lr > 0.0001$ and $lr < 0.00005$.

\begin{figure*}
    \centering
    \includegraphics[width=\linewidth]{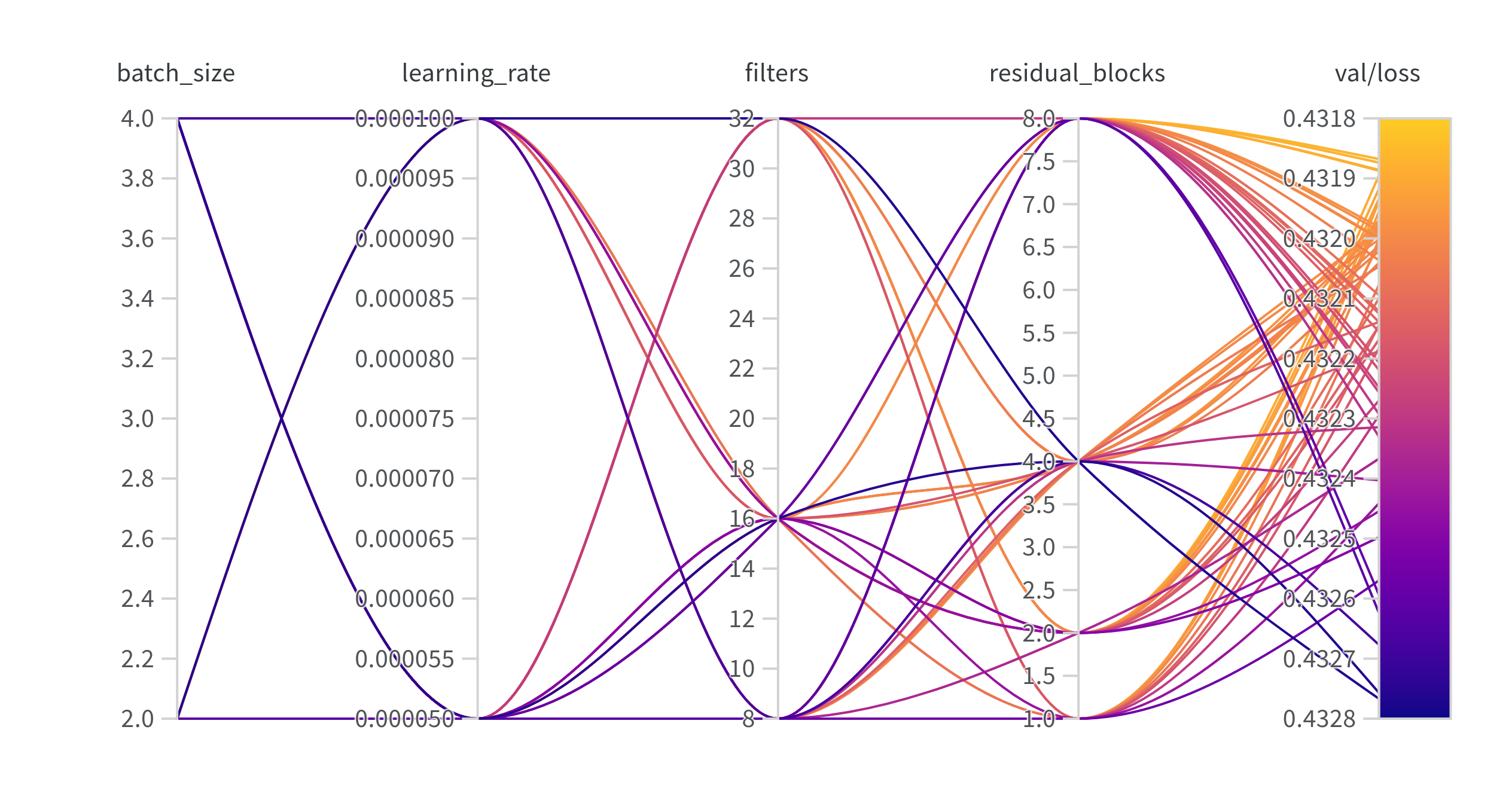} 
    \caption{Results model sweep. With $val/loss >= 0.434$, $batch\_size <= 4$ and $0.0001 >= learning\_rate >=  0.00005$.}
    \label{fig:model_param_lr_filtered}
\end{figure*}

\autoref{fig:model_param_lr_filtered} shows the remaining runs after filtering, all of which result in very similar validation loss values. Since these models are trained on a cropped image and reduced dataset size, we can assume that there is more to learn from the data and that a larger model would be more suitable for the real run. However, bigger models also take longer to train and use more GPU memory, which is a limiting factor when processing full-size images. For the final model (both SR and DN), we opt for 32 RRDB convolutional filters and 4 RRDB blocks to leave room to learn more complex data without hitting our GPU memory limitation. We opt for a batch size of 1 to similarly reduce the computational strain on the GPU and a learning rate of 0.0001. 

\section{Data Hyper-parameter tuning}\label{app:dataparameters}
To tune the data hyper-parameters, the loss function and data scaling, we fix the model hyper-parameters to the values determined in \autoref{sec:modelparameters} but with a batch size of 4 to increase the training speed. Next, we train a model with every possible combination of loss functions (\autoref{sec:loss_functions}) and data scalings (\autoref{subsec:datapreprocessing}), this results in 128 models.

The data hyper-parameters directly influence the visual properties of the generated images. Therefore, to determine their optimal values, we select the best-performing models based on the image quality metrics first and then do a qualitative visual inspection.

Each image quality metric emphasizes different visual elements in the generated image so we define our quantitative measure as a metric score that combines all the metrics into a single value. Some metrics are ascending, and others are descending so we invert ascending metrics such that all metrics are descending. Additionally, since the metrics map to different numerical scales, we apply a min-max normalization to their values before they are summed to create the combined metric score. Here, a lower combined metric score is better. Since images with different data scalings have different properties, the metrics' value is also differs. 

\begin{figure}
    \centering
    \includegraphics[width=\linewidth]{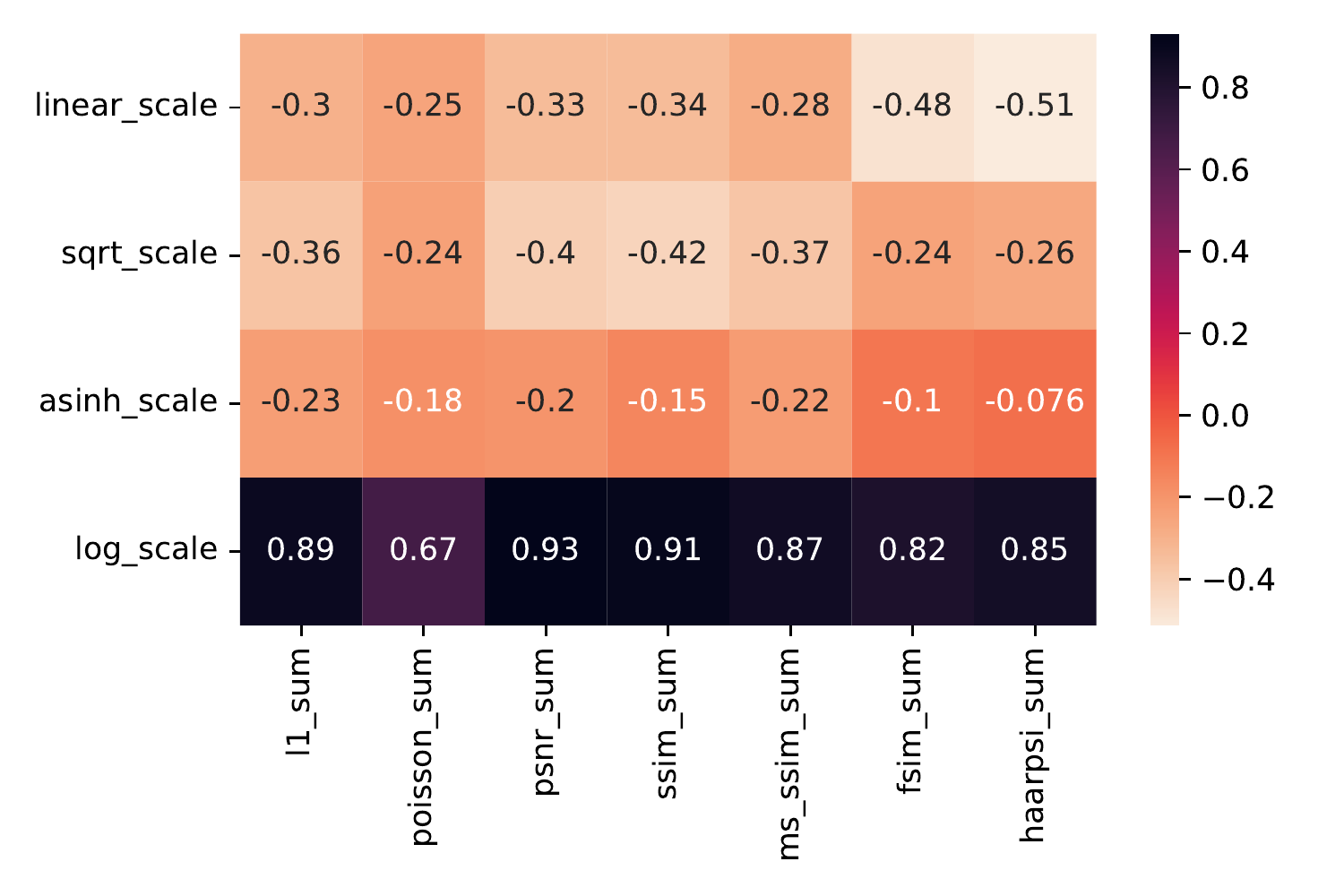} 
    \caption{Correlation matrix data scaling with respect to image quality metrics. Lower correlation indicates better performance.}
    \label{fig:overview_datascale_corr}
\end{figure}

We correlate each hyper-parameter with the combined metric score (\autoref{fig:overview_datascale_corr}) and find that the logarithmic scale correlates heavily with bad performance on all metrics. The asinh scale also under-performs with respect to the sqrt and linear scale. The sup-par performance of asinh and log is likely due to the noise level getting pushed close to the structure level, making it difficult to distinguish between the unpredictable background noise and any real features. The linear and sqrt data scalings perform the best.

Visual inspection of the linear data-scaling models show the tendency of generating patchy images that are not present in the ground truth image. These artefacts are barely visible on linear scales, but they become problematic when the image is stretched. Models trained with a sqrt data scaling suffer less from this problem, see \autoref{fig:lin_scaling_prob}, which motivates our choice of sqrt data scaling for our final model.

\begin{figure}
    \centering
    \includegraphics[width=\linewidth]{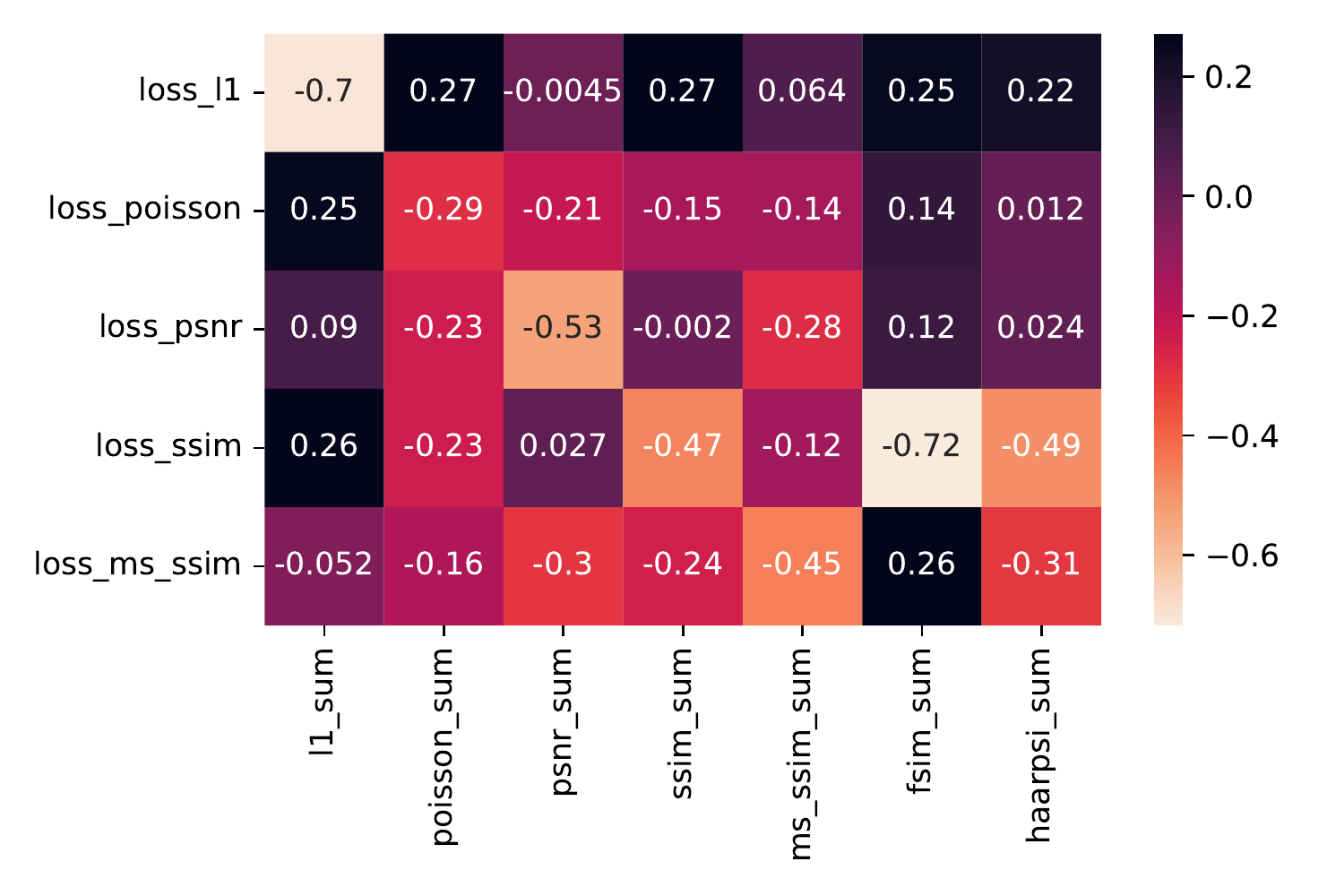} 
    \caption{Correlation of the loss function elements with respect to the summed normalized metrics, given sqrt data scaling. Lower correlation indicates better performance.}
    \label{fig:loss_corr_sqrt}
\end{figure}

Fixing the image scaling to sqrt, we then determine the optimal loss function. Again we correlate the loss function with respect to the combined metric score (\autoref{fig:loss_corr_sqrt}), and find that L1 loss only performs well with respect to the L1 evaluation metric and performs poorly with respect to all other metrics. As one might expect, the best correlations occur where the chosen loss function is also used as the evaluation metric. 

When visually inspecting the generated images we observed that models trained with the SSIM loss tend to contain overly defined structures and AGNs in comparison to the target image.
While models trained with Poisson, PSNR and MS\_SSIM where visually closer to the target images.
Models trained with L1 loss seem to suffer from a quantization problem, where there are distinct regions visible in what should be continuous distributed area.
Based on these observations and the models performance on the image quality metrics we chose to train the final model with the PSNR combined with MS\_SSIM loss function. 

\end{appendices}

\bsp	
\label{lastpage}
\end{document}